\documentclass[aps,twocolumn,longbibliography,notitlepage,superscriptaddress]{revtex4-2}
\pdfoutput=1
\usepackage{graphicx}
\usepackage{dcolumn}
\usepackage{bm}
\usepackage[version=4]{mhchem}
\usepackage{graphicx}
\usepackage{bm}
\usepackage{amssymb}
\usepackage{amsmath}
\usepackage{dcolumn}
\usepackage[caption=false]{subfig}
\usepackage{floatrow}
\usepackage{gensymb}
\usepackage{threeparttable}
\usepackage{multirow}
\usepackage[usenames,dvipsnames]{color}
\usepackage{wasysym}
\usepackage{mathtools}
\usepackage[dvipsnames]{xcolor}
\usepackage[normalem]{ulem}
\usepackage[colorlinks]{hyperref}
\hypersetup{
    colorlinks=true,
    citecolor=blue,
    linkcolor=blue,
    filecolor=magenta,      
    urlcolor=blue,
    }
\usepackage{lineno}

\newcommand{\bs}{\boldsymbol}
\makeatletter
\def\l@subsection#1#2{}
\def\l@subsubsection#1#2{}
\makeatother

    \makeatletter
\def\@fnsymbol#1{\ensuremath{\ifcase#1\or \dagger\or *\or \ddagger\or
   \mathsection\or \mathparagraph\or \|\or **\or \dagger\dagger
   \or \ddagger\ddagger \else\@ctrerr\fi}}
    \makeatother
\begin{document}

\title[title]{Unidirectional Magnetoresistance in Antiferromagnet$|$Heavy-Metal Bilayers} 

\author{Soho Shim}
\thanks{These authors contributed equally.}
\affiliation{Department of Physics, University of Illinois at Urbana-Champaign, Urbana, IL 61801, USA}
\affiliation{Materials Research Laboratory, University of Illinois at Urbana-Champaign, Urbana, IL 61801, USA}
\author{M. Mehraeen}
\thanks{These authors contributed equally.}
\affiliation{Department of Physics, Case Western Reserve University, Cleveland, OH 44106, USA}
\author{Joseph Sklenar}
\affiliation{Department of Physics and Astronomy, Wayne State University, Detroit, MI 48201, USA}
\affiliation{Department of Physics, University of Illinois at Urbana-Champaign, Urbana, IL 61801, USA}
\affiliation{Materials Research Laboratory, University of Illinois at Urbana-Champaign, Urbana, IL 61801, USA}
\author{Junseok Oh}
\affiliation{Department of Physics, University of Illinois at Urbana-Champaign, Urbana, IL 61801, USA}
\affiliation{Materials Research Laboratory, University of Illinois at Urbana-Champaign, Urbana, IL 61801, USA}
\author{Jonathan Gibbons}
\affiliation{Department of Physics, University of Illinois at Urbana-Champaign, Urbana, IL 61801, USA}
\affiliation{Materials Research Laboratory, University of Illinois at Urbana-Champaign, Urbana, IL 61801, USA}
\affiliation{Materials Science Division, Argonne National Laboratory, Lemont IL 60439, USA}
\affiliation{Department of Physics, University of California San Diego, La Jolla, CA 92093, USA}
\author{Hilal Saglam}
\affiliation{Materials Science Division, Argonne National Laboratory, Lemont IL 60439, USA}
\affiliation{Department of Applied Physics, Yale University, New Haven, CT 06511, USA}
\author{Axel Hoffmann}
\affiliation{Department of Materials Science and Engineering, University of Illinois at Urbana-Champaign, Urbana, IL 61801, USA}
\affiliation{Materials Science Division, Argonne National Laboratory, Lemont IL 60439, USA}
\affiliation{Materials Research Laboratory, University of Illinois at Urbana-Champaign, Urbana, IL 61801, USA}
\affiliation{Department of Physics, University of Illinois at Urbana-Champaign, Urbana, IL 61801, USA}
\author{Steven S.-L. Zhang}
\affiliation{Department of Physics, Case Western Reserve University, Cleveland, OH 44106, USA}
\author{Nadya Mason}
\email{nadya@illinois.edu}
\affiliation{Department of Physics, University of Illinois at Urbana-Champaign, Urbana, IL 61801, USA}
\affiliation{Materials Research Laboratory, University of Illinois at Urbana-Champaign, Urbana, IL 61801, USA}

\date{\today}

\begin{abstract}
The interplay between electronic transport and antiferromagnetic order has attracted a surge of interest as recent studies have shown that a moderate change in the spin orientation of a collinear antiferromagnet may have a significant effect on the electronic band structure. Among numerous electrical probes to read out such magnetic order, unidirectional magnetoresistance (UMR), where the resistance changes under the reversal of the current direction, can provide rich insights into the transport properties of spin-orbit coupled systems. However, UMR has never been observed in antiferromagnets before, given the absence of intrinsic spin-dependent scattering. Here, we report a UMR in the antiferromagnetic phase of a FeRh$|$Pt bilayer, which undergoes a sign change and then increases strongly with an increasing external magnetic field, in contrast to UMRs in ferromagnetic and nonmagnetic systems. We show that Rashba spin-orbit coupling alone cannot explain the sizable UMR in the antiferromagnetic bilayer and that field-induced spin canting distorts the Fermi contours to greatly enhance the UMR by two orders of magnitude. Our results can motivate the growing field of antiferromagnetic spintronics, and suggest a route to the development of tunable antiferromagnet-based spintronics devices.
\end{abstract}


\maketitle

\title{FeRhPt_UMR}
\author{jsshim}
\date{May 2021}
\floatsetup[figure]{style=plain,subcapbesideposition=top}

\section{Introduction}
Antiferromagnets$-$a broad class of magnetically ordered materials$-$possess a variety of appealing properties~\cite{jungwirth2016antiferromagnetic,baltz2018antiferromagnetic, siddiqui2020metallic} including sublattice degree of freedom, terahertz resonance, and the lack of stray field, of which their ferromagnetic counterparts are naturally devoid.  Recently, the interplay between electronic transport and the N\'{e}el order of metallic antiferromagnets has attracted a surge of interest, partly stimulated by the realization of electric control of sublattice magnetization utilizing strong spin-orbit coupling (SOC) and space inversion symmetry breaking in collinear antiferromagnetic metals~\cite{wadley2016electrical,Bodnar18NatCommn_SOT-Mn2Au,Wadley18_CuMnAs-AF-domain-current,Meinert18_AF-switching_Mn2Au,cSong18PRAppl_SOT-AF-Mn2Au,hOhno20NatCommn_switching-PtMn-Pt}. A moderate change in the spin orientation of a collinear antiferromagnet (such as spin canting) may have a significant effect on the electronic band structure and subsequently manifest itself in transport properties~\cite{Suzuki16NP_AHE-Heusler-AF,Takahashi18SciAdv_AHE-ETO-AF,arxiv19_hChen_QAHE-cAF, rYang20PRL_sc-AF-DSM,Kipp21ComPhys_CHE-cAF}. There have also been reviving efforts to develop new optical~\cite{kimel2004laser,saidl2017optical} or electrical~\cite{marti2014room,moriyama2018spin,baldrati2018full} probes to read out antiferromagnetic order. Among these efforts, two types of linear-response magnetoresistances are commonly used in transport measurements: the anisotropic magnetoresistance in metallic antiferromagnets~\cite{marti2014room, moriyama2015sequential,wadley2016electrical, moriyama2018spin} and spin Hall magnetoresistance in bilayers consisting of an insulating antiferromagnet and a heavy-metal~\cite{hoogeboom2017negative, fischer2018spin, baldrati2018full,manchon2017spin}, both of which have analogs in ferromagnetic systems~\cite{mcguire1975anisotropic,nakayama2013spin}. 

\begin{figure*}[bht]
    \sidesubfloat[c]{\includegraphics[width=0.3\linewidth,trim={.5cm 1cm 2cm 0cm}]{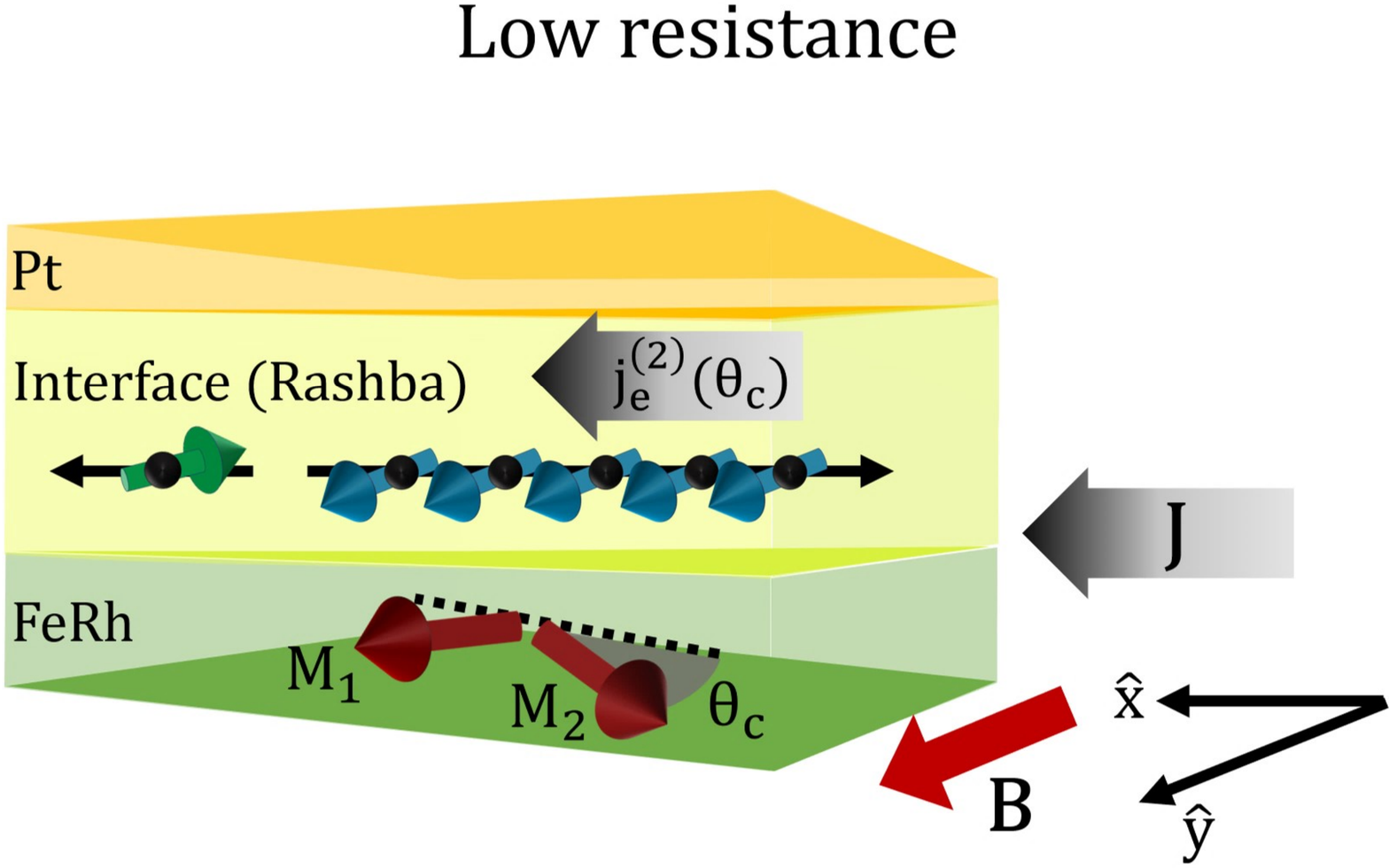}\label{fig:1_a}}\quad%
    \sidesubfloat[c]{\includegraphics[width=0.3\linewidth,trim={.5cm 1cm 2cm 0cm}]{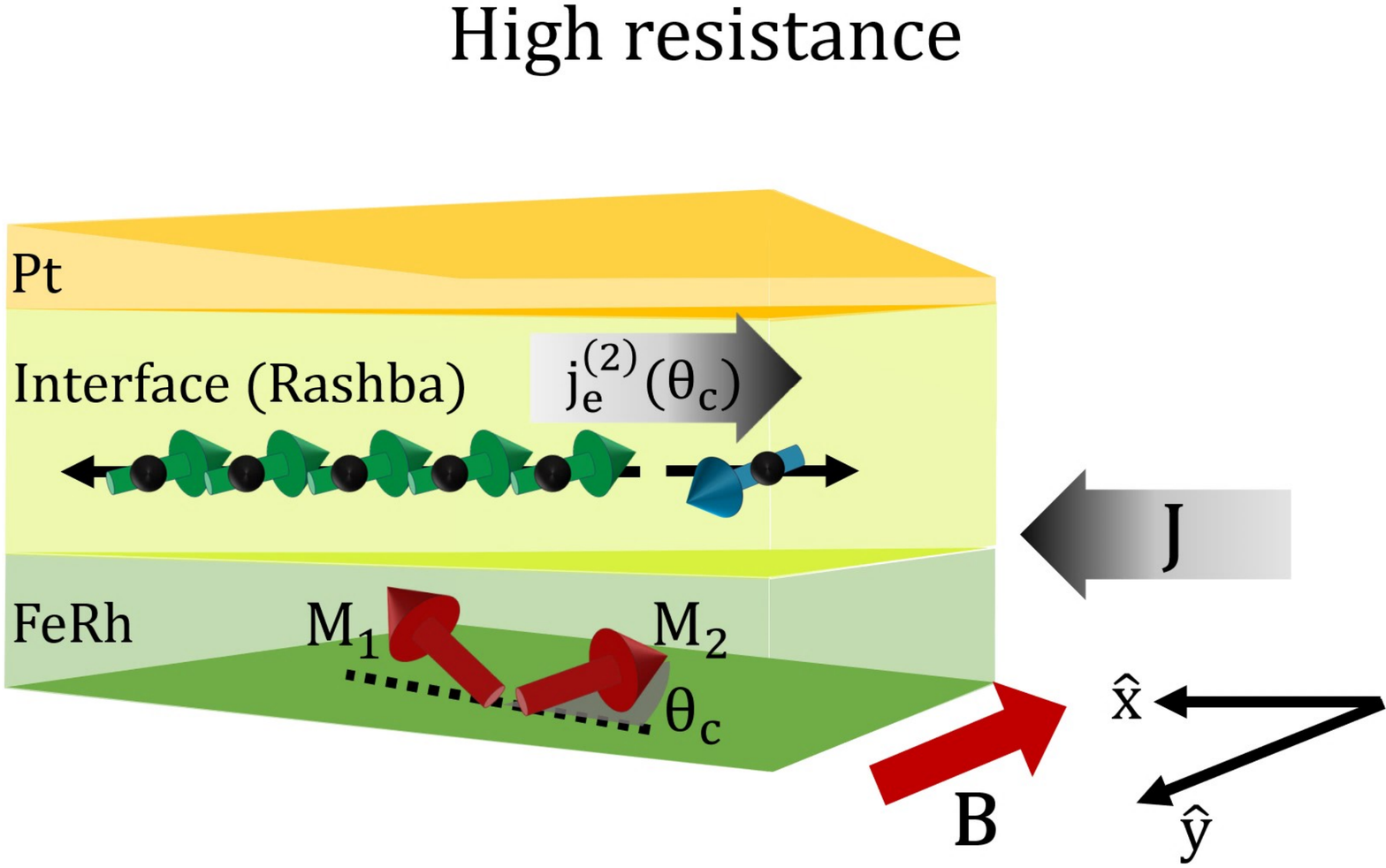}\label{fig:1_b}}\quad
    \sidesubfloat[c]{\includegraphics[width=0.3\linewidth,trim={-1cm 0cm -1cm -3cm}]{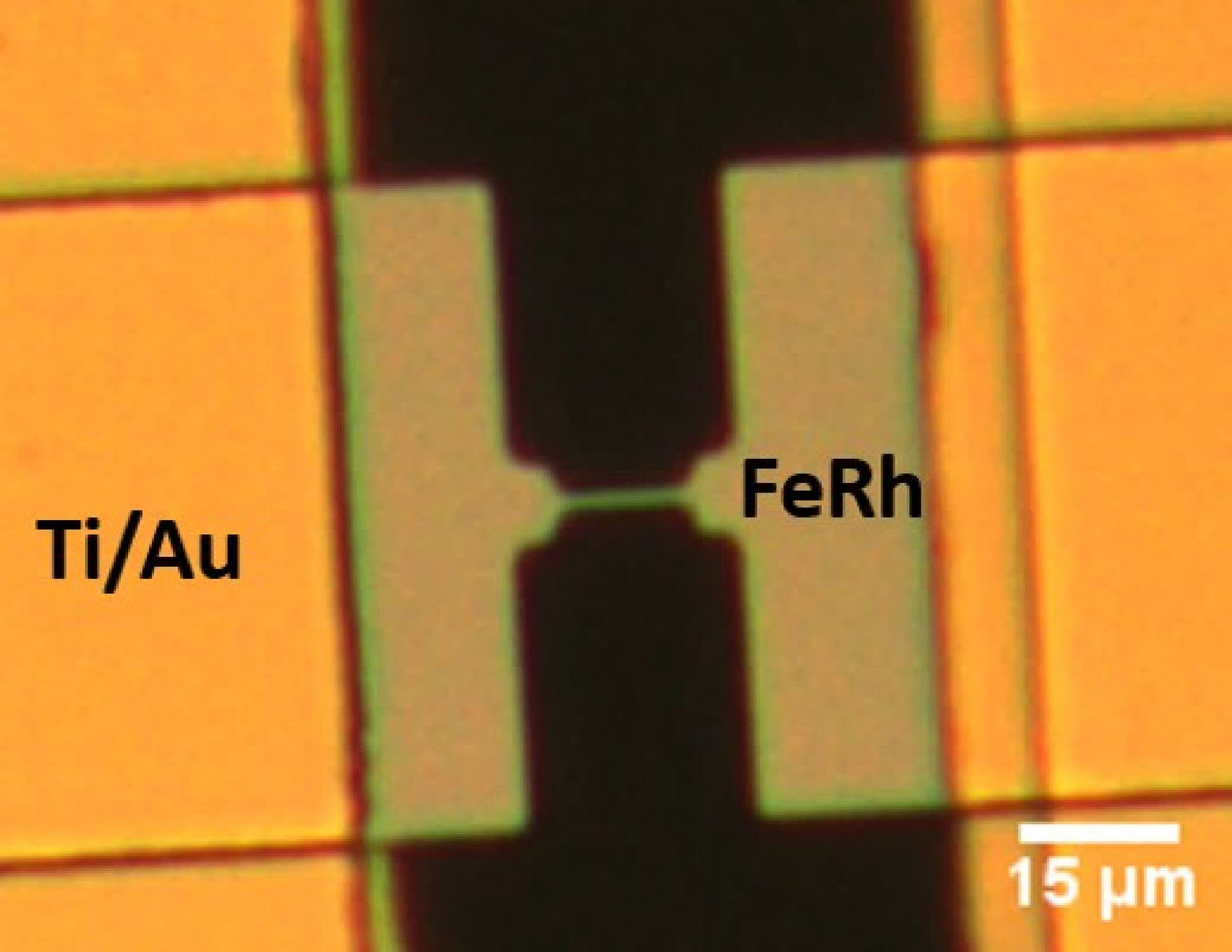}\label{fig:1_c}}\quad%
    \sidesubfloat[c]{\includegraphics[width=0.3\linewidth,trim={1cm 1cm 4cm 1cm}]{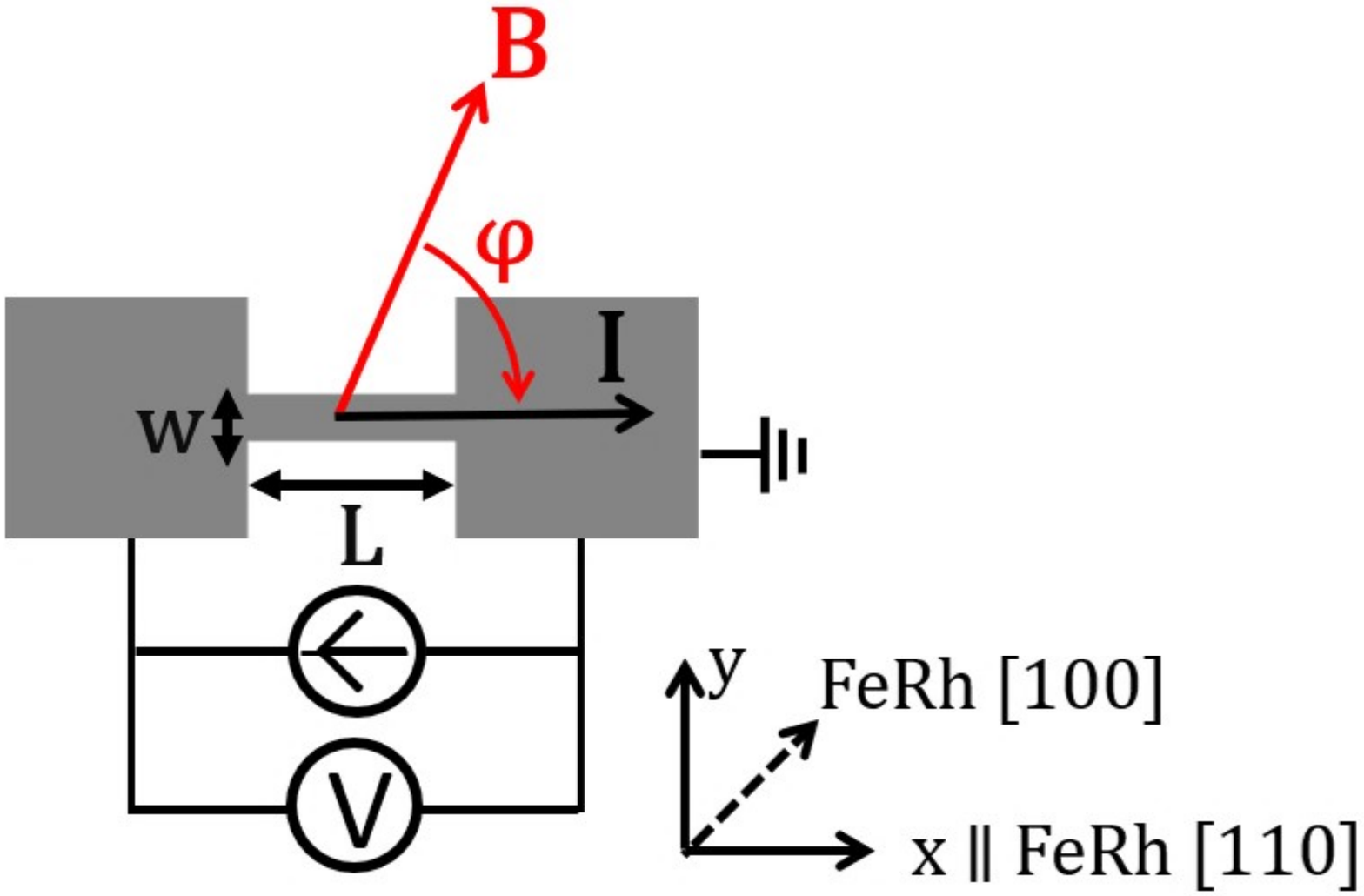}\label{fig:1_d}}%
    \caption{UMR effect in canted antiferromagnet and sample layout: \normalfont{(a-b) Rashba effect-induced nonlinear charge current $\textbf{j}_{e}^{(2)}(\theta_c)$ at the antiferromagnet$\mid$heavy-metal interface for $\mathbf{J} \parallel \hat{\mathbf{x}}$ $(\mathbf{J} \parallel -\hat{\mathbf{x}})$ at large field $\mathbf{B}$. The green and blue arrows in (a-b) indicate the direction of the spin polarizations. Resistance is low (high) when $\textbf{j}_{e}^{(2)}(\theta_c)$ is aligned (anti-aligned) with $\textbf{J}$. (c) Optical microscope image of a FeRh$\mid$Pt microwire device. (d) Schematic of the longitudinal resistance measurements. A large DC current of the order of ${10}^7$~A/$\textup{cm}^2$ is applied to induce a UMR while a small AC current is applied to perform the lock-in amplifier measurement. The angle $\varphi$ is defined as the clockwise angle from the field $\mathbf{B}$ to current $\mathbf{I}$, where $\mathbf{I}$ = $I \hat{\mathbf{x}}$.}\label{fig:1}}
\end{figure*}

In recent years, a new member of the family of magnetoresistances$-$now known as unidirectional magnetoresistance (UMR)$-$has been identified in various ferromagnetic heterostructures having structural inversion asymmetry~\cite{avci2015unidirectional, zhang2016prb, yasuda2016PRL_TI-UMR,langenfeld2016exchange, li2017origin, yin2017thickness, olejnik2015electrical,Lv18NC_UMR-TI-FM,avci2018origins,guillet2021prb}. As opposed to the aforementioned linear-response magnetoresistances, which are current-independent, the UMR is linearly proportional to the applied current, and changes sign when either the direction of the current or the in-plane magnetization (which needs to be aligned perpendicularly to the current) is reversed. The analogous effect in antiferromagnetic layered structures, however, has not been explored. It is of particular interest to discover a mechanism that generates this effect in antiferromagnetic metals, given the absence of intrinsic spin-dependent scattering~\cite{nunez2006theory}$-$a key ingredient in creating the UMR effect in conducting ferromagnets~\cite{zhang2016prb,Zhangs&Vignaleg17spie_UMR}.

In this work, we examine the nonlinear magnetotransport in a FeRh$\mid$Pt bilayer system. It is known that FeRh is a magnetic metal, which undergoes a meta-magnetic transition near room temperature from a ferromagnetic to an antiferromagnetic phase~\cite{kouvel1962anomalous,thiele2003ferh,cherifi2014electric,marti2014room,lee2015large,chirkova2016giant,barker2015higher,chen2017tunneling,mcgrath2020PRBself}. When an in-plane magnetic field is applied perpendicularly to the current direction, a UMR is observed in the antiferromagnetic phase, which changes sign when one switches either the current or the magnetic field orientation. Furthermore, the UMR evolves non-linearly with the external magnetic field and undergoes a sign change as the magnetic field is increased, in stark contrast to the behaviour typically observed in nonmagnetic \cite{ideue2017bulk,guillet2020observation,fan2019unidirectional,he2018observation,he2018bilinear} and ferromagnetic~\cite{avci2015unidirectional, langenfeld2016exchange, avci2018origins, li2017origin, yin2017thickness, olejnik2015electrical,Lv18NC_UMR-TI-FM, guillet2021prb} materials. We attribute the UMR effect to the combined actions of the Rashba SOC at the FeRh$\mid$Pt interface and the antiferromagnetic spin canting, as illustrated in Figs.~\ref{fig:1_a} and~\ref{fig:1_b}. In what follows, we first present our main experimental results and then compare them to theoretical calculations of the UMR effect based on a tight-binding model Hamiltonian that encapsulates both interfacial Rashba SOC and the spin canting effect, demonstrating excellent agreement between theory and experiment. 

\section{Sample layout and transport measurement}

We perform magnetotransport measurements on a FeRh (15~nm)$\mid$Pt (5~nm) bilayer, where the FeRh current path is oriented along the [110] direction, using lithographically defined microwires (length $L = 11$~\textmu m, width $w = 1.4$~\textmu m), as shown in Figs.~\ref{fig:1_c} and~\ref{fig:1_d}. 

\begin{figure}[hbt!]
    \sidesubfloat[]{\includegraphics[width=0.41\linewidth,trim={2.2cm 0cm 0.5cm 0}]{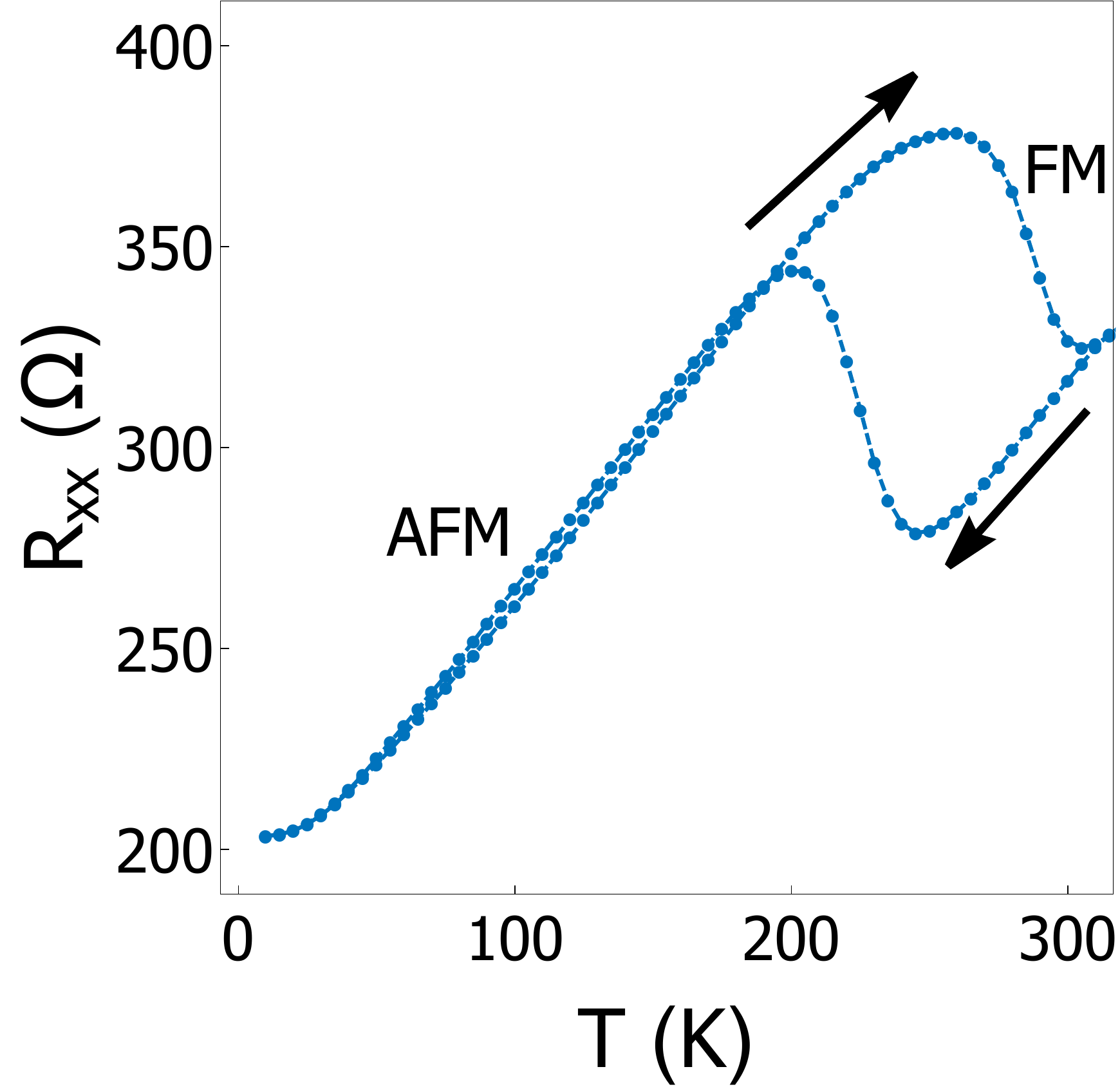}\label{fig:2_a}}\quad%
    \sidesubfloat[]{\includegraphics[width=0.41\linewidth,trim={2.8cm 0 0.4cm 0}]{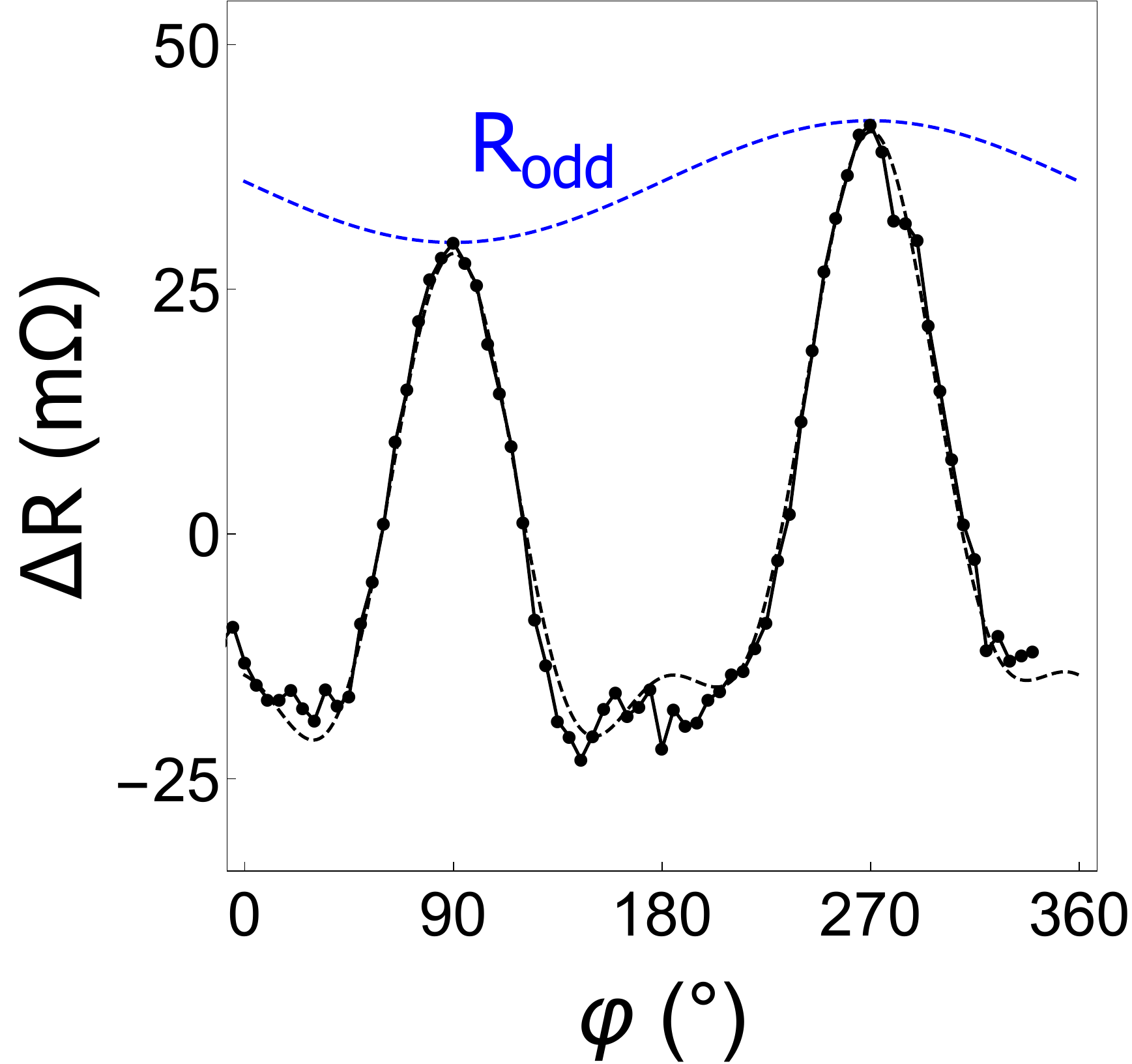}\label{fig:2_b}}\quad%
    \sidesubfloat[]{\includegraphics[width=0.41\linewidth,trim={2.2cm 0 0.5cm 0}]{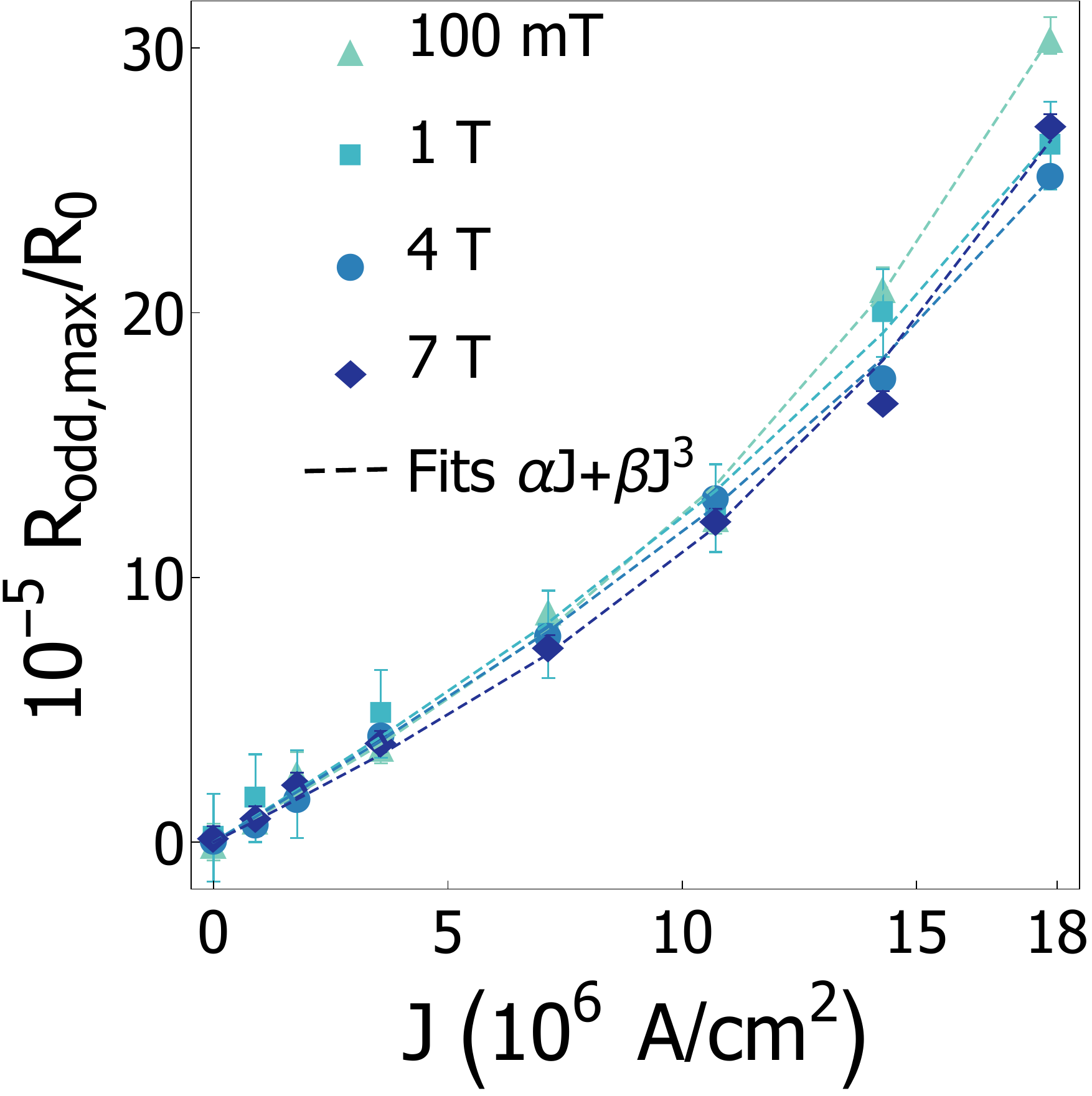}\label{fig:2_c}}\quad%
    \sidesubfloat[]{\includegraphics[width=0.41\linewidth,trim={2.5cm 0 0cm 0}]{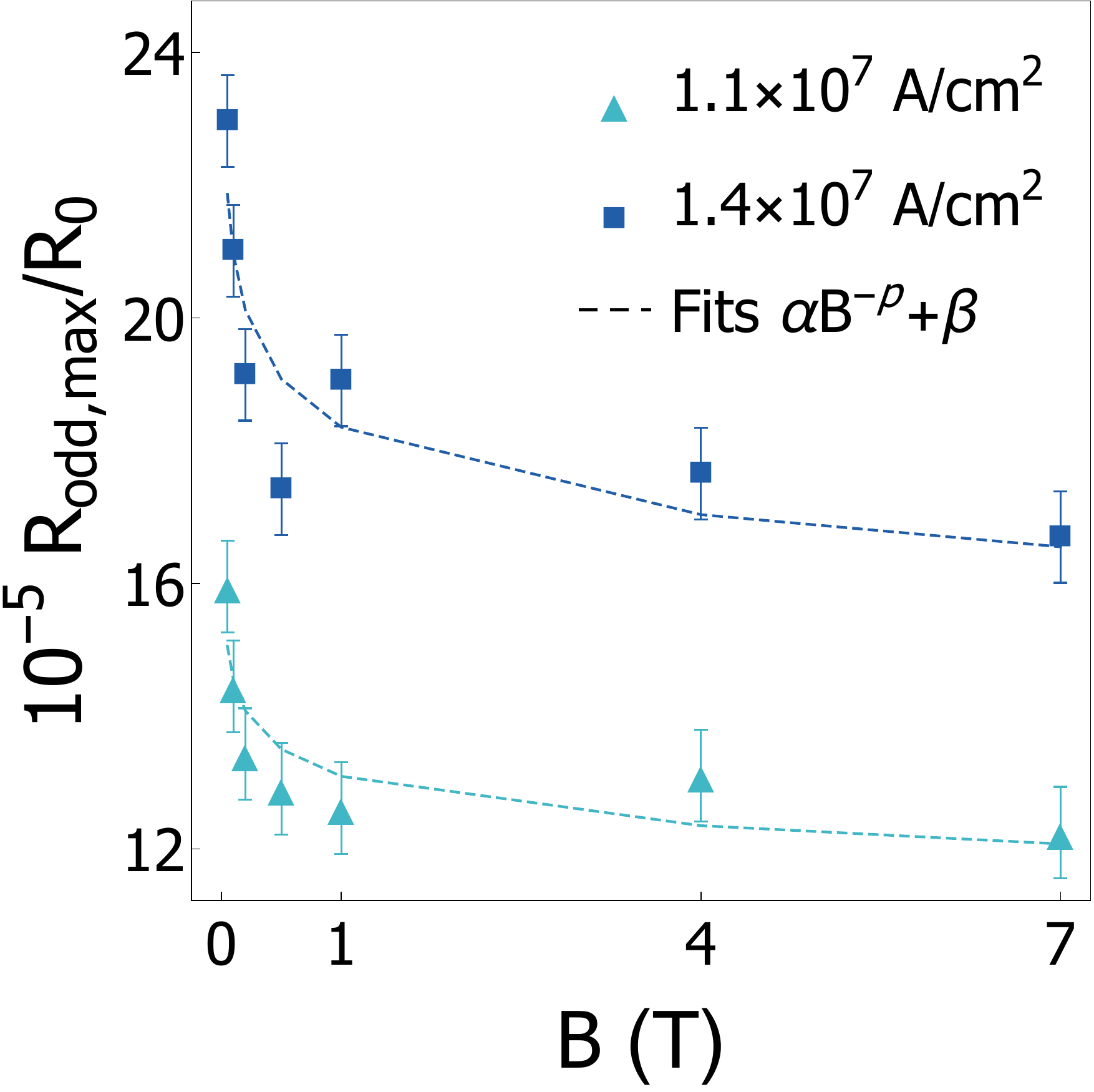}\label{fig:2_d}}%
    \caption{Magnetoresistance characterization and UMR in the ferromagnetic phase of FeRh$\mid$Pt:\normalfont{ (a) Temperature sweep of the longitudinal resistance under an external in-plane field of $B = 4$~T. Clear transition between the ferromagnetic (FM) phase and the antiferromagnetic (AFM) phase is observed. (b) Angular sweep at $T = 10$~K (AFM phase) and $B = 4$~T, under $J = 1.43\times{10}^7$~A/$\textup{cm}^2$. $~\Delta R$ is defined as $\Delta R$ = $R-R_{\text{avg}}$. A magnetoresistance which is odd under the current polarity, $R_{\text{odd}}$, manifests under high current density. (c) $R_{\text{odd,max}}$ in the FM phase of FeRh [110]$\mid$Pt, as a function of $\mathit{J}$ at various field magnitudes. $R_{\text{odd,max}}$ is normalized with the base resistance $R_0$ (measured at $J=0$ and $B=0$). Dashed lines show the fitting curves $\alpha J+\beta J^3$. (d) $R_{\text{odd,max}}$ in the FM phase of FeRh [110]$\mid$Pt, as a function of $B$ at various current densities $\mathit{J}$. Dashed lines show the fitting curves $\alpha B^{-p}+\beta$.}\label{fig:2}}
\end{figure}

We apply a DC charge current and measure the longitudinal resistances with an additional small AC charge current of 10~\textmu A from a lock-in amplifier, under the application of an external magnetic field $\mathbf{B}$. The direction of the $\mathbf{B}$ field is determined by its azimuthal angle $\varphi$ as shown in Fig.~\ref{fig:1_d}. 
We can assume that the N\'{e}el order in the FeRh is aligned perpendicular to the in-plane magnetic field and smoothly rotates with it, as a result of having antiferromagnetic domains large enough to be able to define a single antiferromagnetic spin axis on average~\cite{marti2014room} and a small in-plane anisotropy which corresponds to an effective field of less than 1 T~\cite{mancini2013magnetic,mcgrath2020PRBself}. We note that this method allows for the UMR to be extracted with the first harmonic output of the lock-in amplifier, and that without a DC current bias, the UMR must be measured using the second harmonic output of the lock-in~\cite{guillet2020observation}. As shown in Fig.~\ref{fig:2_b}, a resistance which is odd under the current polarity, defined as $R_{\text{odd}} = [R(I)-R(-I)]/2 = -R_{\text{odd,max}}\sin\varphi$, manifests in the first harmonic measurement on top of the linear-response magnetoresistance background (see Supplementary Section~S2). This odd resistance  only appears for $J > {10}^6$~A/$\textup{cm}^2$. 

All the measurements are carried out at the base temperature of either 10 K (antiferromagnetic phase) or 310~K (ferromagnetic phase) (Fig.~\ref{fig:2_a}). Base temperature refers to the ambient temperature in the sample space. The increase in the device temperature due to Joule heating from the applied current is estimated to be about 4.35~K for $J = 10^6$~A/$\textup{cm}^2$, by comparing the longitudinal resistance value under finite current density with the corresponding temperature value in Fig.~\ref{fig:2_a}. Even with the Joule heating present at a base temperature of 10 K, our device remains strictly within the antiferromagnetic phase (see Supplementary Section~S1).

In Figs.~\ref{fig:2_c} and~\ref{fig:2_d}, we show the dependence of $R_{\text{odd,max}}$ in the ferromagnetic phase on the applied current and magnetic field, and observe a monotonic increase with respect to the applied current. $R_{\text{odd,max}}$ in the ferromagnetic phase can be fitted with the curve $\alpha J+\beta J^3$ (Fig.~\ref{fig:2_c}), which includes the contributions from spin-dependent and magnon scattering mechanisms, in agreement with previous reports on UMR generated in ferromagnetic-metal$\mid$normal-metal bilayers~\cite{avci2018origins}. Furthermore, we observe a suppression in $R_{\text{odd,max}}$ for fields above 1~T, scaling with the power law $B^{-p}$ (Fig.~\ref{fig:2_d}). This amplitude suppression is also consistent with previous studies of UMR in ferromagnetic-metal$\mid$normal-metal bilayers~\cite{avci2018origins}, where the UMR originates from the field-induced gap in the magnon excitation spectrum, which reduces the electron-magnon scattering at high fields.

\section{Observation of UMR in an antiferromagnet}

\begin{figure*}[tph]
    \sidesubfloat[]{\includegraphics[width=0.2\linewidth,trim={1.5cm 0 1cm 0}]{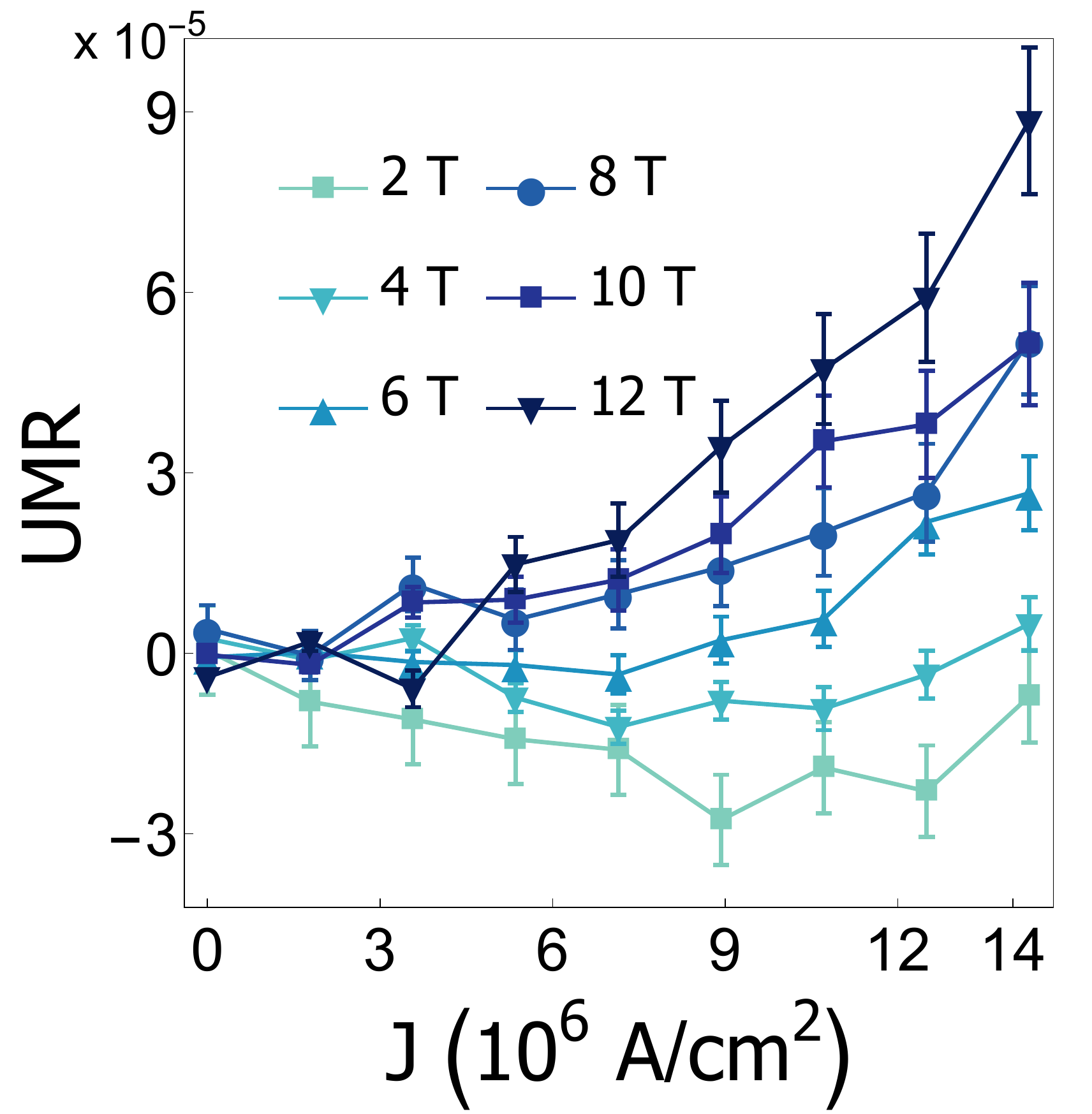}\label{fig:3_a}}\quad%
    \sidesubfloat[]{\includegraphics[width=0.2\linewidth,trim={1.5cm 0 1cm 0}]{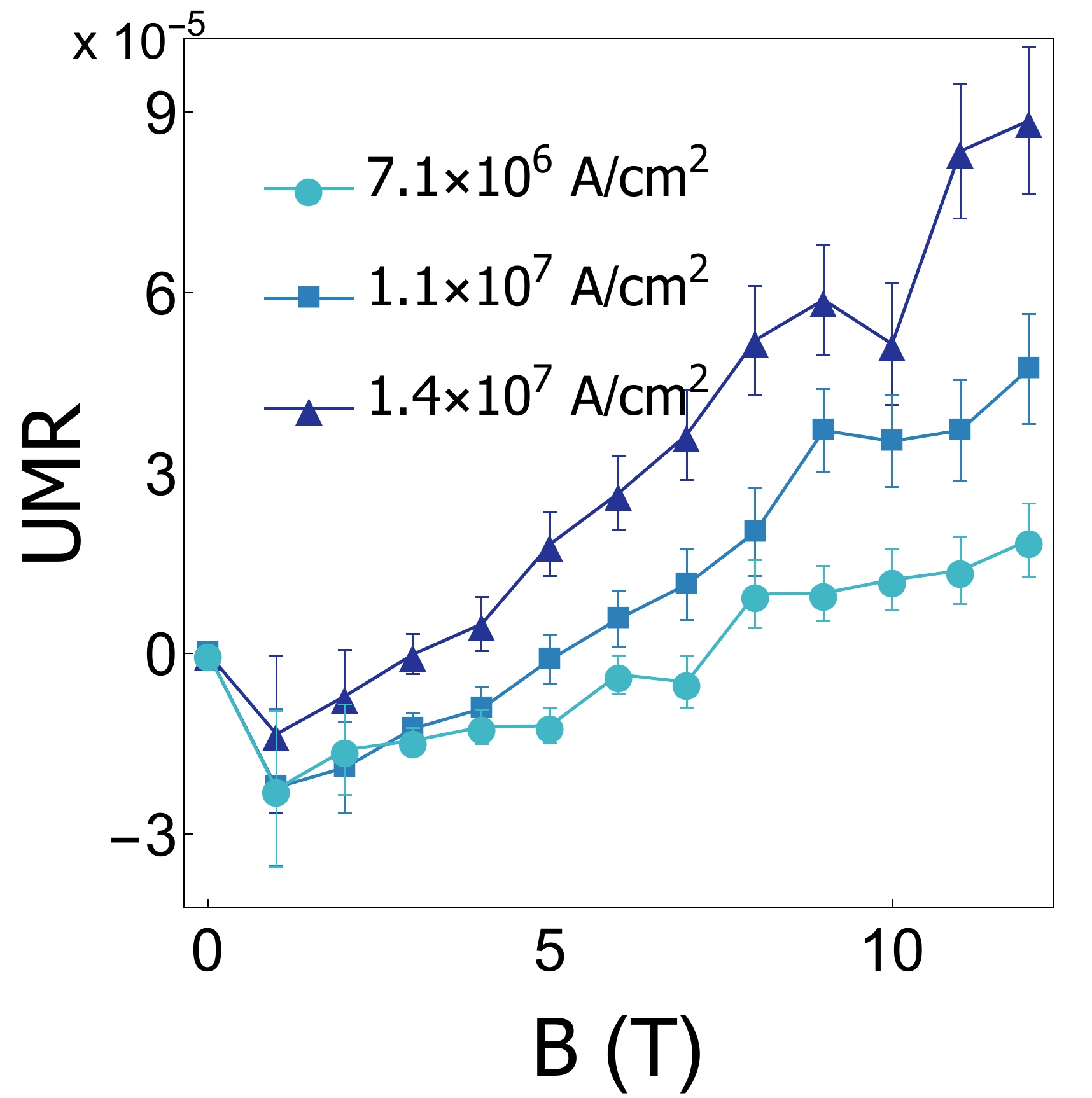}\label{fig:3_b}}\quad%
    \sidesubfloat[]{\includegraphics[width=0.2\linewidth,trim={1.2cm 0 1cm -0.4cm}]{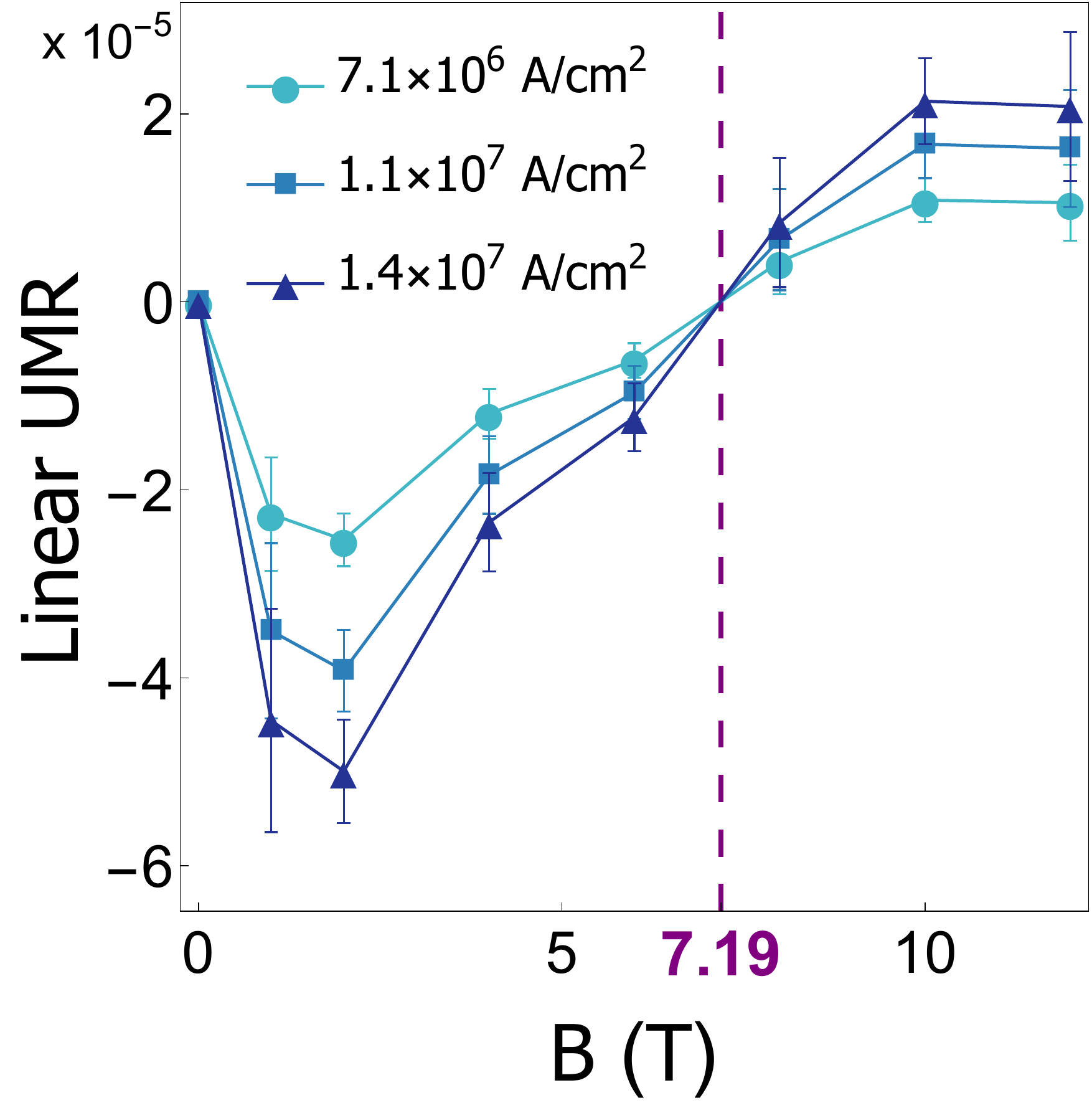}\label{fig:3_c}}\quad%
    \sidesubfloat[]{\includegraphics[width=0.2\linewidth,trim={1.5cm 1cm 1cm 0.3cm}]{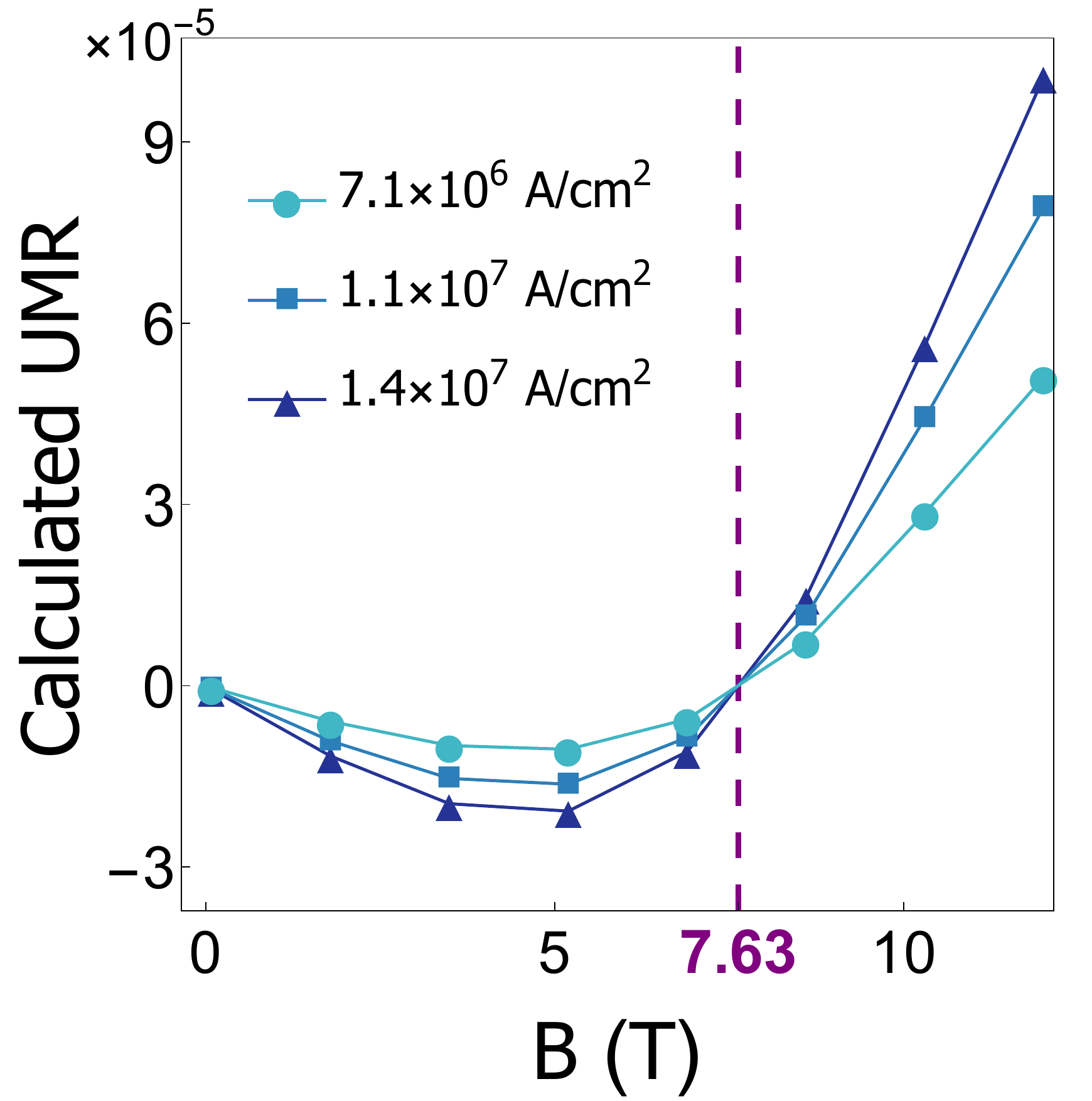}\label{fig:3_d}}%
    \caption{UMR in the antiferromagnetic phase of FeRh$\mid$Pt:\normalfont{ (a) UMR in the antiferromagnetic phase of FeRh [110]$\mid$Pt, as a function of $\mathit{J}$ for various field magnitudes. (b) UMR in the antiferromagnetic phase of FeRh [110]$\mid$Pt, as a function of field $\mathit{B}$ at various current densities $\mathit{J}$. (c) Extracted linear UMR with respect to the field $\mathit{B}$, with the purple dashed line indicating the observed field value at which the UMR undergoes a sign change. The observed sign change field value is $7.19 \pm 0.17$ T (see Supplementary Section~S8) (d) Calculated UMR in the antiferromagnetic phase as a function of $\mathit{B}$ for different values of the applied current, with the purple dashed line indicating the theoretically predicted field value at which the UMR undergoes a sign change. Parameters used: $a=3$ \AA , $\protect%
\tau =10^{-14}$ s, $g=2$, $\protect%
\epsilon _{0}=10$ eV, $\protect\epsilon _{F}=0.617\protect\epsilon _{0}$, $%
t=0.1\protect\epsilon _{0}$ \cite{vzelezny2014relativistic}, $\Delta_{\text{ex}}=0.05\protect\epsilon _{0}$ \cite{vzelezny2014relativistic,saidaoui2017robust}, $%
\tilde{\protect\alpha}_{R}=\protect\alpha _{R}/a=0.05\protect\epsilon _{0}$ \cite{vzelezny2014relativistic,haney2013current}, 
$H_{J}=11.83$ T (corresponding to $\protect\theta _{c}=25^{\circ }$ at $%
B=10$ T) and width of spectral function $=0.002\protect\epsilon _{0}$.}
\label{fig:3}}
\end{figure*}

The behavior of the UMR in the antiferromagnetic phase is shown in Figs.~\ref{fig:3_a} and~\ref{fig:3_b}. In order to extract the UMR from $R_{\text{odd}}$ in the antiferromagnetic phase, we analyzed the contribution from the thermal gradient $\bs{\nabla}T$ and related thermoelectric effects. The anomalous Nernst effect and spin Seebeck effect, both producing a longitudinal voltage proportional to $\mathbf{M}\times \bs{\nabla} T$ (where $\mathbf{M}$ is the net magnetization and $\nabla T \propto J^2$), can give rise to the same angular dependence as observed for the UMR when $\bs{\nabla} T\parallel\hat{\mathbf{z}}$~\cite{weiler2012local,kikkawa2013longitudinal,avci2015unidirectional}. Here, by measuring the transverse (Hall) counterpart scaled by the geometric factor $L/w$ in a similar FeRh$\mid$Pt microwire device, we find that even the maximum thermoelectric voltage contribution in the antiferromagnetic phase is less than 30\% (see Supplementary Section~S3). We also find that $R_{\text{odd,max}}$, the amplitude of $R_{\text{odd}}$, has no dependence on current or external field strength in a control sample of FeRh~(20~nm) without Pt (see Supplementary Section~S3), which indicates a negligible anomalous Nernst effect contribution, in agreement with the literature. We conclude that there is an additional magnetoresistance effect in the FeRh$\mid$Pt bilayers other than thermoelectric voltages. For the antiferromagnetic phase of the FeRh$\mid$Pt bilayer, we extract the UMR from the amplitude of the measured sinusoidal signal (shown in Fig.~\ref{fig:2_b}) using UMR$\sin\varphi = -(R_{\text{odd}}-R_{xx}^{\nabla T})/R_{0}$, following the convention for ferromagnetic-metal$\mid$normal-metal bilayers, in which the UMR of the bilayer increases when the direction of the majority spins in the ferromagnet and the spin accumulation vector are parallel to each other, and decreases when they are antiparallel~\cite{avci2015unidirectional}.

In the antiferromagnetic phase of the FeRh$\mid$Pt bilayer (Figs.~\ref{fig:3_a},~\ref{fig:3_b}), we observe a different trend from what was observed in the ferromagnetic phase. In the antiferromagnetic phase, the UMR is not strongly suppressed for fields higher than 1~T. This can be attributed to the lack of magnon-dependent scattering in antiferromagnets~\cite{nunez2006theory}. Thus, the antiferromagnetic UMR increases approximately linearly with field and current, for large field and current values. The most striking feature of the UMR in the antiferromagnetic phase is the sign change from negative to positive UMR as the magnetic field is increased, which, to our knowledge, has not been observed in either ferromagnetic or nonmagnetic systems.

The UMR observed in the antiferromagnetic phase cannot be due to the spin Hall effect, since the scattering in a metallic antiferromagnet is independent of the spin polarization~\cite{nunez2006theory} and even the canted spin configuration under the external field is not large enough to induce the necessary spin-dependent scattering for a significant spin-Hall UMR (see Supplementary Section~S7). Moreover, the observed UMR is isotropic with respect to the direction of the current flow in the crystalline plane (see Supplementary Section~S4). Thus, it is unlikely that the UMR in the antiferromagnetic phase originates from strong crystal field effects. A lack of field and current dependence of $R_{\text{odd,max}}$ in the control sample of FeRh~(20~nm) without a Pt layer (see Supplementary Section~S3) also implies that the observed UMR cannot be attributed to the intrinsic properties of FeRh. 

As shall be discussed in the next section, we attribute the UMR in FeRh$\mid$Pt to the combined effects of the Rashba SOC at the interface of FeRh and Pt, and the spin canting in the antiferromagnetic spin sublattices. A calculated UMR based on this theory, which is linear in the applied current, is shown in Fig.~\ref{fig:3_d}; this can be compared to the linear component of the experimental UMR (i.e., to first order in the applied current) shown in Fig.~\ref{fig:3_c}. As can be seen, in the intermediate magnetic field range$-$where the UMR sign reversal occurs$-$there is excellent qualitative and quantitative agreement between theory and experiment. In the low field limit, however, there is some quantitative disagreement. This is most likely attributable to the effect of thermal magnons, which was not considered in the theoretical model. As an additional scattering source of conduction electrons, the thermal magnons may modify the magnetoresistance at low magnetic fields, but their contribution is expected to diminish as the external field is increased$-$a trend which is nicely captured by comparing Figs.~\ref{fig:3_c} and \ref{fig:3_d}. 

\section{Physical origin of UMR}

To obtain physical insight into the observed UMR effect arising from the FeRh$\mid$Pt interface, we theoretically investigate the nonlinear magnetotransport by restricting ourselves to the interfacial layer of the FeRh adjacent to the Pt layer, which may be described by the following two-dimensional tight-binding Hamiltonian with broken inversion symmetry of a collinear antiferromagnetic metal with Rashba SOC \cite{vzelezny2014relativistic,baltz2018antiferromagnetic} 
\begin{equation}
\label{hamiltonian}
\hat{H}=\epsilon _{0}+\gamma _{\mathbf{k}}\hat{\tau}_{x}+\Delta _{\text{ex}}\hat{%
\tau}_{z}\hat{\boldsymbol{\sigma }}\cdot \mathbf{m}+\alpha _{R}\hat{\tau}_{x}%
\hat{\boldsymbol{\sigma }}\cdot \hat{\mathbf{z}}\times \mathbf{k}+g\mu _{B}\hat{%
\bs{\sigma}}\cdot \mathbf{B},  
\end{equation}%
where $\epsilon _{0}$ is the on-site energy, $\Delta _{\text{ex}}$ is the s-d
exchange constant between the local moments and the electron, $\alpha _{R}$
is the Rashba SOC constant, and $\hat{\tau}_{i}$ and $\hat{%
\sigma}_{i}$ are Pauli matrices which signify the sublattice and spin
degrees of freedom, respectively. The nearest-neighbor hopping is
represented by $\gamma _{\mathbf{k}}=-2t\left( \cos k_{x}a+\cos
k_{y}a\right) $, where $t$ is the hopping term and $a$ the lattice constant.

\begin{figure}[tph]
    \sidesubfloat[]{\includegraphics[width=0.41\linewidth,trim={1.0cm 0 1.5cm 0}]{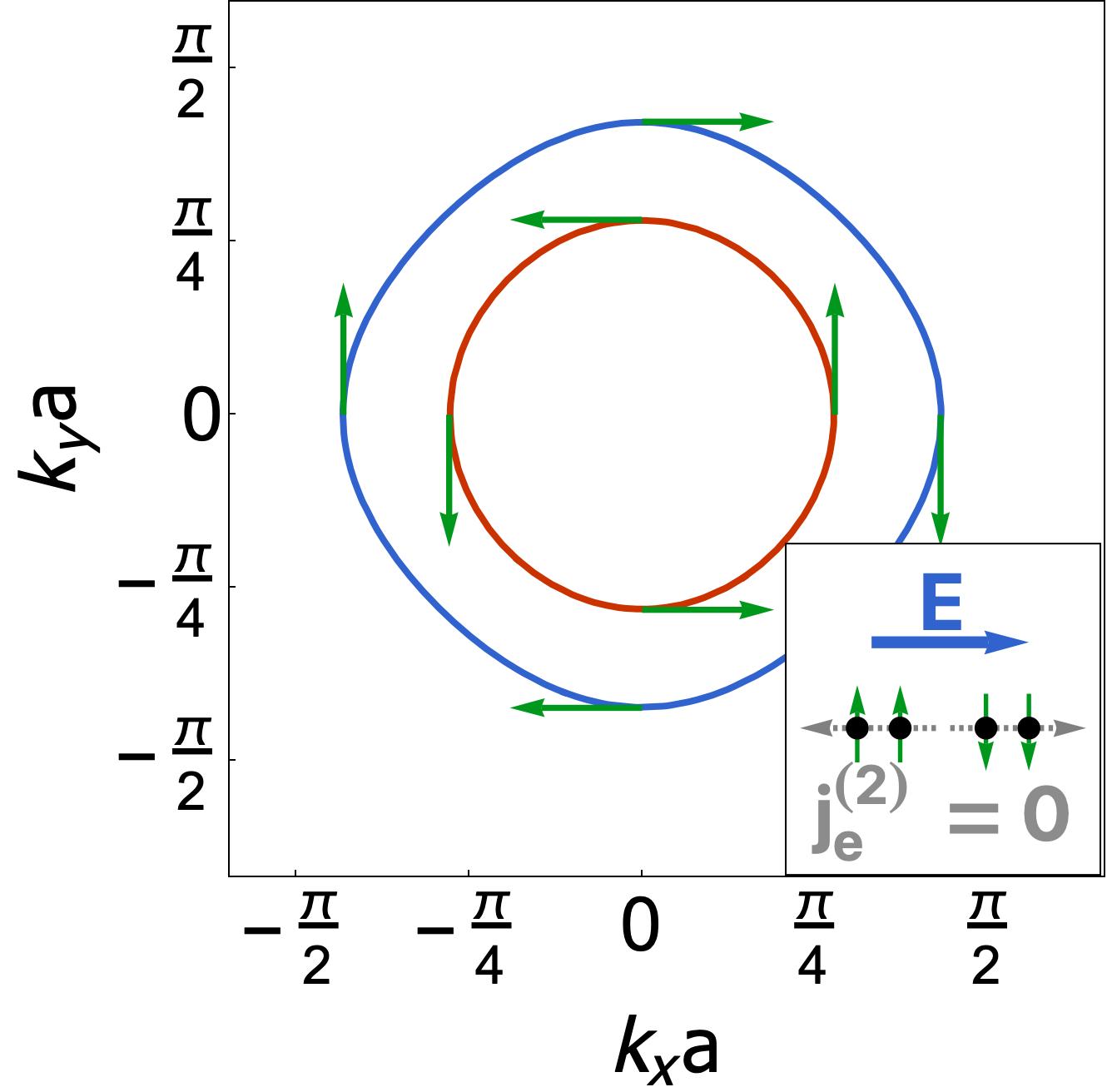}\label{fig:4_a}}\quad%
    \sidesubfloat[]{\includegraphics[width=0.41\linewidth,trim={3cm 0 -0.2cm 0}]{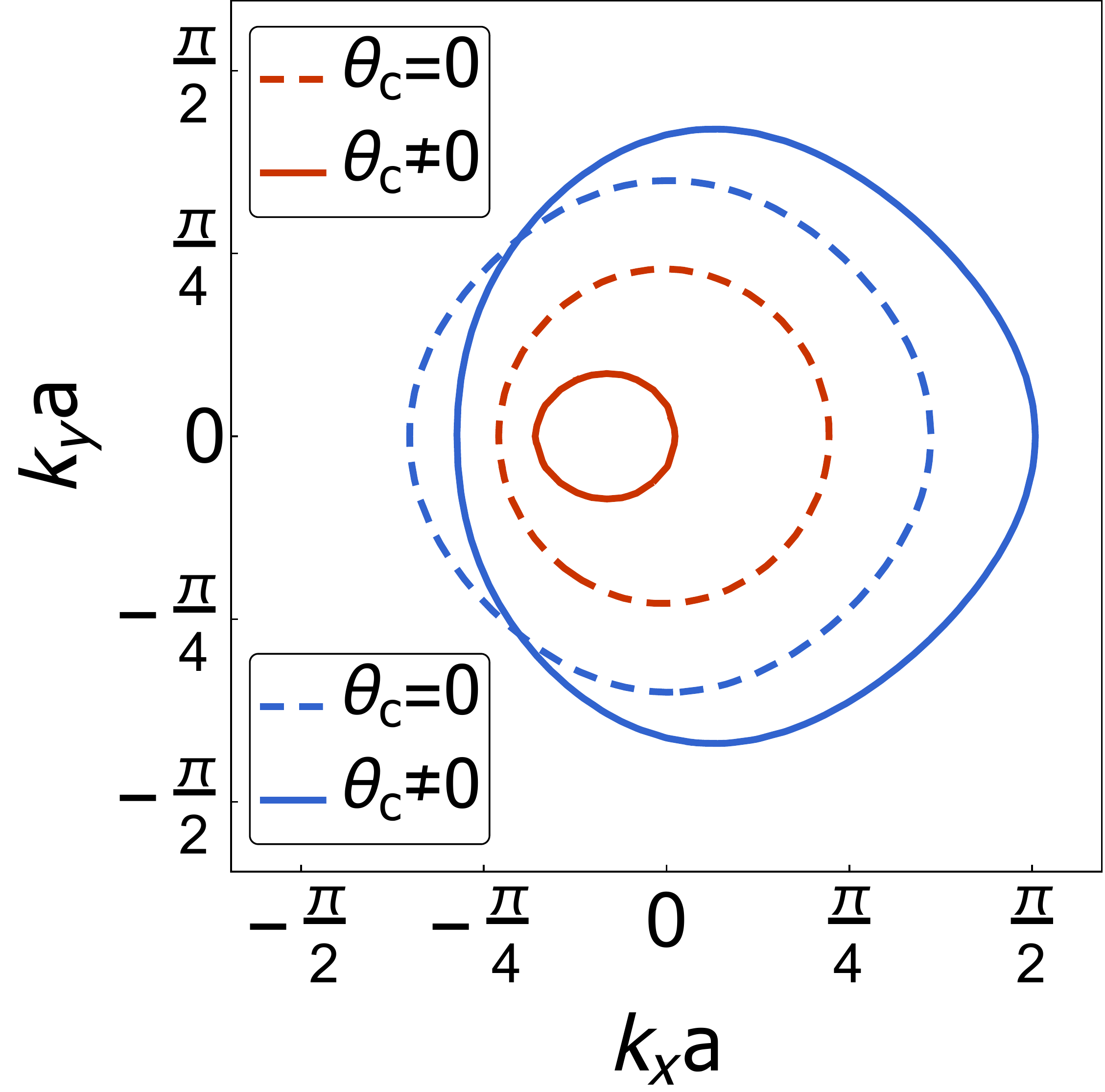}\label{fig:4_b}}\quad%
    \sidesubfloat[]{\includegraphics[width=0.41\linewidth,trim= {0.5cm 0cm 0cm -0cm}]{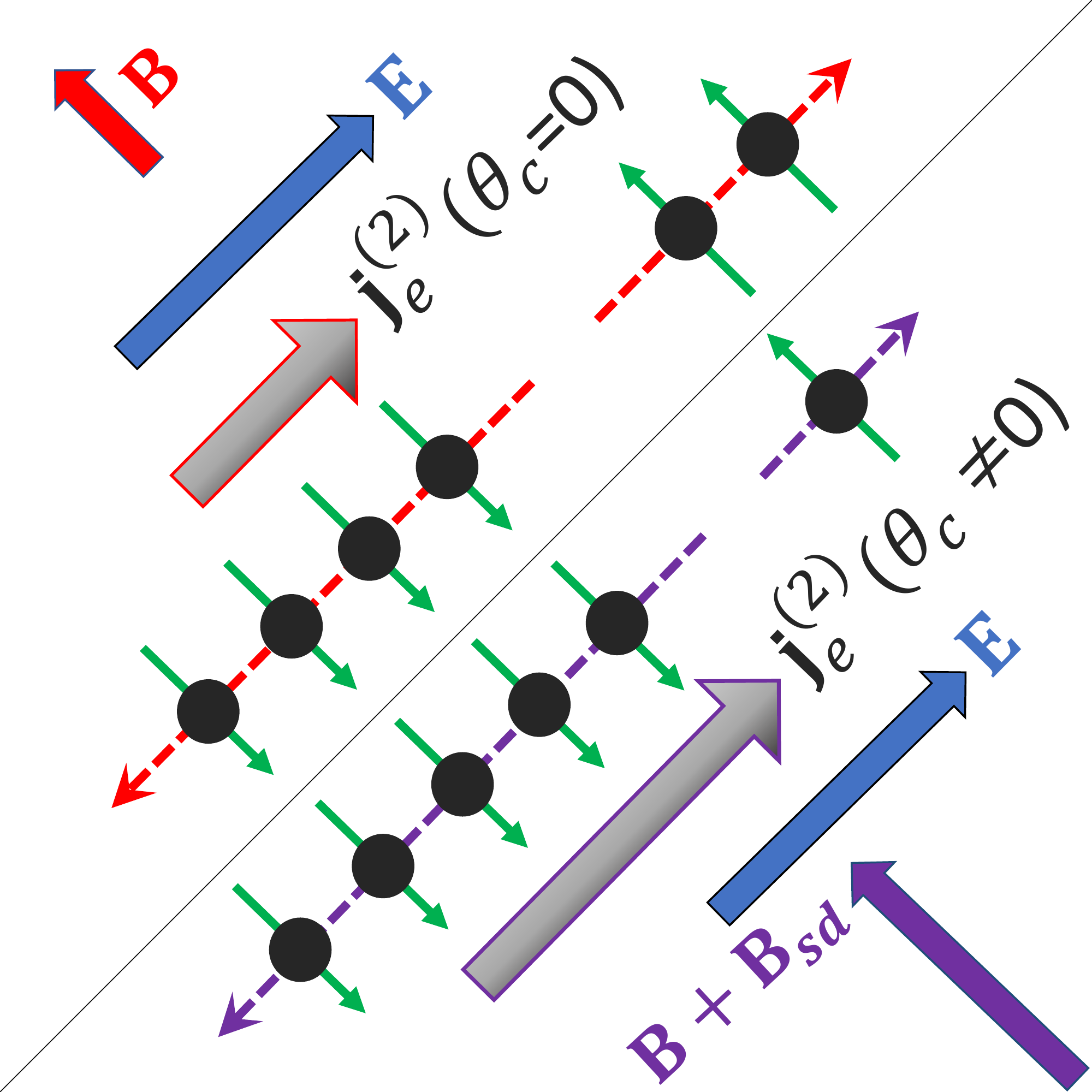}\label{fig:4_c}}\quad%
    \sidesubfloat[]{\includegraphics[width=0.41\linewidth,trim={2cm 0 0.6cm +0cm}]{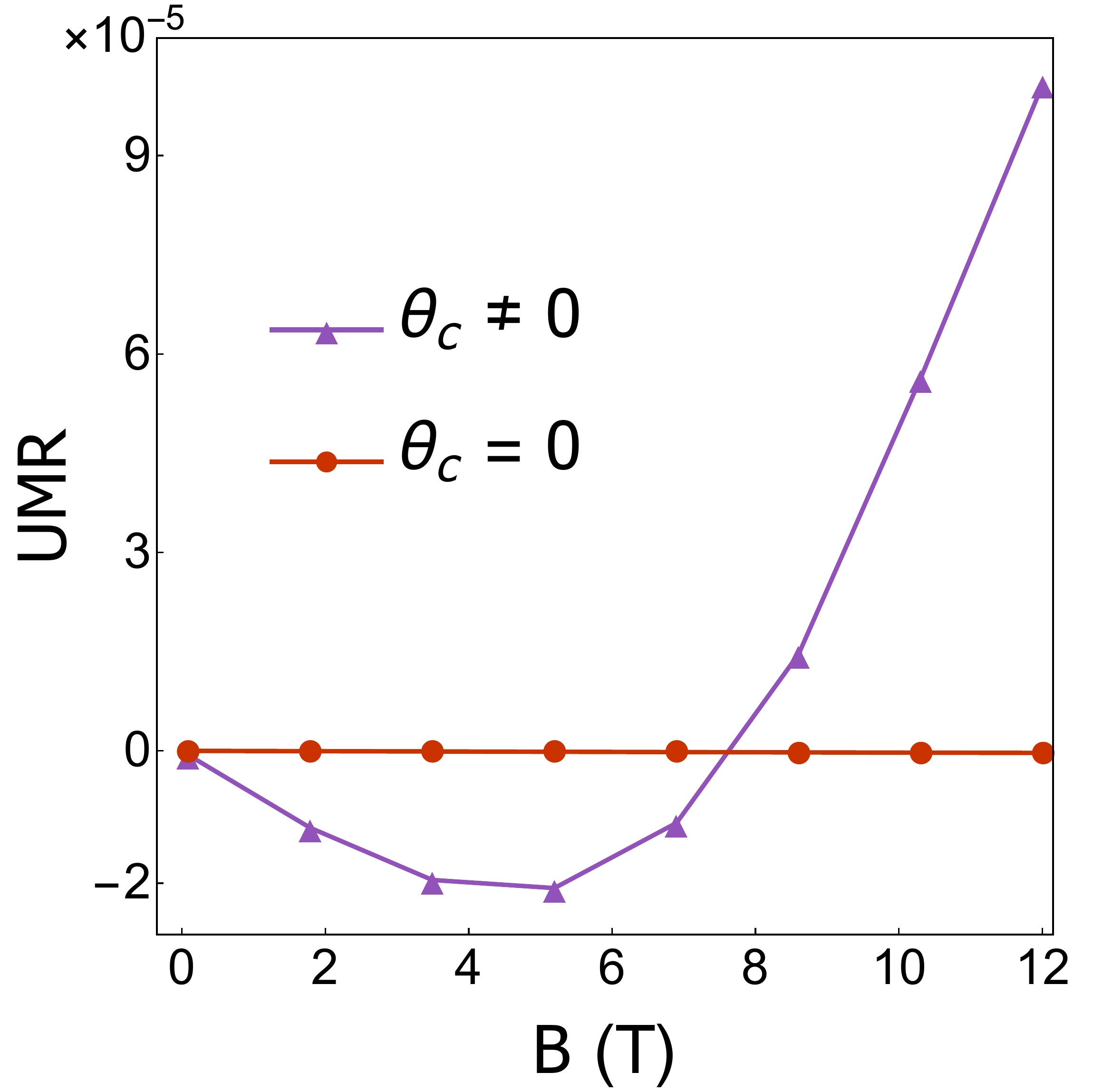}\label{fig:4_d}}%
    \caption{Enhancement of UMR by spin canting:\normalfont{ (a) Spin texture of the conduction bands in the absence of a magnetic field. Due to spin-momentum locking arising from the Rashba effect, when an electric field is applied, a pure spin current but no nonlinear charge current is produced, as depicted in the inset. (b) Spin-dependent distortion of the bands in the presence of a magnetic field at both zero canting (dashed curves) and finite canting (solid curves) The two bands are shown here in red and blue. (c) Generation of a nonlinear charge current transverse to the applied $\mathbf{B}$ field (top left), which increases considerably in the presence of canting (bottom right). (d) Plots of the calculated UMR at finite and vanishing canting as functions of the magnetic field $B$. In the presence of canting, the UMR is stronger by two orders of magnitude.}\label{fig:4}}
\end{figure}

In the absence of an external magnetic field, the Rashba splitting leads to
two Fermi contours with opposite spin chirality in equilibrium, as shown in
Fig.~\ref{fig:4_a}. By applying an in-plane electric field $\mathbf{E}$, a pure
nonlinear spin current (with no corresponding charge current) can be induced in the system due to spin-momentum locking as well as the symmetric electron
distribution in momentum space in the second order of the $\mathbf{E}$ field~\cite{hamamoto2017prb,he2018bilinear,he2019prl}. The nonlinear spin current is converted to a nonlinear charge current in the presence of an in-plane magnetic field perpendicular to the current direction (which establishes an imbalance between the two electron fluxes with opposite spin orientations, see dashed bands in Fig.~\ref{fig:4_b}), leading to the UMR effect~\cite{he2018bilinear,he2019prl}. 

The Hamiltonian, Eq.~(\ref{hamiltonian}), is modified by spin canting when an external
magnetic field is applied perpendicular to the N\'{e}el vector. More
specifically, the sublattice magnetizations tilt toward the applied field
by an angle $\theta _{c}$ relative to the N\'{e}el vector, resulting in a
net magnetization along the field direction. The canting angle $\theta _{c}$
can be determined by minimizing the magnetic energy density of a collinear
antiferromagnet as given below  
\begin{equation}
\epsilon _{m}=-\mathbf{B}\cdot \left( \mathbf{M}_{A}+\mathbf{M}_{B}\right) +%
\frac{H_{J}}{M_{s}}\mathbf{M}_{A}\cdot \mathbf{M}_{B}.
\end{equation}%
Here, $\mathbf{M}_{A}$ and $\mathbf{M}_{B}$ are the magnetizations of
sublattices A and B, $H_{J}$ is the effective exchange field measuring the
interaction between the two sublattices and $M_{s}$ is the saturation
magnetization, where we take $M_{A}=M_{B}=M_{s}$. For an external magnetic
field applied along the $y$-direction, we find $\theta _{c}=\arcsin \left(
B/2H_{J}\right)$ (see Supplementary Section~S5). To capture the spin
canting effect, we make the substitution $\Delta _{\text{ex}}\hat{\tau}_{z}\hat{%
\sigma}_{x}\rightarrow \Delta _{\text{ex}}\left( \hat{\tau}_{z}\hat{\sigma}_{x}\cos
\theta _{c}+\hat{\sigma}_{y}\sin \theta _{c}\right) $ for the s-d exchange
term in the Hamiltonian. Note that the net magnetization that is
parallel to $\mathbf{B}$ now couples with the same sign to the electronic
spin, giving rise to an effective magnetic field $B_{\text{sd}}=\Delta _{\text{ex}}B/\left(
2g\mu _{B}H_{J}\right) $.

This strong effective magnetic field due to spin canting greatly enhances the amount of distortion of the Fermi contours when an external magnetic field is applied, as shown by the solid bands in Fig.~\ref{fig:4_b}, which
leads to a more profound UMR, as shown schematically in Fig.~\ref{fig:4_c}. To confirm this, we calculate the nonlinear
longitudinal charge current density $j_{e,x}^{(2)}=-\frac{e^{3}\tau
^{2}E_{x}^{2}}{\hbar ^{2}}\sum_{n}\int_{\mathbf{k}}\left( \frac{\partial
^{2}f_{n}}{\partial k_{x}^{2}}\right) v_{n,x}$ using the Boltzmann transport
formalism (see Supplementary Section~S5), where $n$ is the band-index, $%
\tau $ is the momentum relaxation time, $\int_{\mathbf{k}}\equiv \int_{\text{%
BZ}}\frac{d^{2}\mathbf{k}}{(2\pi )^{2}},$ and $\mathbf{v}_{n}=\frac{\partial
\epsilon _{n}(\mathbf{k})}{\hbar \partial \mathbf{k}}$ is the group velocity
of the $n$-th band. From the total longitudinal resistivity $\rho_{xx}=E_x/j_{e,x}$, we calculate the UMR as $\text{UMR}\equiv -[\rho_{xx}(E_x)-\rho_{xx}(-E_x)]/\rho_{xx}(E_x)\approx 2\sigma_{xx}^{(1)}/\sigma_D$, with $\sigma_{xx}^{(1)}=j_{e,x}^{(2)}/E_x$ and $\sigma_D$ the Drude conductivity. 
The calculated UMR, both with and without canting, as a function of the applied magnetic field is plotted in Fig.~\ref{fig:4_d}, from which it is evident that spin canting indeed plays an important role in the enhancement of the UMR effect. 

\section{Mechanism of UMR sign change}

In order to understand the origin of the sign change in the UMR as the magnetic field intensity is tuned, we first note that this is made possible by spin canting. More precisely, in the presence of canting, the Hamiltonian acquires a significant nonlinear dependence on the magnetic field through the effective field $B_{\text{sd}}$, which, in turn, allows for the UMR to evolve nonlinearly with respect to $B$. Another element that plays an important role is the asymmetry of the band structure as the magnetic field and, by extension, $B_{\text{sd}}$, are switched on.

\begin{figure*}[tph]
    \sidesubfloat[]{\includegraphics[width=0.2\linewidth,trim={1.5cm 0.3cm 0.5cm 0}]{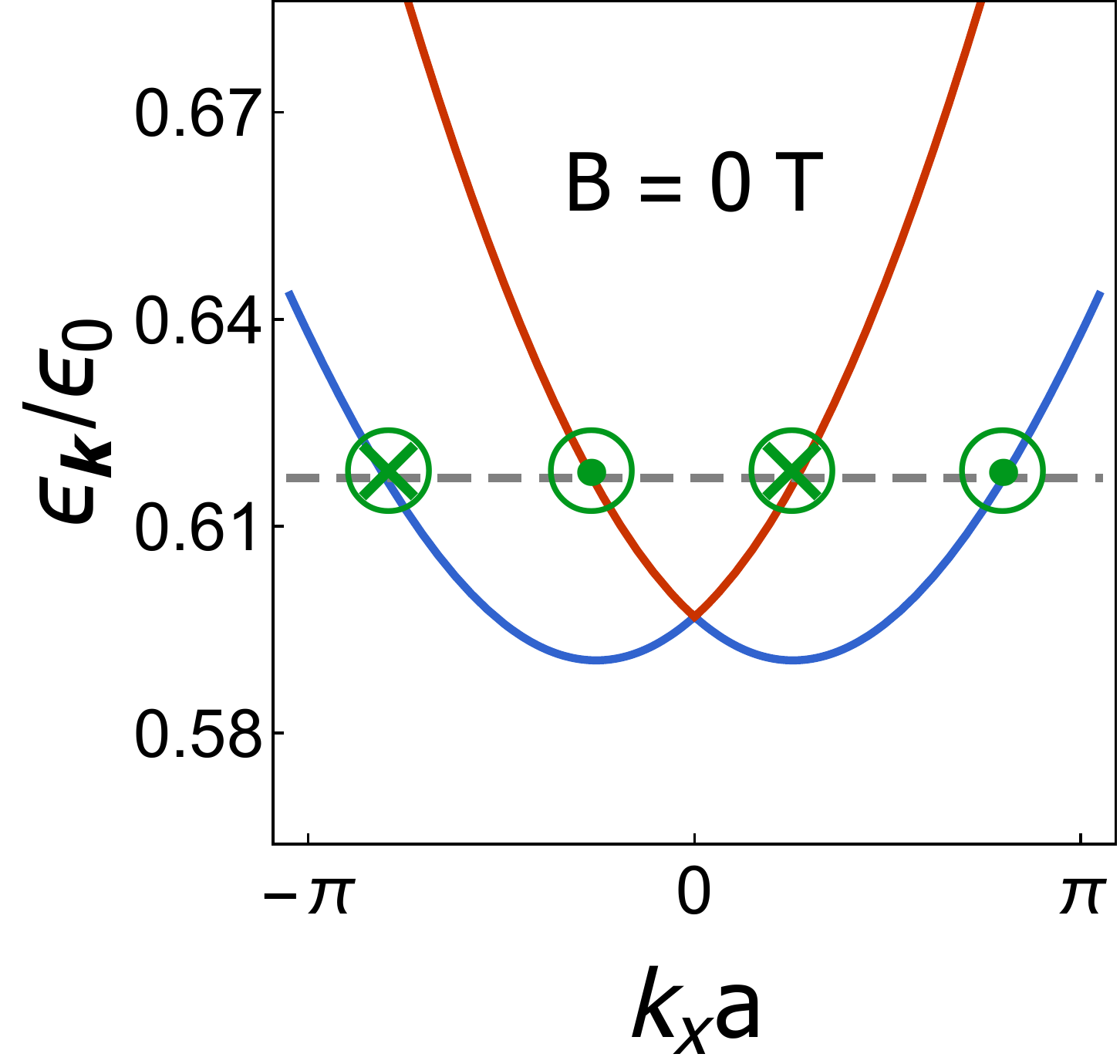}\label{fig:5_a}}\quad%
    \sidesubfloat[]{\includegraphics[width=0.2\linewidth,trim={1.5cm 0.3cm 0.5cm 0}]{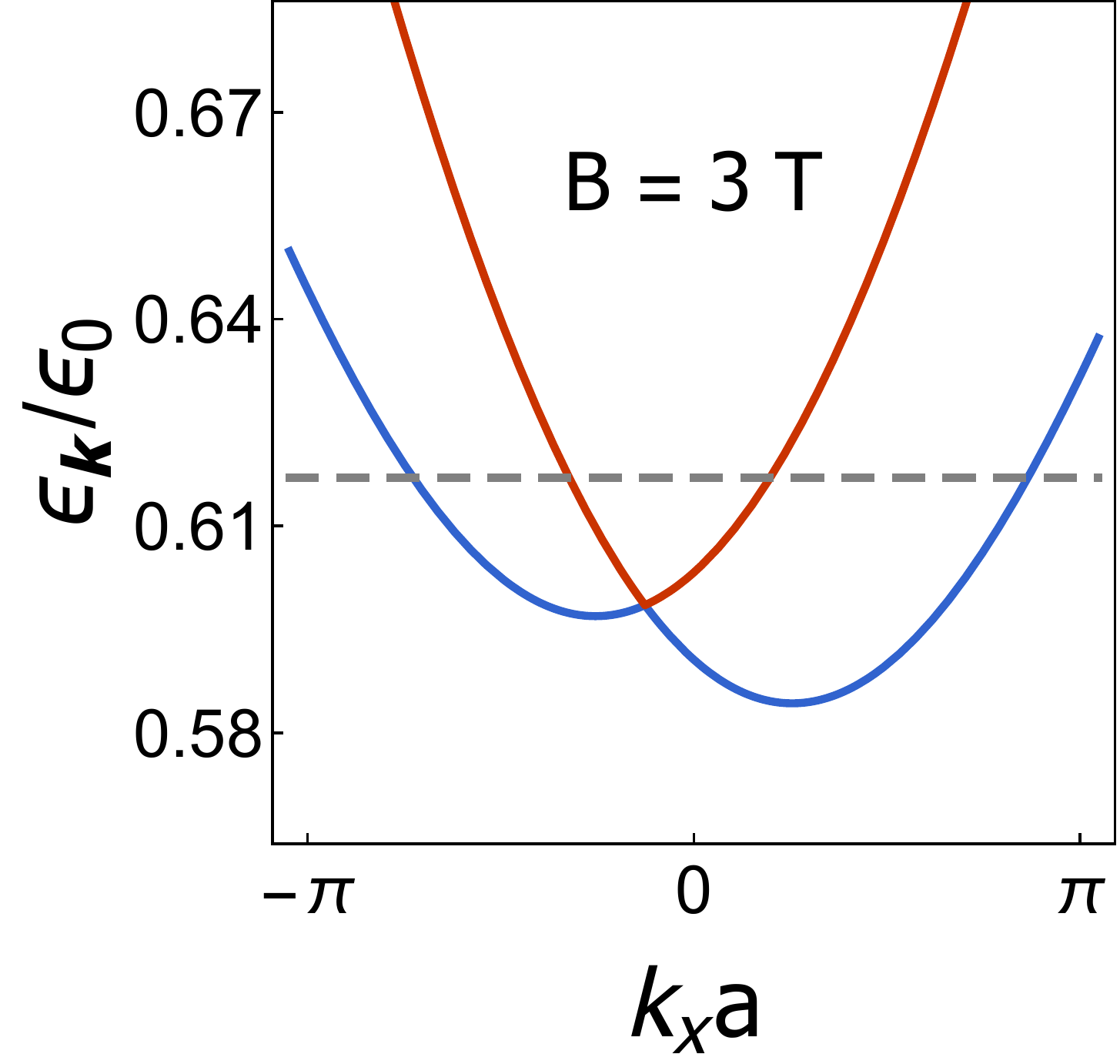}\label{fig:5_b}}\quad%
    \sidesubfloat[]{\includegraphics[width=0.2\linewidth,trim={1.5cm 0.3cm 0.5cm 0cm}]{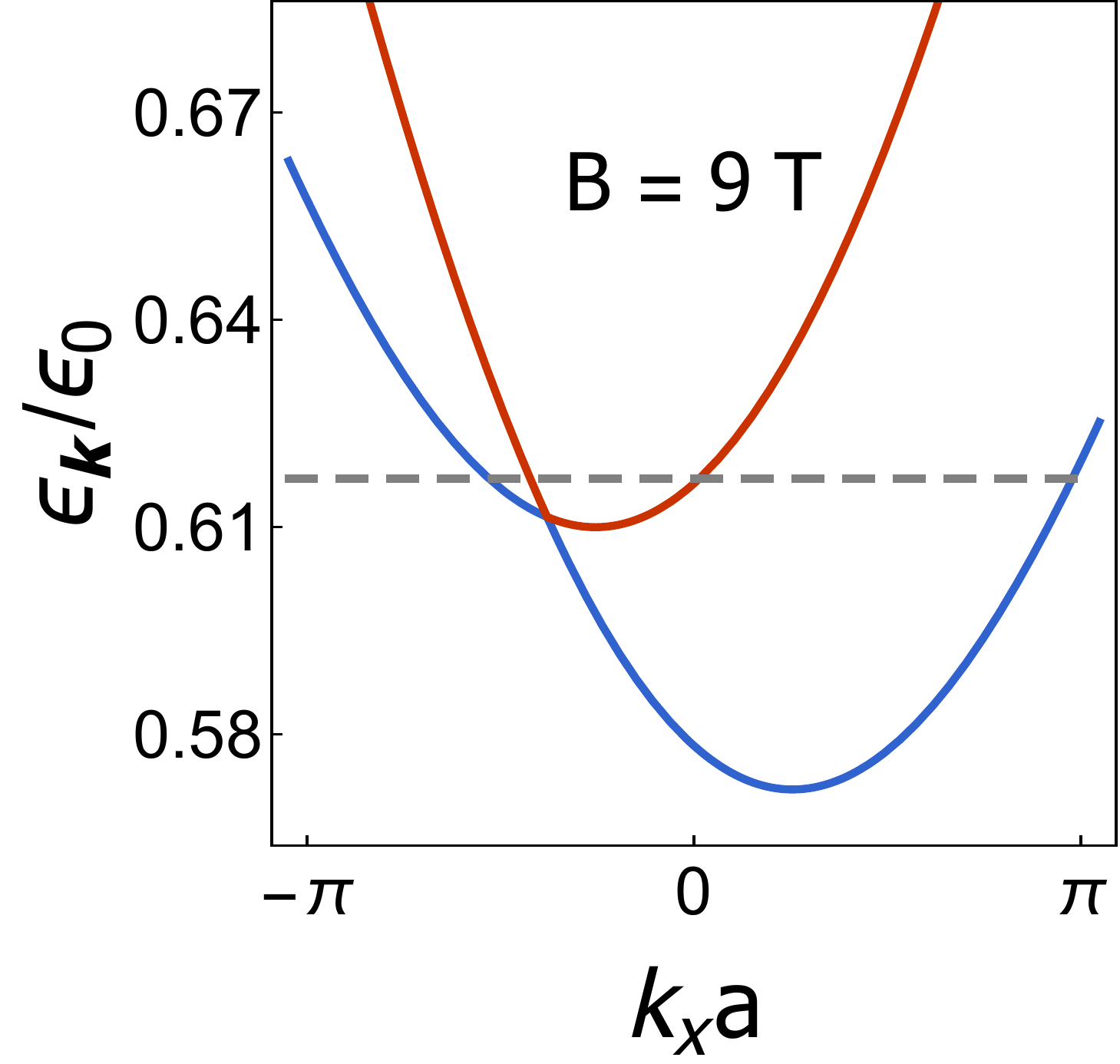}\label{fig:5_c}}\quad%
    \sidesubfloat[]{\includegraphics[width=0.2\linewidth,trim={1.5cm 1cm 0.3cm 0.7cm}]{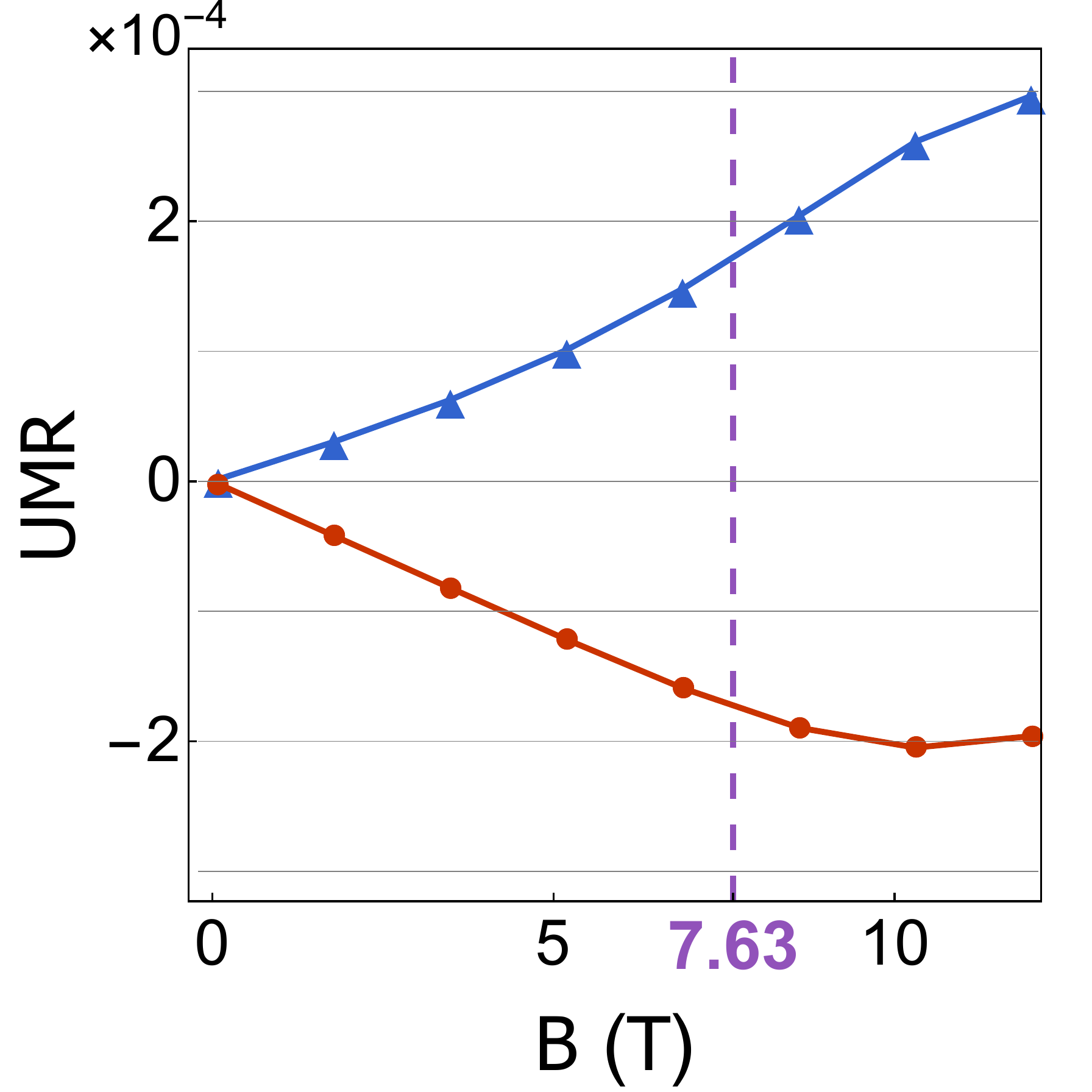}\label{fig:5_d}}%
    \caption{Mechanism of UMR sign change:\normalfont{ Structure of the conduction bands at $k_y=0$ and a) 0 T, b) 3 T and c) 9 T, indicating the breaking of time reversal symmetry induced by the magnetic field. The two bands are shown here in red and blue, with the dashed line indicating the Fermi level. d) Opposite contributions of the inner (red dots) and outer (blue triangles) conduction bands to the UMR, with the purple dashed line indicating the field at which the sign change in the overall UMR occurs.}\label{fig:5}}
\end{figure*}

Initially, at zero magnetic field, as depicted in Fig.~\ref{fig:5_a}, the conduction bands are symmetric about the $\Gamma$ point and the UMR is absent. When the magnetic field is turned on, a small spin-dependent asymmetry develops in the band structure, which results in a positive UMR contribution from the outer band (blue in Fig.~\ref{fig:5}) and a negative contribution from the inner band (red). As shown in Figs.~\ref{fig:5_b} and~\ref{fig:5_d}, at low fields such as 3 T, the asymmetry in the outer band is not enough to counter the dominant presence of the inner band near the Fermi level and the overall UMR is negative.

As the magnetic field strength is increased, the inner band continues its ascent from the Fermi sea, while the asymmetry of the outer band near the Fermi level continues to grow. At around 7-8 T, the contribution of the two bands becomes equal, at which point the overall UMR undergoes a sign change. As depicted in Fig.~\ref{fig:5_c}, at large fields, the contribution from the outer band is dominant near the Fermi level and the UMR is positive. The dependence of the UMR for each conduction band on the magnetic field for a given current value is displayed in Fig.~\ref{fig:5_d}, along with the theoretically predicted field value of 7.63 T at which the sign change occurs. This value may be compared with the sign-change field of the linear components of the observed UMR (Fig.~\ref{fig:3_c}). The observed magnetic field value of $7.19 \pm 0.17$ T is in very good agreement with the theoretically predicted field value. 

Based on this picture, it is also possible to understand why such a magnetic-field-dependent sign change in the UMR is unique to antiferromagnetic systems and is not expected to occur in ferromagnetic and nonmagnetic systems. In the latter, one could still have a positive or negative UMR depending on the relative position of the Fermi level with respect to the conduction bands; However, the UMR would not change sign in these systems due to the lack of sublattice degree of freedom and the resulting antiferromagnetic spin canting (see Supplementary Section~S6).

\section{Conclusion}
In summary, we observe a UMR in the antiferromagnetic phase of a FeRh$|$Pt bilayer, which undergoes a sign change and then increases strongly with an increasing external magnetic field, largely different from UMRs in ferromagnetic and nonmagnetic systems. We show that the Rashba SOC alone, a mechanism known for UMRs in ferromagnetic and nonmagnetic systems, cannot explain the sizable UMR in the antiferromagnetic bilayer. Antiferromagnetic spin canting also plays a crucial role in enhancing the UMR by inducing a strong effective magnetic field that significantly distorts  the band structure.

The UMR effect we have observed is not exclusive to FeRh$\mid$Pt bilayers and is expected to exist in antiferromagnetic heterostructures satisfying two conditions: an easy-plane antiferromagnet which displays magnetic domains large enough to define a single antiferromagnetic spin axis on average, and small in-plane anisotropy to ensure coherent rotation of the spin axis with an applied in-plane field.  Thus, in the context of antiferromagnetism, this effect has both fundamental and practical implications. First, the new mechanism for UMR in the antiferromagnetic bilayer can advance fundamental understanding of non-collinear antiferromagnetic systems. Antiferromagnets have recently attracted great interest as systems that host emergent phenomena, competing orders, and symmetry-tunable band-structures. Therefore, our findings can motivate future studies to further explore the interplay between spin texture, electronic band structure, and the associated emergent phenomena in antiferromagnets. 

Moreover, key parameters of antiferromagnets (difficult to measure in thin films) can be back-engineered from the UMR effect.  For example, susceptibility and magnetocrystalline anisotropy parameters have not been directily measured in FeRh thin films \cite{mcgrath2020PRBself}.  Based on the field-dependence of the UMR that we report, we can estimate the magnetic susceptibility of the 15 nm thick FeRh thin films to be \textit{five} times greater than what has been reported for bulk samples (see Supplementary Section~S5). In addition, our analysis reveals that while the value of the UMR sign reversal field is, in general, a complicated function of the materials parameters of the bilayer system, we expect that it will scale linearly with the magnetic susceptibility (see Supplementary Section S9)$-$a prediction which may prove useful in future studies of UMR in other antiferromagnetic systems. From the perspective of applications, our findings may also lead toward the development of antiferromagnet-based spintronics, such as two-terminal devices \cite{miron2011perpendicular,liu2012spin}, where the spin information can be controlled by both electric voltage and magnetic field.

\section*{Acknowledgements}

We thank David G. Cahill and Matthew J. Gilbert for valuable discussions. This research was primarily supported by the NSF through the University of Illinois at Urbana-Champaign Materials Research Science and Engineering Center DMR-1720633 and was carried out in part in the Materials Research Laboratory Central Research Facilities, University of Illinois.  Thin-film growth at Argonne National Laboratory was supported by the U.S. Department of Energy, Office of Science, Materials Science and Engineering Division. J.\,O. acknowledges support from the Army under W911NF-20-1-0024. Work by M.\,M. and S.\,S.-L.\,Z. was supported by the College of Arts and Sciences, Case Western Reserve University.

\section*{Author Contributions}
S.\,S., J.\,S. and N.\,M. designed the magnetotransport experiments.  J.\,G., H.\,S. and A.\,H. were responsible for synthesizing FeRh films.  S.\,S. and J.\,S. performed magnetic characterization of FeRh.  S.\,S., J.\,S., J.\,O., J.\,G. and H.\,S. fabricated devices.  S.\,S., J.\,S. and J.\,O. performed magnetotransport experiments.  A.\,H. and N.\,M. supervised the experimental work. M.\,M. and S.\,S.-L.\,Z. devised the theory. S.\,S. and M.\,M. analyzed the results with J.\,S., S.\,S.-L.\,Z., and N.\,M. providing input. S.\,S., M.\,M., J.\,S., S.\,S.-L.\,Z., and N.\,M. wrote the manuscript with input from all other authors.

\section*{Materials and Methods}
The FeRh films used in this work were deposited onto (100) MgO substrates using DC magnetron sputtering.  Before sputtering, the substrates were annealed within the sputter deposition system at 850$^\circ$ C for one hour in order to desorb the potential contaminants on the surface.  After the substrates were cleaned, the temperature was lowered to 450$^\circ$ C for deposition.  The sputter target used for deposition was an equiatomic FeRh source.  During growth, 6.5~sccm of Ar gas was introduced into the chamber, and the pressure was set to 6~mTorr.  The DC sputtering power used was 50 W, and the growth rate was 0.7 $\mathrm{\AA}$/s. After one hour post-annealing of FeRh films at 650$^\circ$ C, the heater was turned off and the films were cooled down to room temperature before depositing the Pt layer, which was deposited at room temperature, with the pressure of 5~mTorr, power of 7~W, and the growth rate of 0.6 $\mathrm{\AA}$/s~\cite{saglam2019spin}. The films were subsequently patterned into microwires by direct writing laser lithography and Ar milling. The microwires were patterned so that the current flowed along the [110] crystalline orientations of FeRh.

For thin FeRh films deposited onto MgO substrates, it has been reported that there exist residual ferromagnetic moments, confined to within 6-8~nm of the interface of FeRh and MgO~\cite{fan2010ferromagnetism}. To determine the magnitude of the residual ferromagnetic moment, magnetic characterization on FeRh~(15~nm) with an in-plane field was carried out in a vibrating sample magnetometer by Quantum Design. With an in-plane field of 0.5~T, the residual ferromagnetic moment at $T = 10$~K was estimated to be 4~\% of the saturated magnetization in the ferromagnetic phase (1260~emu/$\textup{cm}^3$) (see Supplementary Section~S10). The residual ferromagnetic moment, however, does not affect the UMR because no UMR is observed for the single layer FeRh thin film on MgO (see Supplementary Section~S3) and the field scale for the observed antiferromagnetic UMR is very large compared to the ferromagnetic case (see Figures 2 and 3).
Transport experiments were four-probe resistance measurements carried out using the Physical Property Measurement System by Quantum Design equipped with a horizontal rotator module. A DC current of the order of mA, generated by a Keithley sourcemeter 2400, was applied to achieve the desired current density $\mathit{J}$ and a 13~Hz AC current of 10~\textmu A, generated by a Stanford Research Systems SR830 lock-in amplifier, was applied to probe the magnetoresistance. To accommodate different device sizes, the current was adapted to have the same current density. The rotation of the sample was provided by a motorized stage with a precision of 0.0133\degree.

\bibliography{mainref2}

\begin{thebibliography}{69}%
\makeatletter
\providecommand \@ifxundefined [1]{%
 \@ifx{#1\undefined}
}%
\providecommand \@ifnum [1]{%
 \ifnum #1\expandafter \@firstoftwo
 \else \expandafter \@secondoftwo
 \fi
}%
\providecommand \@ifx [1]{%
 \ifx #1\expandafter \@firstoftwo
 \else \expandafter \@secondoftwo
 \fi
}%
\providecommand \natexlab [1]{#1}%
\providecommand \enquote  [1]{``#1''}%
\providecommand \bibnamefont  [1]{#1}%
\providecommand \bibfnamefont [1]{#1}%
\providecommand \citenamefont [1]{#1}%
\providecommand \href@noop [0]{\@secondoftwo}%
\providecommand \href [0]{\begingroup \@sanitize@url \@href}%
\providecommand \@href[1]{\@@startlink{#1}\@@href}%
\providecommand \@@href[1]{\endgroup#1\@@endlink}%
\providecommand \@sanitize@url [0]{\catcode `\\12\catcode `\$12\catcode
  `\&12\catcode `\#12\catcode `\^12\catcode `\_12\catcode `\%12\relax}%
\providecommand \@@startlink[1]{}%
\providecommand \@@endlink[0]{}%
\providecommand \url  [0]{\begingroup\@sanitize@url \@url }%
\providecommand \@url [1]{\endgroup\@href {#1}{\urlprefix }}%
\providecommand \urlprefix  [0]{URL }%
\providecommand \Eprint [0]{\href }%
\providecommand \doibase [0]{https://doi.org/}%
\providecommand \selectlanguage [0]{\@gobble}%
\providecommand \bibinfo  [0]{\@secondoftwo}%
\providecommand \bibfield  [0]{\@secondoftwo}%
\providecommand \translation [1]{[#1]}%
\providecommand \BibitemOpen [0]{}%
\providecommand \bibitemStop [0]{}%
\providecommand \bibitemNoStop [0]{.\EOS\space}%
\providecommand \EOS [0]{\spacefactor3000\relax}%
\providecommand \BibitemShut  [1]{\csname bibitem#1\endcsname}%
\let\auto@bib@innerbib\@empty
\bibitem [{\citenamefont {{Jungwirth}}\ \emph {et~al.}(2016)\citenamefont
  {{Jungwirth}}, \citenamefont {{Marti}}, \citenamefont {{Wadley}},\ and\
  \citenamefont {{Wunderlich}}}]{jungwirth2016antiferromagnetic}%
  \BibitemOpen
  \bibfield  {author} {\bibinfo {author} {\bibfnamefont {T.}~\bibnamefont
  {{Jungwirth}}}, \bibinfo {author} {\bibfnamefont {X.}~\bibnamefont
  {{Marti}}}, \bibinfo {author} {\bibfnamefont {P.}~\bibnamefont {{Wadley}}},\
  and\ \bibinfo {author} {\bibfnamefont {J.}~\bibnamefont {{Wunderlich}}},\
  }\bibfield  {title} {\bibinfo {title} {{\textit{Antiferromagnetic
  Spintronics}}},\ }\href {https://doi.org/10.1038/nnano.2016.18} {\bibfield
  {journal} {\bibinfo  {journal} {Nat. Nanotechnol.}\ }\textbf {\bibinfo
  {volume} {11}},\ \bibinfo {pages} {231} (\bibinfo {year} {2016})}\BibitemShut
  {NoStop}%
\bibitem [{\citenamefont {Baltz}\ \emph {et~al.}(2018)\citenamefont {Baltz},
  \citenamefont {Manchon}, \citenamefont {Tsoi}, \citenamefont {Moriyama},
  \citenamefont {Ono},\ and\ \citenamefont
  {Tserkovnyak}}]{baltz2018antiferromagnetic}%
  \BibitemOpen
  \bibfield  {author} {\bibinfo {author} {\bibfnamefont {V.}~\bibnamefont
  {Baltz}}, \bibinfo {author} {\bibfnamefont {A.}~\bibnamefont {Manchon}},
  \bibinfo {author} {\bibfnamefont {M.}~\bibnamefont {Tsoi}}, \bibinfo {author}
  {\bibfnamefont {T.}~\bibnamefont {Moriyama}}, \bibinfo {author}
  {\bibfnamefont {T.}~\bibnamefont {Ono}},\ and\ \bibinfo {author}
  {\bibfnamefont {Y.}~\bibnamefont {Tserkovnyak}},\ }\bibfield  {title}
  {\bibinfo {title} {\textit{Antiferromagnetic Spintronics}},\ }\href
  {https://doi.org/10.1103/RevModPhys.90.015005} {\bibfield  {journal}
  {\bibinfo  {journal} {Rev. Mod. Phys.}\ }\textbf {\bibinfo {volume} {90}},\
  \bibinfo {pages} {015005} (\bibinfo {year} {2018})}\BibitemShut {NoStop}%
\bibitem [{\citenamefont {Siddiqui}\ \emph {et~al.}(2020)\citenamefont
  {Siddiqui}, \citenamefont {Sklenar}, \citenamefont {Kang}, \citenamefont
  {Gilbert}, \citenamefont {Schleife}, \citenamefont {Mason},\ and\
  \citenamefont {Hoffmann}}]{siddiqui2020metallic}%
  \BibitemOpen
  \bibfield  {author} {\bibinfo {author} {\bibfnamefont {S.~A.}\ \bibnamefont
  {Siddiqui}}, \bibinfo {author} {\bibfnamefont {J.}~\bibnamefont {Sklenar}},
  \bibinfo {author} {\bibfnamefont {K.}~\bibnamefont {Kang}}, \bibinfo {author}
  {\bibfnamefont {M.~J.}\ \bibnamefont {Gilbert}}, \bibinfo {author}
  {\bibfnamefont {A.}~\bibnamefont {Schleife}}, \bibinfo {author}
  {\bibfnamefont {N.}~\bibnamefont {Mason}},\ and\ \bibinfo {author}
  {\bibfnamefont {A.}~\bibnamefont {Hoffmann}},\ }\bibfield  {title} {\bibinfo
  {title} {\textit{Metallic Antiferromagnets}},\ }\href
  {https://doi.org/10.1063/5.0009445} {\bibfield  {journal} {\bibinfo
  {journal} {J. Appl. Phys.}\ }\textbf {\bibinfo {volume} {128}},\ \bibinfo
  {pages} {040904} (\bibinfo {year} {2020})}\BibitemShut {NoStop}%
\bibitem [{\citenamefont {Wadley}\ \emph {et~al.}(2016)\citenamefont {Wadley},
  \citenamefont {Howells}, \citenamefont {{\v Z}elezn{\'y}}, \citenamefont
  {Andrews}, \citenamefont {Hills}, \citenamefont {Campion}, \citenamefont
  {Nov{\'a}k}, \citenamefont {Olejn{\'\i}k}, \citenamefont {Maccherozzi},
  \citenamefont {Dhesi} \emph {et~al.}}]{wadley2016electrical}%
  \BibitemOpen
  \bibfield  {author} {\bibinfo {author} {\bibfnamefont {P.}~\bibnamefont
  {Wadley}}, \bibinfo {author} {\bibfnamefont {B.}~\bibnamefont {Howells}},
  \bibinfo {author} {\bibfnamefont {J.}~\bibnamefont {{\v Z}elezn{\'y}}},
  \bibinfo {author} {\bibfnamefont {C.}~\bibnamefont {Andrews}}, \bibinfo
  {author} {\bibfnamefont {V.}~\bibnamefont {Hills}}, \bibinfo {author}
  {\bibfnamefont {R.~P.}\ \bibnamefont {Campion}}, \bibinfo {author}
  {\bibfnamefont {V.}~\bibnamefont {Nov{\'a}k}}, \bibinfo {author}
  {\bibfnamefont {K.}~\bibnamefont {Olejn{\'\i}k}}, \bibinfo {author}
  {\bibfnamefont {F.}~\bibnamefont {Maccherozzi}}, \bibinfo {author}
  {\bibfnamefont {S.~S.}\ \bibnamefont {Dhesi}}, \emph {et~al.},\ }\bibfield
  {title} {\bibinfo {title} {\textit{Electrical Switching of an
  Antiferromagnet}},\ }\href {https://doi.org/10.1126/science.aab1031}
  {\bibfield  {journal} {\bibinfo  {journal} {Science}\ }\textbf {\bibinfo
  {volume} {351}},\ \bibinfo {pages} {587} (\bibinfo {year}
  {2016})}\BibitemShut {NoStop}%
\bibitem [{\citenamefont {Bodnar}\ \emph {et~al.}(2018)\citenamefont {Bodnar},
  \citenamefont {{\v{S}}mejkal}, \citenamefont {Turek}, \citenamefont
  {Jungwirth}, \citenamefont {Gomonay}, \citenamefont {Sinova}, \citenamefont
  {Sapozhnik}, \citenamefont {Elmers}, \citenamefont {Kl{\"a}ui},\ and\
  \citenamefont {Jourdan}}]{Bodnar18NatCommn_SOT-Mn2Au}%
  \BibitemOpen
  \bibfield  {author} {\bibinfo {author} {\bibfnamefont {S.~Y.}\ \bibnamefont
  {Bodnar}}, \bibinfo {author} {\bibfnamefont {L.}~\bibnamefont
  {{\v{S}}mejkal}}, \bibinfo {author} {\bibfnamefont {I.}~\bibnamefont
  {Turek}}, \bibinfo {author} {\bibfnamefont {T.}~\bibnamefont {Jungwirth}},
  \bibinfo {author} {\bibfnamefont {O.}~\bibnamefont {Gomonay}}, \bibinfo
  {author} {\bibfnamefont {J.}~\bibnamefont {Sinova}}, \bibinfo {author}
  {\bibfnamefont {A.}~\bibnamefont {Sapozhnik}}, \bibinfo {author}
  {\bibfnamefont {H.-J.}\ \bibnamefont {Elmers}}, \bibinfo {author}
  {\bibfnamefont {M.}~\bibnamefont {Kl{\"a}ui}},\ and\ \bibinfo {author}
  {\bibfnamefont {M.}~\bibnamefont {Jourdan}},\ }\bibfield  {title} {\bibinfo
  {title} {\textit{Writing and Reading Antiferromagnetic
  $\textnormal{Mn}_2\textnormal{Au}$ by \text{N}{\'e}el Spin-Orbit Torques and
  Large Anisotropic Magnetoresistance}},\ }\href
  {https://doi.org/10.1038/s41467-017-02780-x} {\bibfield  {journal} {\bibinfo
  {journal} {Nat. Commun.}\ }\textbf {\bibinfo {volume} {9}},\ \bibinfo {pages}
  {1} (\bibinfo {year} {2018})}\BibitemShut {NoStop}%
\bibitem [{\citenamefont {Wadley}\ \emph {et~al.}(2018)\citenamefont {Wadley},
  \citenamefont {Reimers}, \citenamefont {Grzybowski}, \citenamefont {Andrews},
  \citenamefont {Wang}, \citenamefont {Chauhan}, \citenamefont {Gallagher},
  \citenamefont {Campion}, \citenamefont {Edmonds}, \citenamefont {Dhesi} \emph
  {et~al.}}]{Wadley18_CuMnAs-AF-domain-current}%
  \BibitemOpen
  \bibfield  {author} {\bibinfo {author} {\bibfnamefont {P.}~\bibnamefont
  {Wadley}}, \bibinfo {author} {\bibfnamefont {S.}~\bibnamefont {Reimers}},
  \bibinfo {author} {\bibfnamefont {M.~J.}\ \bibnamefont {Grzybowski}},
  \bibinfo {author} {\bibfnamefont {C.}~\bibnamefont {Andrews}}, \bibinfo
  {author} {\bibfnamefont {M.}~\bibnamefont {Wang}}, \bibinfo {author}
  {\bibfnamefont {J.~S.}\ \bibnamefont {Chauhan}}, \bibinfo {author}
  {\bibfnamefont {B.~L.}\ \bibnamefont {Gallagher}}, \bibinfo {author}
  {\bibfnamefont {R.~P.}\ \bibnamefont {Campion}}, \bibinfo {author}
  {\bibfnamefont {K.~W.}\ \bibnamefont {Edmonds}}, \bibinfo {author}
  {\bibfnamefont {S.~S.}\ \bibnamefont {Dhesi}}, \emph {et~al.},\ }\bibfield
  {title} {\bibinfo {title} {\textit{Current Polarity-Dependent Manipulation of
  Antiferromagnetic Domains}},\ }\href
  {https://doi.org/10.1038/s41565-018-0079-1} {\bibfield  {journal} {\bibinfo
  {journal} {Nat. Nanotechnol.}\ }\textbf {\bibinfo {volume} {13}},\ \bibinfo
  {pages} {362} (\bibinfo {year} {2018})}\BibitemShut {NoStop}%
\bibitem [{\citenamefont {Meinert}\ \emph {et~al.}(2018)\citenamefont
  {Meinert}, \citenamefont {Graulich},\ and\ \citenamefont
  {Matalla-Wagner}}]{Meinert18_AF-switching_Mn2Au}%
  \BibitemOpen
  \bibfield  {author} {\bibinfo {author} {\bibfnamefont {M.}~\bibnamefont
  {Meinert}}, \bibinfo {author} {\bibfnamefont {D.}~\bibnamefont {Graulich}},\
  and\ \bibinfo {author} {\bibfnamefont {T.}~\bibnamefont {Matalla-Wagner}},\
  }\bibfield  {title} {\bibinfo {title} {\textit{Electrical Switching of
  Antiferromagnetic $\textnormal{Mn}_2\textnormal{Au}$ and the Role of Thermal
  Activation}},\ }\href {https://doi.org/10.1103/PhysRevApplied.9.064040}
  {\bibfield  {journal} {\bibinfo  {journal} {Phys. Rev. Applied}\ }\textbf
  {\bibinfo {volume} {9}},\ \bibinfo {pages} {064040} (\bibinfo {year}
  {2018})}\BibitemShut {NoStop}%
\bibitem [{\citenamefont {Zhou}\ \emph {et~al.}(2018)\citenamefont {Zhou},
  \citenamefont {Zhang}, \citenamefont {Li}, \citenamefont {Chen},
  \citenamefont {Shi}, \citenamefont {Tan}, \citenamefont {Gu}, \citenamefont
  {Saleem}, \citenamefont {Wu}, \citenamefont {Pan},\ and\ \citenamefont
  {Song}}]{cSong18PRAppl_SOT-AF-Mn2Au}%
  \BibitemOpen
  \bibfield  {author} {\bibinfo {author} {\bibfnamefont {X.~F.}\ \bibnamefont
  {Zhou}}, \bibinfo {author} {\bibfnamefont {J.}~\bibnamefont {Zhang}},
  \bibinfo {author} {\bibfnamefont {F.}~\bibnamefont {Li}}, \bibinfo {author}
  {\bibfnamefont {X.~Z.}\ \bibnamefont {Chen}}, \bibinfo {author}
  {\bibfnamefont {G.~Y.}\ \bibnamefont {Shi}}, \bibinfo {author} {\bibfnamefont
  {Y.~Z.}\ \bibnamefont {Tan}}, \bibinfo {author} {\bibfnamefont {Y.~D.}\
  \bibnamefont {Gu}}, \bibinfo {author} {\bibfnamefont {M.~S.}\ \bibnamefont
  {Saleem}}, \bibinfo {author} {\bibfnamefont {H.~Q.}\ \bibnamefont {Wu}},
  \bibinfo {author} {\bibfnamefont {F.}~\bibnamefont {Pan}},\ and\ \bibinfo
  {author} {\bibfnamefont {C.}~\bibnamefont {Song}},\ }\bibfield  {title}
  {\bibinfo {title} {\textit{Strong Orientation-Dependent Spin-Orbit Torque in
  Thin Films of the Antiferromagnet $\textnormal{Mn}_2\textnormal{Au}$}},\
  }\href {https://doi.org/10.1103/PhysRevApplied.9.054028} {\bibfield
  {journal} {\bibinfo  {journal} {Phys. Rev. Applied}\ }\textbf {\bibinfo
  {volume} {9}},\ \bibinfo {pages} {054028} (\bibinfo {year}
  {2018})}\BibitemShut {NoStop}%
\bibitem [{\citenamefont {DuttaGupta}\ \emph {et~al.}(2020)\citenamefont
  {DuttaGupta}, \citenamefont {Kurenkov}, \citenamefont {Tretiakov},
  \citenamefont {Krishnaswamy}, \citenamefont {Sala}, \citenamefont
  {Krizakova}, \citenamefont {Maccherozzi}, \citenamefont {Dhesi},
  \citenamefont {Gambardella}, \citenamefont {Fukami} \emph
  {et~al.}}]{hOhno20NatCommn_switching-PtMn-Pt}%
  \BibitemOpen
  \bibfield  {author} {\bibinfo {author} {\bibfnamefont {S.}~\bibnamefont
  {DuttaGupta}}, \bibinfo {author} {\bibfnamefont {A.}~\bibnamefont
  {Kurenkov}}, \bibinfo {author} {\bibfnamefont {O.~A.}\ \bibnamefont
  {Tretiakov}}, \bibinfo {author} {\bibfnamefont {G.}~\bibnamefont
  {Krishnaswamy}}, \bibinfo {author} {\bibfnamefont {G.}~\bibnamefont {Sala}},
  \bibinfo {author} {\bibfnamefont {V.}~\bibnamefont {Krizakova}}, \bibinfo
  {author} {\bibfnamefont {F.}~\bibnamefont {Maccherozzi}}, \bibinfo {author}
  {\bibfnamefont {S.}~\bibnamefont {Dhesi}}, \bibinfo {author} {\bibfnamefont
  {P.}~\bibnamefont {Gambardella}}, \bibinfo {author} {\bibfnamefont
  {S.}~\bibnamefont {Fukami}}, \emph {et~al.},\ }\bibfield  {title} {\bibinfo
  {title} {\textit{Spin-Orbit Torque Switching of an Antiferromagnetic Metallic
  Heterostructure}},\ }\href {https://doi.org/10.1038/s41467-020-19511-4}
  {\bibfield  {journal} {\bibinfo  {journal} {Nat. Commun.}\ }\textbf {\bibinfo
  {volume} {11}},\ \bibinfo {pages} {1} (\bibinfo {year} {2020})}\BibitemShut
  {NoStop}%
\bibitem [{\citenamefont {{Suzuki}}\ \emph {et~al.}(2016)\citenamefont
  {{Suzuki}}, \citenamefont {{Chisnell}}, \citenamefont {{Devarakonda}},
  \citenamefont {{Liu}}, \citenamefont {{Feng}}, \citenamefont {{Xiao}},
  \citenamefont {{Lynn}},\ and\ \citenamefont
  {{Checkelsky}}}]{Suzuki16NP_AHE-Heusler-AF}%
  \BibitemOpen
  \bibfield  {author} {\bibinfo {author} {\bibfnamefont {T.}~\bibnamefont
  {{Suzuki}}}, \bibinfo {author} {\bibfnamefont {R.}~\bibnamefont
  {{Chisnell}}}, \bibinfo {author} {\bibfnamefont {A.}~\bibnamefont
  {{Devarakonda}}}, \bibinfo {author} {\bibfnamefont {Y.~T.}\ \bibnamefont
  {{Liu}}}, \bibinfo {author} {\bibfnamefont {W.}~\bibnamefont {{Feng}}},
  \bibinfo {author} {\bibfnamefont {D.}~\bibnamefont {{Xiao}}}, \bibinfo
  {author} {\bibfnamefont {J.~W.}\ \bibnamefont {{Lynn}}},\ and\ \bibinfo
  {author} {\bibfnamefont {J.~G.}\ \bibnamefont {{Checkelsky}}},\ }\bibfield
  {title} {\bibinfo {title} {{\textit{Large Anomalous Hall Effect in a
  Half-Heusler Antiferromagnet}}},\ }\href {https://doi.org/10.1038/nphys3831}
  {\bibfield  {journal} {\bibinfo  {journal} {Nat. Phys.}\ }\textbf {\bibinfo
  {volume} {12}},\ \bibinfo {pages} {1119} (\bibinfo {year}
  {2016})}\BibitemShut {NoStop}%
\bibitem [{\citenamefont {Takahashi}\ \emph {et~al.}(2018)\citenamefont
  {Takahashi}, \citenamefont {Ishizuka}, \citenamefont {Murata}, \citenamefont
  {Wang}, \citenamefont {Tokura}, \citenamefont {Nagaosa},\ and\ \citenamefont
  {Kawasaki}}]{Takahashi18SciAdv_AHE-ETO-AF}%
  \BibitemOpen
  \bibfield  {author} {\bibinfo {author} {\bibfnamefont {K.~S.}\ \bibnamefont
  {Takahashi}}, \bibinfo {author} {\bibfnamefont {H.}~\bibnamefont {Ishizuka}},
  \bibinfo {author} {\bibfnamefont {T.}~\bibnamefont {Murata}}, \bibinfo
  {author} {\bibfnamefont {Q.~Y.}\ \bibnamefont {Wang}}, \bibinfo {author}
  {\bibfnamefont {Y.}~\bibnamefont {Tokura}}, \bibinfo {author} {\bibfnamefont
  {N.}~\bibnamefont {Nagaosa}},\ and\ \bibinfo {author} {\bibfnamefont
  {M.}~\bibnamefont {Kawasaki}},\ }\bibfield  {title} {\bibinfo {title}
  {\textit{Anomalous \text{H}all Effect Derived from Multiple \text{W}eyl Nodes
  in High-Mobility $\textnormal{EuTiO}_3$ Films}},\ }\href
  {https://doi.org/10.1126/sciadv.aar7880} {\bibfield  {journal} {\bibinfo
  {journal} {Sci. Adv.}\ }\textbf {\bibinfo {volume} {4}} (\bibinfo {year}
  {2018})}\BibitemShut {NoStop}%
\bibitem [{\citenamefont {Li}\ \emph {et~al.}()\citenamefont {Li},
  \citenamefont {MacDonald},\ and\ \citenamefont
  {Chen}}]{arxiv19_hChen_QAHE-cAF}%
  \BibitemOpen
  \bibfield  {author} {\bibinfo {author} {\bibfnamefont {X.}~\bibnamefont
  {Li}}, \bibinfo {author} {\bibfnamefont {A.~H.}\ \bibnamefont {MacDonald}},\
  and\ \bibinfo {author} {\bibfnamefont {H.}~\bibnamefont {Chen}},\ }\bibfield
  {title} {\bibinfo {title} {\textit{Quantum Anomalous \text{H}all Effect
  through Canted Antiferromagnetism}},\ }\href@noop {} {\ }\Eprint
  {https://arxiv.org/abs/1902.10650} {arXiv:1902.10650 [cond-mat.mes-hall]}
  \BibitemShut {NoStop}%
\bibitem [{\citenamefont {Yang}\ \emph {et~al.}(2020)\citenamefont {Yang},
  \citenamefont {Corasaniti}, \citenamefont {Le}, \citenamefont {Liao},
  \citenamefont {Wang}, \citenamefont {Du}, \citenamefont {Petrovic},
  \citenamefont {Qiu}, \citenamefont {Hu},\ and\ \citenamefont
  {Degiorgi}}]{rYang20PRL_sc-AF-DSM}%
  \BibitemOpen
  \bibfield  {author} {\bibinfo {author} {\bibfnamefont {R.}~\bibnamefont
  {Yang}}, \bibinfo {author} {\bibfnamefont {M.}~\bibnamefont {Corasaniti}},
  \bibinfo {author} {\bibfnamefont {C.~C.}\ \bibnamefont {Le}}, \bibinfo
  {author} {\bibfnamefont {Z.~Y.}\ \bibnamefont {Liao}}, \bibinfo {author}
  {\bibfnamefont {A.~F.}\ \bibnamefont {Wang}}, \bibinfo {author}
  {\bibfnamefont {Q.}~\bibnamefont {Du}}, \bibinfo {author} {\bibfnamefont
  {C.}~\bibnamefont {Petrovic}}, \bibinfo {author} {\bibfnamefont {X.~G.}\
  \bibnamefont {Qiu}}, \bibinfo {author} {\bibfnamefont {J.~P.}\ \bibnamefont
  {Hu}},\ and\ \bibinfo {author} {\bibfnamefont {L.}~\bibnamefont {Degiorgi}},\
  }\bibfield  {title} {\bibinfo {title} {\textit{Spin-Canting-Induced Band
  Reconstruction in the \text{D}irac Material
  $\textnormal{Ca}_{1\ensuremath{-}x}\textnormal{Na}_{x}\textnormal{MnBi}_{2}$}},\
  }\href {https://doi.org/10.1103/PhysRevLett.124.137201} {\bibfield  {journal}
  {\bibinfo  {journal} {Phys. Rev. Lett.}\ }\textbf {\bibinfo {volume} {124}},\
  \bibinfo {pages} {137201} (\bibinfo {year} {2020})}\BibitemShut {NoStop}%
\bibitem [{\citenamefont {{Kipp}}\ \emph {et~al.}(2021)\citenamefont {{Kipp}},
  \citenamefont {{Samanta}}, \citenamefont {{Lux}}, \citenamefont {{Merte}},
  \citenamefont {{Go}}, \citenamefont {{Hanke}}, \citenamefont {{Redies}},
  \citenamefont {{Freimuth}}, \citenamefont {{Bl{\"u}gel}}, \citenamefont
  {{Le{\v{z}}ai{\'c}}},\ and\ \citenamefont
  {{Mokrousov}}}]{Kipp21ComPhys_CHE-cAF}%
  \BibitemOpen
  \bibfield  {author} {\bibinfo {author} {\bibfnamefont {J.}~\bibnamefont
  {{Kipp}}}, \bibinfo {author} {\bibfnamefont {K.}~\bibnamefont {{Samanta}}},
  \bibinfo {author} {\bibfnamefont {F.~R.}\ \bibnamefont {{Lux}}}, \bibinfo
  {author} {\bibfnamefont {M.}~\bibnamefont {{Merte}}}, \bibinfo {author}
  {\bibfnamefont {D.}~\bibnamefont {{Go}}}, \bibinfo {author} {\bibfnamefont
  {J.-P.}\ \bibnamefont {{Hanke}}}, \bibinfo {author} {\bibfnamefont
  {M.}~\bibnamefont {{Redies}}}, \bibinfo {author} {\bibfnamefont
  {F.}~\bibnamefont {{Freimuth}}}, \bibinfo {author} {\bibfnamefont
  {S.}~\bibnamefont {{Bl{\"u}gel}}}, \bibinfo {author} {\bibfnamefont
  {M.}~\bibnamefont {{Le{\v{z}}ai{\'c}}}},\ and\ \bibinfo {author}
  {\bibfnamefont {Y.}~\bibnamefont {{Mokrousov}}},\ }\bibfield  {title}
  {\bibinfo {title} {{\textit{The Chiral Hall Effect in Canted Ferromagnets and
  Antiferromagnets}}},\ }\href {https://doi.org/10.1038/s42005-021-00587-3}
  {\bibfield  {journal} {\bibinfo  {journal} {Commun. Phys.}\ }\textbf
  {\bibinfo {volume} {4}},\ \bibinfo {eid} {99} (\bibinfo {year}
  {2021})}\BibitemShut {NoStop}%
\bibitem [{\citenamefont {{Kimel}}\ \emph {et~al.}(2004)\citenamefont
  {{Kimel}}, \citenamefont {{Kirilyuk}}, \citenamefont {{Tsvetkov}},
  \citenamefont {{Pisarev}},\ and\ \citenamefont {{Rasing}}}]{kimel2004laser}%
  \BibitemOpen
  \bibfield  {author} {\bibinfo {author} {\bibfnamefont {A.~V.}\ \bibnamefont
  {{Kimel}}}, \bibinfo {author} {\bibfnamefont {A.}~\bibnamefont {{Kirilyuk}}},
  \bibinfo {author} {\bibfnamefont {A.}~\bibnamefont {{Tsvetkov}}}, \bibinfo
  {author} {\bibfnamefont {R.~V.}\ \bibnamefont {{Pisarev}}},\ and\ \bibinfo
  {author} {\bibfnamefont {T.}~\bibnamefont {{Rasing}}},\ }\bibfield  {title}
  {\bibinfo {title} {{\textit{Laser-Induced Ultrafast Spin Reorientation in the
  Antiferromagnet \textnormal{TmFeO}$_{3}$}}},\ }\href
  {https://doi.org/10.1038/nature02659} {\bibfield  {journal} {\bibinfo
  {journal} {Nature}\ }\textbf {\bibinfo {volume} {429}},\ \bibinfo {pages}
  {850} (\bibinfo {year} {2004})}\BibitemShut {NoStop}%
\bibitem [{\citenamefont {{Saidl}}\ \emph {et~al.}(2017)\citenamefont
  {{Saidl}}, \citenamefont {{N{\v{e}}mec}}, \citenamefont {{Wadley}},
  \citenamefont {{Hills}}, \citenamefont {{Campion}}, \citenamefont
  {{Nov{\'a}k}}, \citenamefont {{Edmonds}}, \citenamefont {{Maccherozzi}},
  \citenamefont {{Dhesi}}, \citenamefont {{Gallagher}} \emph
  {et~al.}}]{saidl2017optical}%
  \BibitemOpen
  \bibfield  {author} {\bibinfo {author} {\bibfnamefont {V.}~\bibnamefont
  {{Saidl}}}, \bibinfo {author} {\bibfnamefont {P.}~\bibnamefont
  {{N{\v{e}}mec}}}, \bibinfo {author} {\bibfnamefont {P.}~\bibnamefont
  {{Wadley}}}, \bibinfo {author} {\bibfnamefont {V.}~\bibnamefont {{Hills}}},
  \bibinfo {author} {\bibfnamefont {R.~P.}\ \bibnamefont {{Campion}}}, \bibinfo
  {author} {\bibfnamefont {V.}~\bibnamefont {{Nov{\'a}k}}}, \bibinfo {author}
  {\bibfnamefont {K.~W.}\ \bibnamefont {{Edmonds}}}, \bibinfo {author}
  {\bibfnamefont {F.}~\bibnamefont {{Maccherozzi}}}, \bibinfo {author}
  {\bibfnamefont {S.~S.}\ \bibnamefont {{Dhesi}}}, \bibinfo {author}
  {\bibfnamefont {B.~L.}\ \bibnamefont {{Gallagher}}}, \emph {et~al.},\
  }\bibfield  {title} {\bibinfo {title} {{\textit{Optical Determination of the
  N{\'e}el Vector in a \textnormal{CuMnAs} Thin-Film Antiferromagnet}}},\
  }\href {https://doi.org/10.1038/nphoton.2016.255} {\bibfield  {journal}
  {\bibinfo  {journal} {Nat. Photon.}\ }\textbf {\bibinfo {volume} {11}},\
  \bibinfo {pages} {91} (\bibinfo {year} {2017})}\BibitemShut {NoStop}%
\bibitem [{\citenamefont {Marti}\ \emph {et~al.}(2014)\citenamefont {Marti},
  \citenamefont {Fina}, \citenamefont {Frontera}, \citenamefont {Liu},
  \citenamefont {Wadley}, \citenamefont {He}, \citenamefont {Paull},
  \citenamefont {Clarkson}, \citenamefont {Kudrnovsk{\`y}}, \citenamefont
  {Turek} \emph {et~al.}}]{marti2014room}%
  \BibitemOpen
  \bibfield  {author} {\bibinfo {author} {\bibfnamefont {X.}~\bibnamefont
  {Marti}}, \bibinfo {author} {\bibfnamefont {I.}~\bibnamefont {Fina}},
  \bibinfo {author} {\bibfnamefont {C.}~\bibnamefont {Frontera}}, \bibinfo
  {author} {\bibfnamefont {J.}~\bibnamefont {Liu}}, \bibinfo {author}
  {\bibfnamefont {P.}~\bibnamefont {Wadley}}, \bibinfo {author} {\bibfnamefont
  {Q.}~\bibnamefont {He}}, \bibinfo {author} {\bibfnamefont {R.}~\bibnamefont
  {Paull}}, \bibinfo {author} {\bibfnamefont {J.}~\bibnamefont {Clarkson}},
  \bibinfo {author} {\bibfnamefont {J.}~\bibnamefont {Kudrnovsk{\`y}}},
  \bibinfo {author} {\bibfnamefont {I.}~\bibnamefont {Turek}}, \emph {et~al.},\
  }\bibfield  {title} {\bibinfo {title} {\textit{Room-Temperature
  Antiferromagnetic Memory Resistor}},\ }\href
  {https://doi.org/10.1038/nmat3861} {\bibfield  {journal} {\bibinfo  {journal}
  {Nat. Mater.}\ }\textbf {\bibinfo {volume} {13}},\ \bibinfo {pages} {367}
  (\bibinfo {year} {2014})}\BibitemShut {NoStop}%
\bibitem [{\citenamefont {{Moriyama}}\ \emph {et~al.}(2018)\citenamefont
  {{Moriyama}}, \citenamefont {{Oda}}, \citenamefont {{Ohkochi}}, \citenamefont
  {{Kimata}},\ and\ \citenamefont {{Ono}}}]{moriyama2018spin}%
  \BibitemOpen
  \bibfield  {author} {\bibinfo {author} {\bibfnamefont {T.}~\bibnamefont
  {{Moriyama}}}, \bibinfo {author} {\bibfnamefont {K.}~\bibnamefont {{Oda}}},
  \bibinfo {author} {\bibfnamefont {T.}~\bibnamefont {{Ohkochi}}}, \bibinfo
  {author} {\bibfnamefont {M.}~\bibnamefont {{Kimata}}},\ and\ \bibinfo
  {author} {\bibfnamefont {T.}~\bibnamefont {{Ono}}},\ }\bibfield  {title}
  {\bibinfo {title} {{\textit{Spin Torque Control of Antiferromagnetic Moments
  in \textnormal{NiO}}}},\ }\href {https://doi.org/10.1038/s41598-018-32508-w}
  {\bibfield  {journal} {\bibinfo  {journal} {Sci. Rep.}\ }\textbf {\bibinfo
  {volume} {8}},\ \bibinfo {eid} {14167} (\bibinfo {year} {2018})}\BibitemShut
  {NoStop}%
\bibitem [{\citenamefont {Baldrati}\ \emph {et~al.}(2018)\citenamefont
  {Baldrati}, \citenamefont {Ross}, \citenamefont {Niizeki}, \citenamefont
  {Schneider}, \citenamefont {Ramos}, \citenamefont {Cramer}, \citenamefont
  {Gomonay}, \citenamefont {Filianina}, \citenamefont {Savchenko},
  \citenamefont {Heinze} \emph {et~al.}}]{baldrati2018full}%
  \BibitemOpen
  \bibfield  {author} {\bibinfo {author} {\bibfnamefont {L.}~\bibnamefont
  {Baldrati}}, \bibinfo {author} {\bibfnamefont {A.}~\bibnamefont {Ross}},
  \bibinfo {author} {\bibfnamefont {T.}~\bibnamefont {Niizeki}}, \bibinfo
  {author} {\bibfnamefont {C.}~\bibnamefont {Schneider}}, \bibinfo {author}
  {\bibfnamefont {R.}~\bibnamefont {Ramos}}, \bibinfo {author} {\bibfnamefont
  {J.}~\bibnamefont {Cramer}}, \bibinfo {author} {\bibfnamefont
  {O.}~\bibnamefont {Gomonay}}, \bibinfo {author} {\bibfnamefont
  {M.}~\bibnamefont {Filianina}}, \bibinfo {author} {\bibfnamefont
  {T.}~\bibnamefont {Savchenko}}, \bibinfo {author} {\bibfnamefont
  {D.}~\bibnamefont {Heinze}}, \emph {et~al.},\ }\bibfield  {title} {\bibinfo
  {title} {\textit{Full Angular Dependence of the Spin \text{H}all and Ordinary
  Magnetoresistance in Epitaxial Antiferromagnetic \textnormal{NiO(001)/Pt}
  Thin Films}},\ }\href {https://doi.org/10.1103/PhysRevB.98.024422} {\bibfield
   {journal} {\bibinfo  {journal} {Phys. Rev. B}\ }\textbf {\bibinfo {volume}
  {98}},\ \bibinfo {pages} {024422} (\bibinfo {year} {2018})}\BibitemShut
  {NoStop}%
\bibitem [{\citenamefont {{Moriyama}}\ \emph {et~al.}(2015)\citenamefont
  {{Moriyama}}, \citenamefont {{Matsuzaki}}, \citenamefont {{Kim}},
  \citenamefont {{Suzuki}}, \citenamefont {{Taniyama}},\ and\ \citenamefont
  {{Ono}}}]{moriyama2015sequential}%
  \BibitemOpen
  \bibfield  {author} {\bibinfo {author} {\bibfnamefont {T.}~\bibnamefont
  {{Moriyama}}}, \bibinfo {author} {\bibfnamefont {N.}~\bibnamefont
  {{Matsuzaki}}}, \bibinfo {author} {\bibfnamefont {K.-J.}\ \bibnamefont
  {{Kim}}}, \bibinfo {author} {\bibfnamefont {I.}~\bibnamefont {{Suzuki}}},
  \bibinfo {author} {\bibfnamefont {T.}~\bibnamefont {{Taniyama}}},\ and\
  \bibinfo {author} {\bibfnamefont {T.}~\bibnamefont {{Ono}}},\ }\bibfield
  {title} {\bibinfo {title} {{\textit{Sequential Write-Read Operations in
  \textnormal{FeRh} Antiferromagnetic Memory}}},\ }\href
  {https://doi.org/10.1063/1.4931567} {\bibfield  {journal} {\bibinfo
  {journal} {Appl. Phys. Lett.}\ }\textbf {\bibinfo {volume} {107}},\ \bibinfo
  {eid} {122403} (\bibinfo {year} {2015})}\BibitemShut {NoStop}%
\bibitem [{\citenamefont {{Hoogeboom}}\ \emph {et~al.}(2017)\citenamefont
  {{Hoogeboom}}, \citenamefont {{Aqeel}}, \citenamefont {{Kuschel}},
  \citenamefont {{Palstra}},\ and\ \citenamefont {{van
  Wees}}}]{hoogeboom2017negative}%
  \BibitemOpen
  \bibfield  {author} {\bibinfo {author} {\bibfnamefont {G.~R.}\ \bibnamefont
  {{Hoogeboom}}}, \bibinfo {author} {\bibfnamefont {A.}~\bibnamefont
  {{Aqeel}}}, \bibinfo {author} {\bibfnamefont {T.}~\bibnamefont {{Kuschel}}},
  \bibinfo {author} {\bibfnamefont {T.~T.~M.}\ \bibnamefont {{Palstra}}},\ and\
  \bibinfo {author} {\bibfnamefont {B.~J.}\ \bibnamefont {{van Wees}}},\
  }\bibfield  {title} {\bibinfo {title} {{\textit{Negative Spin Hall
  Magnetoresistance of \textnormal{Pt} on the Bulk Easy-Plane Antiferromagnet
  \textnormal{NiO}}}},\ }\href {https://doi.org/10.1063/1.4997588} {\bibfield
  {journal} {\bibinfo  {journal} {Appl. Phys. Lett.}\ }\textbf {\bibinfo
  {volume} {111}},\ \bibinfo {eid} {052409} (\bibinfo {year}
  {2017})}\BibitemShut {NoStop}%
\bibitem [{\citenamefont {Fischer}\ \emph {et~al.}(2018)\citenamefont
  {Fischer}, \citenamefont {Gomonay}, \citenamefont {Schlitz}, \citenamefont
  {Ganzhorn}, \citenamefont {Vlietstra}, \citenamefont {Althammer},
  \citenamefont {Huebl}, \citenamefont {Opel}, \citenamefont {Gross},
  \citenamefont {Goennenwein},\ and\ \citenamefont
  {Gepr\"ags}}]{fischer2018spin}%
  \BibitemOpen
  \bibfield  {author} {\bibinfo {author} {\bibfnamefont {J.}~\bibnamefont
  {Fischer}}, \bibinfo {author} {\bibfnamefont {O.}~\bibnamefont {Gomonay}},
  \bibinfo {author} {\bibfnamefont {R.}~\bibnamefont {Schlitz}}, \bibinfo
  {author} {\bibfnamefont {K.}~\bibnamefont {Ganzhorn}}, \bibinfo {author}
  {\bibfnamefont {N.}~\bibnamefont {Vlietstra}}, \bibinfo {author}
  {\bibfnamefont {M.}~\bibnamefont {Althammer}}, \bibinfo {author}
  {\bibfnamefont {H.}~\bibnamefont {Huebl}}, \bibinfo {author} {\bibfnamefont
  {M.}~\bibnamefont {Opel}}, \bibinfo {author} {\bibfnamefont {R.}~\bibnamefont
  {Gross}}, \bibinfo {author} {\bibfnamefont {S.~T.~B.}\ \bibnamefont
  {Goennenwein}},\ and\ \bibinfo {author} {\bibfnamefont {S.}~\bibnamefont
  {Gepr\"ags}},\ }\bibfield  {title} {\bibinfo {title} {\textit{Spin
  \text{H}all Magnetoresistance in Antiferromagnet/Heavy-Metal
  Heterostructures}},\ }\href {https://doi.org/10.1103/PhysRevB.97.014417}
  {\bibfield  {journal} {\bibinfo  {journal} {Phys. Rev. B}\ }\textbf {\bibinfo
  {volume} {97}},\ \bibinfo {pages} {014417} (\bibinfo {year}
  {2018})}\BibitemShut {NoStop}%
\bibitem [{\citenamefont {Manchon}(2017)}]{manchon2017spin}%
  \BibitemOpen
  \bibfield  {author} {\bibinfo {author} {\bibfnamefont {A.}~\bibnamefont
  {Manchon}},\ }\bibfield  {title} {\bibinfo {title} {\textit{Spin \text{H}all
  Magnetoresistance in Antiferromagnet/Normal Metal Bilayers}},\ }\href
  {https://doi.org/10.1002/pssr.201600409} {\bibfield  {journal} {\bibinfo
  {journal} {Phys. Status Solidi (RRL)--Rapid Research Letters}\ }\textbf
  {\bibinfo {volume} {11}},\ \bibinfo {pages} {1600409} (\bibinfo {year}
  {2017})}\BibitemShut {NoStop}%
\bibitem [{\citenamefont {{McGuire}}\ and\ \citenamefont
  {{Potter}}(1975)}]{mcguire1975anisotropic}%
  \BibitemOpen
  \bibfield  {author} {\bibinfo {author} {\bibfnamefont {T.}~\bibnamefont
  {{McGuire}}}\ and\ \bibinfo {author} {\bibfnamefont {R.}~\bibnamefont
  {{Potter}}},\ }\bibfield  {title} {\bibinfo {title} {{\textit{Anisotropic
  Magnetoresistance in Ferromagnetic 3d Alloys}}},\ }\href
  {https://doi.org/10.1109/TMAG.1975.1058782} {\bibfield  {journal} {\bibinfo
  {journal} {IEEE Trans. Magn.}\ }\textbf {\bibinfo {volume} {11}},\ \bibinfo
  {pages} {1018} (\bibinfo {year} {1975})}\BibitemShut {NoStop}%
\bibitem [{\citenamefont {Nakayama}\ \emph {et~al.}(2013)\citenamefont
  {Nakayama}, \citenamefont {Althammer}, \citenamefont {Chen}, \citenamefont
  {Uchida}, \citenamefont {Kajiwara}, \citenamefont {Kikuchi}, \citenamefont
  {Ohtani}, \citenamefont {Gepr\"ags}, \citenamefont {Opel}, \citenamefont
  {Takahashi} \emph {et~al.}}]{nakayama2013spin}%
  \BibitemOpen
  \bibfield  {author} {\bibinfo {author} {\bibfnamefont {H.}~\bibnamefont
  {Nakayama}}, \bibinfo {author} {\bibfnamefont {M.}~\bibnamefont {Althammer}},
  \bibinfo {author} {\bibfnamefont {Y.-T.}\ \bibnamefont {Chen}}, \bibinfo
  {author} {\bibfnamefont {K.}~\bibnamefont {Uchida}}, \bibinfo {author}
  {\bibfnamefont {Y.}~\bibnamefont {Kajiwara}}, \bibinfo {author}
  {\bibfnamefont {D.}~\bibnamefont {Kikuchi}}, \bibinfo {author} {\bibfnamefont
  {T.}~\bibnamefont {Ohtani}}, \bibinfo {author} {\bibfnamefont
  {S.}~\bibnamefont {Gepr\"ags}}, \bibinfo {author} {\bibfnamefont
  {M.}~\bibnamefont {Opel}}, \bibinfo {author} {\bibfnamefont {S.}~\bibnamefont
  {Takahashi}}, \emph {et~al.},\ }\bibfield  {title} {\bibinfo {title}
  {\textit{Spin \text{H}all Magnetoresistance Induced by a Nonequilibrium
  Proximity Effect}},\ }\href {https://doi.org/10.1103/PhysRevLett.110.206601}
  {\bibfield  {journal} {\bibinfo  {journal} {Phys. Rev. Lett.}\ }\textbf
  {\bibinfo {volume} {110}},\ \bibinfo {pages} {206601} (\bibinfo {year}
  {2013})}\BibitemShut {NoStop}%
\bibitem [{\citenamefont {Avci}\ \emph {et~al.}(2015)\citenamefont {Avci},
  \citenamefont {Garello}, \citenamefont {Ghosh}, \citenamefont {Gabureac},
  \citenamefont {Alvarado},\ and\ \citenamefont
  {Gambardella}}]{avci2015unidirectional}%
  \BibitemOpen
  \bibfield  {author} {\bibinfo {author} {\bibfnamefont {C.~O.}\ \bibnamefont
  {Avci}}, \bibinfo {author} {\bibfnamefont {K.}~\bibnamefont {Garello}},
  \bibinfo {author} {\bibfnamefont {A.}~\bibnamefont {Ghosh}}, \bibinfo
  {author} {\bibfnamefont {M.}~\bibnamefont {Gabureac}}, \bibinfo {author}
  {\bibfnamefont {S.~F.}\ \bibnamefont {Alvarado}},\ and\ \bibinfo {author}
  {\bibfnamefont {P.}~\bibnamefont {Gambardella}},\ }\bibfield  {title}
  {\bibinfo {title} {\textit{Unidirectional Spin \text{H}all Magnetoresistance
  in Ferromagnet/Normal Metal Bilayers}},\ }\href
  {https://doi.org/10.1038/nphys3356} {\bibfield  {journal} {\bibinfo
  {journal} {Nat. Phys.}\ }\textbf {\bibinfo {volume} {11}},\ \bibinfo {pages}
  {570} (\bibinfo {year} {2015})}\BibitemShut {NoStop}%
\bibitem [{\citenamefont {Zhang}\ and\ \citenamefont
  {Vignale}(2016)}]{zhang2016prb}%
  \BibitemOpen
  \bibfield  {author} {\bibinfo {author} {\bibfnamefont {S.~S.-L.}\
  \bibnamefont {Zhang}}\ and\ \bibinfo {author} {\bibfnamefont
  {G.}~\bibnamefont {Vignale}},\ }\bibfield  {title} {\bibinfo {title}
  {\textit{Theory of Unidirectional Spin \text{H}all Magnetoresistance in
  Heavy-Metal/Ferromagnetic-Metal Bilayers}},\ }\href
  {https://doi.org/10.1103/PhysRevB.94.140411} {\bibfield  {journal} {\bibinfo
  {journal} {Phys. Rev. B}\ }\textbf {\bibinfo {volume} {94}},\ \bibinfo
  {pages} {140411} (\bibinfo {year} {2016})}\BibitemShut {NoStop}%
\bibitem [{\citenamefont {Yasuda}\ \emph {et~al.}(2016)\citenamefont {Yasuda},
  \citenamefont {Tsukazaki}, \citenamefont {Yoshimi}, \citenamefont
  {Takahashi}, \citenamefont {Kawasaki},\ and\ \citenamefont
  {Tokura}}]{yasuda2016PRL_TI-UMR}%
  \BibitemOpen
  \bibfield  {author} {\bibinfo {author} {\bibfnamefont {K.}~\bibnamefont
  {Yasuda}}, \bibinfo {author} {\bibfnamefont {A.}~\bibnamefont {Tsukazaki}},
  \bibinfo {author} {\bibfnamefont {R.}~\bibnamefont {Yoshimi}}, \bibinfo
  {author} {\bibfnamefont {K.~S.}\ \bibnamefont {Takahashi}}, \bibinfo {author}
  {\bibfnamefont {M.}~\bibnamefont {Kawasaki}},\ and\ \bibinfo {author}
  {\bibfnamefont {Y.}~\bibnamefont {Tokura}},\ }\bibfield  {title} {\bibinfo
  {title} {\textit{Large Unidirectional Magnetoresistance in a Magnetic
  Topological Insulator}},\ }\href
  {https://doi.org/10.1103/PhysRevLett.117.127202} {\bibfield  {journal}
  {\bibinfo  {journal} {Phys. Rev. Lett.}\ }\textbf {\bibinfo {volume} {117}},\
  \bibinfo {pages} {127202} (\bibinfo {year} {2016})}\BibitemShut {NoStop}%
\bibitem [{\citenamefont {Langenfeld}\ \emph {et~al.}(2016)\citenamefont
  {Langenfeld}, \citenamefont {Tshitoyan}, \citenamefont {Fang}, \citenamefont
  {Wells}, \citenamefont {Moore},\ and\ \citenamefont
  {Ferguson}}]{langenfeld2016exchange}%
  \BibitemOpen
  \bibfield  {author} {\bibinfo {author} {\bibfnamefont {S.}~\bibnamefont
  {Langenfeld}}, \bibinfo {author} {\bibfnamefont {V.}~\bibnamefont
  {Tshitoyan}}, \bibinfo {author} {\bibfnamefont {Z.}~\bibnamefont {Fang}},
  \bibinfo {author} {\bibfnamefont {A.}~\bibnamefont {Wells}}, \bibinfo
  {author} {\bibfnamefont {T.}~\bibnamefont {Moore}},\ and\ \bibinfo {author}
  {\bibfnamefont {A.}~\bibnamefont {Ferguson}},\ }\bibfield  {title} {\bibinfo
  {title} {\textit{Exchange Magnon Induced Resistance Asymmetry in Permalloy
  Spin-\text{H}all Oscillators}},\ }\href {https://doi.org/10.1063/1.4948921}
  {\bibfield  {journal} {\bibinfo  {journal} {Appl. Phys. Lett.}\ }\textbf
  {\bibinfo {volume} {108}},\ \bibinfo {pages} {192402} (\bibinfo {year}
  {2016})}\BibitemShut {NoStop}%
\bibitem [{\citenamefont {Li}\ \emph {et~al.}(2017)\citenamefont {Li},
  \citenamefont {Kim}, \citenamefont {Lee}, \citenamefont {Lee}, \citenamefont
  {Koyama}, \citenamefont {Chiba}, \citenamefont {Moriyama}, \citenamefont
  {Lee}, \citenamefont {Kim},\ and\ \citenamefont {Ono}}]{li2017origin}%
  \BibitemOpen
  \bibfield  {author} {\bibinfo {author} {\bibfnamefont {T.}~\bibnamefont
  {Li}}, \bibinfo {author} {\bibfnamefont {S.}~\bibnamefont {Kim}}, \bibinfo
  {author} {\bibfnamefont {S.-J.}\ \bibnamefont {Lee}}, \bibinfo {author}
  {\bibfnamefont {S.-W.}\ \bibnamefont {Lee}}, \bibinfo {author} {\bibfnamefont
  {T.}~\bibnamefont {Koyama}}, \bibinfo {author} {\bibfnamefont
  {D.}~\bibnamefont {Chiba}}, \bibinfo {author} {\bibfnamefont
  {T.}~\bibnamefont {Moriyama}}, \bibinfo {author} {\bibfnamefont {K.-J.}\
  \bibnamefont {Lee}}, \bibinfo {author} {\bibfnamefont {K.-J.}\ \bibnamefont
  {Kim}},\ and\ \bibinfo {author} {\bibfnamefont {T.}~\bibnamefont {Ono}},\
  }\bibfield  {title} {\bibinfo {title} {\textit{Origin of Threshold Current
  Density for Asymmetric Magnetoresistance in \textnormal{Pt/Py} Bilayers}},\
  }\href {https://doi.org/10.7567/APEX.10.073001} {\bibfield  {journal}
  {\bibinfo  {journal} {Appl. Phys. Express}\ }\textbf {\bibinfo {volume}
  {10}},\ \bibinfo {pages} {073001} (\bibinfo {year} {2017})}\BibitemShut
  {NoStop}%
\bibitem [{\citenamefont {Yin}\ \emph {et~al.}(2017)\citenamefont {Yin},
  \citenamefont {Han}, \citenamefont {de~Jong}, \citenamefont {Lavrijsen},
  \citenamefont {Duine}, \citenamefont {Swagten},\ and\ \citenamefont
  {Koopmans}}]{yin2017thickness}%
  \BibitemOpen
  \bibfield  {author} {\bibinfo {author} {\bibfnamefont {Y.}~\bibnamefont
  {Yin}}, \bibinfo {author} {\bibfnamefont {D.-S.}\ \bibnamefont {Han}},
  \bibinfo {author} {\bibfnamefont {M.~C.}\ \bibnamefont {de~Jong}}, \bibinfo
  {author} {\bibfnamefont {R.}~\bibnamefont {Lavrijsen}}, \bibinfo {author}
  {\bibfnamefont {R.~A.}\ \bibnamefont {Duine}}, \bibinfo {author}
  {\bibfnamefont {H.~J.}\ \bibnamefont {Swagten}},\ and\ \bibinfo {author}
  {\bibfnamefont {B.}~\bibnamefont {Koopmans}},\ }\bibfield  {title} {\bibinfo
  {title} {\textit{Thickness Dependence of Unidirectional Spin-\text{H}all
  Magnetoresistance in Metallic Bilayers}},\ }\href
  {https://doi.org/10.1063/1.5003725} {\bibfield  {journal} {\bibinfo
  {journal} {Appl. Phys. Lett.}\ }\textbf {\bibinfo {volume} {111}},\ \bibinfo
  {pages} {232405} (\bibinfo {year} {2017})}\BibitemShut {NoStop}%
\bibitem [{\citenamefont {Olejn\'{\i}k}\ \emph {et~al.}(2015)\citenamefont
  {Olejn\'{\i}k}, \citenamefont {Nov\'ak}, \citenamefont {Wunderlich},\ and\
  \citenamefont {Jungwirth}}]{olejnik2015electrical}%
  \BibitemOpen
  \bibfield  {author} {\bibinfo {author} {\bibfnamefont {K.}~\bibnamefont
  {Olejn\'{\i}k}}, \bibinfo {author} {\bibfnamefont {V.}~\bibnamefont
  {Nov\'ak}}, \bibinfo {author} {\bibfnamefont {J.}~\bibnamefont
  {Wunderlich}},\ and\ \bibinfo {author} {\bibfnamefont {T.}~\bibnamefont
  {Jungwirth}},\ }\bibfield  {title} {\bibinfo {title} {\textit{Electrical
  Detection of Magnetization Reversal without Auxiliary Magnets}},\ }\href
  {https://doi.org/10.1103/PhysRevB.91.180402} {\bibfield  {journal} {\bibinfo
  {journal} {Phys. Rev. B}\ }\textbf {\bibinfo {volume} {91}},\ \bibinfo
  {pages} {180402} (\bibinfo {year} {2015})}\BibitemShut {NoStop}%
\bibitem [{\citenamefont {Lv}\ \emph {et~al.}(2018)\citenamefont {Lv},
  \citenamefont {Kally}, \citenamefont {Zhang}, \citenamefont {Lee},
  \citenamefont {Jamali}, \citenamefont {Samarth},\ and\ \citenamefont
  {Wang}}]{Lv18NC_UMR-TI-FM}%
  \BibitemOpen
  \bibfield  {author} {\bibinfo {author} {\bibfnamefont {Y.}~\bibnamefont
  {Lv}}, \bibinfo {author} {\bibfnamefont {J.}~\bibnamefont {Kally}}, \bibinfo
  {author} {\bibfnamefont {D.}~\bibnamefont {Zhang}}, \bibinfo {author}
  {\bibfnamefont {J.~S.}\ \bibnamefont {Lee}}, \bibinfo {author} {\bibfnamefont
  {M.}~\bibnamefont {Jamali}}, \bibinfo {author} {\bibfnamefont
  {N.}~\bibnamefont {Samarth}},\ and\ \bibinfo {author} {\bibfnamefont {J.-P.}\
  \bibnamefont {Wang}},\ }\bibfield  {title} {\bibinfo {title}
  {\textit{Unidirectional Spin-\text{H}all and \text{R}ashba\text{-E}delstein
  Magnetoresistance in Topological Insulator-Ferromagnet Layer
  Heterostructures}},\ }\href {https://doi.org/10.1038/s41467-017-02491-3}
  {\bibfield  {journal} {\bibinfo  {journal} {Nat. Commun.}\ }\textbf {\bibinfo
  {volume} {9}},\ \bibinfo {pages} {111} (\bibinfo {year} {2018})}\BibitemShut
  {NoStop}%
\bibitem [{\citenamefont {Avci}\ \emph {et~al.}(2018)\citenamefont {Avci},
  \citenamefont {Mendil}, \citenamefont {Beach},\ and\ \citenamefont
  {Gambardella}}]{avci2018origins}%
  \BibitemOpen
  \bibfield  {author} {\bibinfo {author} {\bibfnamefont {C.~O.}\ \bibnamefont
  {Avci}}, \bibinfo {author} {\bibfnamefont {J.}~\bibnamefont {Mendil}},
  \bibinfo {author} {\bibfnamefont {G.~S.~D.}\ \bibnamefont {Beach}},\ and\
  \bibinfo {author} {\bibfnamefont {P.}~\bibnamefont {Gambardella}},\
  }\bibfield  {title} {\bibinfo {title} {\textit{Origins of the Unidirectional
  Spin \text{H}all Magnetoresistance in Metallic Bilayers}},\ }\href
  {https://doi.org/10.1103/PhysRevLett.121.087207} {\bibfield  {journal}
  {\bibinfo  {journal} {Phys. Rev. Lett.}\ }\textbf {\bibinfo {volume} {121}},\
  \bibinfo {pages} {087207} (\bibinfo {year} {2018})}\BibitemShut {NoStop}%
\bibitem [{\citenamefont {{Guillet}}\ \emph {et~al.}(2021)\citenamefont
  {{Guillet}}, \citenamefont {{Marty}}, \citenamefont {{Vergnaud}},
  \citenamefont {{Jamet}}, \citenamefont {{Zucchetti}}, \citenamefont
  {{Isella}}, \citenamefont {{Barbedienne}}, \citenamefont {{Jaffr{\`e}s}},
  \citenamefont {{Reyren}}, \citenamefont {{George}},\ and\ \citenamefont
  {{Fert}}}]{guillet2021prb}%
  \BibitemOpen
  \bibfield  {author} {\bibinfo {author} {\bibfnamefont {T.}~\bibnamefont
  {{Guillet}}}, \bibinfo {author} {\bibfnamefont {A.}~\bibnamefont {{Marty}}},
  \bibinfo {author} {\bibfnamefont {C.}~\bibnamefont {{Vergnaud}}}, \bibinfo
  {author} {\bibfnamefont {M.}~\bibnamefont {{Jamet}}}, \bibinfo {author}
  {\bibfnamefont {C.}~\bibnamefont {{Zucchetti}}}, \bibinfo {author}
  {\bibfnamefont {G.}~\bibnamefont {{Isella}}}, \bibinfo {author}
  {\bibfnamefont {Q.}~\bibnamefont {{Barbedienne}}}, \bibinfo {author}
  {\bibfnamefont {H.}~\bibnamefont {{Jaffr{\`e}s}}}, \bibinfo {author}
  {\bibfnamefont {N.}~\bibnamefont {{Reyren}}}, \bibinfo {author}
  {\bibfnamefont {J.~M.}\ \bibnamefont {{George}}},\ and\ \bibinfo {author}
  {\bibfnamefont {A.}~\bibnamefont {{Fert}}},\ }\bibfield  {title} {\bibinfo
  {title} {{\textit{Large Rashba Unidirectional Magnetoresistance in the
  \textnormal{Fe/Ge(111)} Interface States}}},\ }\href
  {https://doi.org/10.1103/PhysRevB.103.064411} {\bibfield  {journal} {\bibinfo
   {journal} {Phys. Rev. B}\ }\textbf {\bibinfo {volume} {103}},\ \bibinfo
  {eid} {064411} (\bibinfo {year} {2021})}\BibitemShut {NoStop}%
\bibitem [{\citenamefont {N\'u\~nez}\ \emph {et~al.}(2006)\citenamefont
  {N\'u\~nez}, \citenamefont {Duine}, \citenamefont {Haney},\ and\
  \citenamefont {MacDonald}}]{nunez2006theory}%
  \BibitemOpen
  \bibfield  {author} {\bibinfo {author} {\bibfnamefont {A.~S.}\ \bibnamefont
  {N\'u\~nez}}, \bibinfo {author} {\bibfnamefont {R.~A.}\ \bibnamefont
  {Duine}}, \bibinfo {author} {\bibfnamefont {P.}~\bibnamefont {Haney}},\ and\
  \bibinfo {author} {\bibfnamefont {A.~H.}\ \bibnamefont {MacDonald}},\
  }\bibfield  {title} {\bibinfo {title} {\textit{Theory of Spin Torques and
  Giant Magnetoresistance in Antiferromagnetic Metals}},\ }\href
  {https://doi.org/10.1103/PhysRevB.73.214426} {\bibfield  {journal} {\bibinfo
  {journal} {Phys. Rev. B}\ }\textbf {\bibinfo {volume} {73}},\ \bibinfo
  {pages} {214426} (\bibinfo {year} {2006})}\BibitemShut {NoStop}%
\bibitem [{\citenamefont {Zhang}\ and\ \citenamefont
  {Vignale}(2017)}]{Zhangs&Vignaleg17spie_UMR}%
  \BibitemOpen
  \bibfield  {author} {\bibinfo {author} {\bibfnamefont {S.~S.-L.}\
  \bibnamefont {Zhang}}\ and\ \bibinfo {author} {\bibfnamefont
  {G.}~\bibnamefont {Vignale}},\ }\bibfield  {title} {\bibinfo {title}
  {{\textit{Theory of Unidirectional Magnetoresistance in Magnetic
  Heterostructures}}},\ }\href {https://doi.org/10.1117/12.2275154} {\bibfield
  {journal} {\bibinfo  {journal} {Proc. SPIE}\ }\textbf {\bibinfo {volume}
  {10357}},\ \bibinfo {pages} {1035707} (\bibinfo {year} {2017})}\BibitemShut
  {NoStop}%
\bibitem [{\citenamefont {Barker}\ and\ \citenamefont
  {Chantrell}(2015)}]{barker2015higher}%
  \BibitemOpen
  \bibfield  {author} {\bibinfo {author} {\bibfnamefont {J.}~\bibnamefont
  {Barker}}\ and\ \bibinfo {author} {\bibfnamefont {R.~W.}\ \bibnamefont
  {Chantrell}},\ }\bibfield  {title} {\bibinfo {title} {\textit{Higher-Order
  Exchange Interactions Leading to Metamagnetism in \textnormal{FeRh}}},\
  }\href {https://doi.org/10.1103/PhysRevB.92.094402} {\bibfield  {journal}
  {\bibinfo  {journal} {Phys. Rev. B}\ }\textbf {\bibinfo {volume} {92}},\
  \bibinfo {pages} {094402} (\bibinfo {year} {2015})}\BibitemShut {NoStop}%
\bibitem [{\citenamefont {McGrath}\ \emph {et~al.}(2020)\citenamefont
  {McGrath}, \citenamefont {Camley},\ and\ \citenamefont
  {Livesey}}]{mcgrath2020PRBself}%
  \BibitemOpen
  \bibfield  {author} {\bibinfo {author} {\bibfnamefont {B.~R.}\ \bibnamefont
  {McGrath}}, \bibinfo {author} {\bibfnamefont {R.~E.}\ \bibnamefont
  {Camley}},\ and\ \bibinfo {author} {\bibfnamefont {K.~L.}\ \bibnamefont
  {Livesey}},\ }\bibfield  {title} {\bibinfo {title} {\textit{Self-Consistent
  Local Mean-Field theory for Phase Transitions and Magnetic Properties of
  \textnormal{FeRh}}},\ }\href {https://doi.org/10.1103/PhysRevB.101.014444}
  {\bibfield  {journal} {\bibinfo  {journal} {Phys. Rev. B}\ }\textbf {\bibinfo
  {volume} {101}},\ \bibinfo {pages} {014444} (\bibinfo {year}
  {2020})}\BibitemShut {NoStop}%
\bibitem [{\citenamefont {Ideue}\ \emph {et~al.}(2017)\citenamefont {Ideue},
  \citenamefont {Hamamoto}, \citenamefont {Koshikawa}, \citenamefont {Ezawa},
  \citenamefont {Shimizu}, \citenamefont {Kaneko}, \citenamefont {Tokura},
  \citenamefont {Nagaosa},\ and\ \citenamefont {Iwasa}}]{ideue2017bulk}%
  \BibitemOpen
  \bibfield  {author} {\bibinfo {author} {\bibfnamefont {T.}~\bibnamefont
  {Ideue}}, \bibinfo {author} {\bibfnamefont {K.}~\bibnamefont {Hamamoto}},
  \bibinfo {author} {\bibfnamefont {S.}~\bibnamefont {Koshikawa}}, \bibinfo
  {author} {\bibfnamefont {M.}~\bibnamefont {Ezawa}}, \bibinfo {author}
  {\bibfnamefont {S.}~\bibnamefont {Shimizu}}, \bibinfo {author} {\bibfnamefont
  {Y.}~\bibnamefont {Kaneko}}, \bibinfo {author} {\bibfnamefont
  {Y.}~\bibnamefont {Tokura}}, \bibinfo {author} {\bibfnamefont
  {N.}~\bibnamefont {Nagaosa}},\ and\ \bibinfo {author} {\bibfnamefont
  {Y.}~\bibnamefont {Iwasa}},\ }\bibfield  {title} {\bibinfo {title}
  {\textit{Bulk Rectification Effect in a Polar Semiconductor}},\ }\href
  {https://doi.org/10.1038/nphys4056} {\bibfield  {journal} {\bibinfo
  {journal} {Nat. Phys.}\ }\textbf {\bibinfo {volume} {13}},\ \bibinfo {pages}
  {578} (\bibinfo {year} {2017})}\BibitemShut {NoStop}%
\bibitem [{\citenamefont {Guillet}\ \emph {et~al.}(2020)\citenamefont
  {Guillet}, \citenamefont {Zucchetti}, \citenamefont {Barbedienne},
  \citenamefont {Marty}, \citenamefont {Isella}, \citenamefont {Cagnon},
  \citenamefont {Vergnaud}, \citenamefont {Jaffr\`es}, \citenamefont {Reyren},
  \citenamefont {George} \emph {et~al.}}]{guillet2020observation}%
  \BibitemOpen
  \bibfield  {author} {\bibinfo {author} {\bibfnamefont {T.}~\bibnamefont
  {Guillet}}, \bibinfo {author} {\bibfnamefont {C.}~\bibnamefont {Zucchetti}},
  \bibinfo {author} {\bibfnamefont {Q.}~\bibnamefont {Barbedienne}}, \bibinfo
  {author} {\bibfnamefont {A.}~\bibnamefont {Marty}}, \bibinfo {author}
  {\bibfnamefont {G.}~\bibnamefont {Isella}}, \bibinfo {author} {\bibfnamefont
  {L.}~\bibnamefont {Cagnon}}, \bibinfo {author} {\bibfnamefont
  {C.}~\bibnamefont {Vergnaud}}, \bibinfo {author} {\bibfnamefont
  {H.}~\bibnamefont {Jaffr\`es}}, \bibinfo {author} {\bibfnamefont
  {N.}~\bibnamefont {Reyren}}, \bibinfo {author} {\bibfnamefont {J.-M.}\
  \bibnamefont {George}}, \emph {et~al.},\ }\bibfield  {title} {\bibinfo
  {title} {\textit{Observation of Large Unidirectional \text{R}ashba
  Magnetoresistance in \textnormal{Ge(111)}}},\ }\href
  {https://doi.org/10.1103/PhysRevLett.124.027201} {\bibfield  {journal}
  {\bibinfo  {journal} {Phys. Rev. Lett.}\ }\textbf {\bibinfo {volume} {124}},\
  \bibinfo {pages} {027201} (\bibinfo {year} {2020})}\BibitemShut {NoStop}%
\bibitem [{\citenamefont {Fan}\ \emph {et~al.}(2019)\citenamefont {Fan},
  \citenamefont {Shao}, \citenamefont {Pan}, \citenamefont {Che}, \citenamefont
  {He}, \citenamefont {Yin}, \citenamefont {Zheng}, \citenamefont {Yu},
  \citenamefont {Nie}, \citenamefont {Masir} \emph
  {et~al.}}]{fan2019unidirectional}%
  \BibitemOpen
  \bibfield  {author} {\bibinfo {author} {\bibfnamefont {Y.}~\bibnamefont
  {Fan}}, \bibinfo {author} {\bibfnamefont {Q.}~\bibnamefont {Shao}}, \bibinfo
  {author} {\bibfnamefont {L.}~\bibnamefont {Pan}}, \bibinfo {author}
  {\bibfnamefont {X.}~\bibnamefont {Che}}, \bibinfo {author} {\bibfnamefont
  {Q.}~\bibnamefont {He}}, \bibinfo {author} {\bibfnamefont {G.}~\bibnamefont
  {Yin}}, \bibinfo {author} {\bibfnamefont {C.}~\bibnamefont {Zheng}}, \bibinfo
  {author} {\bibfnamefont {G.}~\bibnamefont {Yu}}, \bibinfo {author}
  {\bibfnamefont {T.}~\bibnamefont {Nie}}, \bibinfo {author} {\bibfnamefont
  {M.~R.}\ \bibnamefont {Masir}}, \emph {et~al.},\ }\bibfield  {title}
  {\bibinfo {title} {\textit{Unidirectional Magneto-Resistance in
  Modulation-Doped Magnetic Topological Insulators}},\ }\href
  {https://doi.org/10.1021/acs.nanolett.8b03702} {\bibfield  {journal}
  {\bibinfo  {journal} {Nano Lett.}\ }\textbf {\bibinfo {volume} {19}},\
  \bibinfo {pages} {692} (\bibinfo {year} {2019})}\BibitemShut {NoStop}%
\bibitem [{\citenamefont {He}\ \emph {et~al.}(2018{\natexlab{a}})\citenamefont
  {He}, \citenamefont {Walker}, \citenamefont {Zhang}, \citenamefont {Bruno},
  \citenamefont {Bahramy}, \citenamefont {Lee}, \citenamefont {Ramaswamy},
  \citenamefont {Cai}, \citenamefont {Heinonen}, \citenamefont {Vignale} \emph
  {et~al.}}]{he2018observation}%
  \BibitemOpen
  \bibfield  {author} {\bibinfo {author} {\bibfnamefont {P.}~\bibnamefont
  {He}}, \bibinfo {author} {\bibfnamefont {S.~M.}\ \bibnamefont {Walker}},
  \bibinfo {author} {\bibfnamefont {S.~S.-L.}\ \bibnamefont {Zhang}}, \bibinfo
  {author} {\bibfnamefont {F.~Y.}\ \bibnamefont {Bruno}}, \bibinfo {author}
  {\bibfnamefont {M.~S.}\ \bibnamefont {Bahramy}}, \bibinfo {author}
  {\bibfnamefont {J.~M.}\ \bibnamefont {Lee}}, \bibinfo {author} {\bibfnamefont
  {R.}~\bibnamefont {Ramaswamy}}, \bibinfo {author} {\bibfnamefont
  {K.}~\bibnamefont {Cai}}, \bibinfo {author} {\bibfnamefont {O.}~\bibnamefont
  {Heinonen}}, \bibinfo {author} {\bibfnamefont {G.}~\bibnamefont {Vignale}},
  \emph {et~al.},\ }\bibfield  {title} {\bibinfo {title} {\textit{Observation
  of Out-of-Plane Spin Texture in a $\textnormal{SrTiO}_{3}(111)$
  Two-Dimensional Electron Gas}},\ }\href
  {https://doi.org/10.1103/PhysRevLett.120.266802} {\bibfield  {journal}
  {\bibinfo  {journal} {Phys. Rev. Lett.}\ }\textbf {\bibinfo {volume} {120}},\
  \bibinfo {pages} {266802} (\bibinfo {year} {2018}{\natexlab{a}})}\BibitemShut
  {NoStop}%
\bibitem [{\citenamefont {He}\ \emph {et~al.}(2018{\natexlab{b}})\citenamefont
  {He}, \citenamefont {Zhang}, \citenamefont {Zhu}, \citenamefont {Liu},
  \citenamefont {Wang}, \citenamefont {Yu}, \citenamefont {Vignale},\ and\
  \citenamefont {Yang}}]{he2018bilinear}%
  \BibitemOpen
  \bibfield  {author} {\bibinfo {author} {\bibfnamefont {P.}~\bibnamefont
  {He}}, \bibinfo {author} {\bibfnamefont {S.~S.-L.}\ \bibnamefont {Zhang}},
  \bibinfo {author} {\bibfnamefont {D.}~\bibnamefont {Zhu}}, \bibinfo {author}
  {\bibfnamefont {Y.}~\bibnamefont {Liu}}, \bibinfo {author} {\bibfnamefont
  {Y.}~\bibnamefont {Wang}}, \bibinfo {author} {\bibfnamefont {J.}~\bibnamefont
  {Yu}}, \bibinfo {author} {\bibfnamefont {G.}~\bibnamefont {Vignale}},\ and\
  \bibinfo {author} {\bibfnamefont {H.}~\bibnamefont {Yang}},\ }\bibfield
  {title} {\bibinfo {title} {\textit{Bilinear Magnetoelectric Resistance as a
  Probe of Three-Dimensional Spin Texture in Topological Surface States}},\
  }\href {https://doi.org/10.1038/s41567-017-0039-y} {\bibfield  {journal}
  {\bibinfo  {journal} {Nat. Phys.}\ }\textbf {\bibinfo {volume} {14}},\
  \bibinfo {pages} {495} (\bibinfo {year} {2018}{\natexlab{b}})}\BibitemShut
  {NoStop}%
\bibitem [{\citenamefont {Mancini}\ \emph {et~al.}(2013)\citenamefont
  {Mancini}, \citenamefont {Pressacco}, \citenamefont {Haertinger},
  \citenamefont {Fullerton}, \citenamefont {Suzuki}, \citenamefont
  {Woltersdorf},\ and\ \citenamefont {Back}}]{mancini2013magnetic}%
  \BibitemOpen
  \bibfield  {author} {\bibinfo {author} {\bibfnamefont {E.}~\bibnamefont
  {Mancini}}, \bibinfo {author} {\bibfnamefont {F.}~\bibnamefont {Pressacco}},
  \bibinfo {author} {\bibfnamefont {M.}~\bibnamefont {Haertinger}}, \bibinfo
  {author} {\bibfnamefont {E.}~\bibnamefont {Fullerton}}, \bibinfo {author}
  {\bibfnamefont {T.}~\bibnamefont {Suzuki}}, \bibinfo {author} {\bibfnamefont
  {G.}~\bibnamefont {Woltersdorf}},\ and\ \bibinfo {author} {\bibfnamefont
  {C.}~\bibnamefont {Back}},\ }\bibfield  {title} {\bibinfo {title}
  {\textit{Magnetic Phase Transition in Iron--Rhodium Thin Films Probed by
  Ferromagnetic Resonance}},\ }\href
  {http://dx.doi.org/10.1088/0022-3727/46/24/245302} {\bibfield  {journal}
  {\bibinfo  {journal} {J. Phys. D: Appl. Phys.}\ }\textbf {\bibinfo {volume}
  {46}},\ \bibinfo {pages} {245302} (\bibinfo {year} {2013})}\BibitemShut
  {NoStop}%
\bibitem [{\citenamefont {Weiler}\ \emph {et~al.}(2012)\citenamefont {Weiler},
  \citenamefont {Althammer}, \citenamefont {Czeschka}, \citenamefont {Huebl},
  \citenamefont {Wagner}, \citenamefont {Opel}, \citenamefont {Imort},
  \citenamefont {Reiss}, \citenamefont {Thomas}, \citenamefont {Gross},\ and\
  \citenamefont {Goennenwein}}]{weiler2012local}%
  \BibitemOpen
  \bibfield  {author} {\bibinfo {author} {\bibfnamefont {M.}~\bibnamefont
  {Weiler}}, \bibinfo {author} {\bibfnamefont {M.}~\bibnamefont {Althammer}},
  \bibinfo {author} {\bibfnamefont {F.~D.}\ \bibnamefont {Czeschka}}, \bibinfo
  {author} {\bibfnamefont {H.}~\bibnamefont {Huebl}}, \bibinfo {author}
  {\bibfnamefont {M.~S.}\ \bibnamefont {Wagner}}, \bibinfo {author}
  {\bibfnamefont {M.}~\bibnamefont {Opel}}, \bibinfo {author} {\bibfnamefont
  {I.-M.}\ \bibnamefont {Imort}}, \bibinfo {author} {\bibfnamefont
  {G.}~\bibnamefont {Reiss}}, \bibinfo {author} {\bibfnamefont
  {A.}~\bibnamefont {Thomas}}, \bibinfo {author} {\bibfnamefont
  {R.}~\bibnamefont {Gross}},\ and\ \bibinfo {author} {\bibfnamefont
  {S.~T.~B.}\ \bibnamefont {Goennenwein}},\ }\bibfield  {title} {\bibinfo
  {title} {\textit{Local Charge and Spin Currents in Magnetothermal
  Landscapes}},\ }\href {https://doi.org/10.1103/PhysRevLett.108.106602}
  {\bibfield  {journal} {\bibinfo  {journal} {Phys. Rev. Lett.}\ }\textbf
  {\bibinfo {volume} {108}},\ \bibinfo {pages} {106602} (\bibinfo {year}
  {2012})}\BibitemShut {NoStop}%
\bibitem [{\citenamefont {Kikkawa}\ \emph {et~al.}(2013)\citenamefont
  {Kikkawa}, \citenamefont {Uchida}, \citenamefont {Shiomi}, \citenamefont
  {Qiu}, \citenamefont {Hou}, \citenamefont {Tian}, \citenamefont {Nakayama},
  \citenamefont {Jin},\ and\ \citenamefont {Saitoh}}]{kikkawa2013longitudinal}%
  \BibitemOpen
  \bibfield  {author} {\bibinfo {author} {\bibfnamefont {T.}~\bibnamefont
  {Kikkawa}}, \bibinfo {author} {\bibfnamefont {K.}~\bibnamefont {Uchida}},
  \bibinfo {author} {\bibfnamefont {Y.}~\bibnamefont {Shiomi}}, \bibinfo
  {author} {\bibfnamefont {Z.}~\bibnamefont {Qiu}}, \bibinfo {author}
  {\bibfnamefont {D.}~\bibnamefont {Hou}}, \bibinfo {author} {\bibfnamefont
  {D.}~\bibnamefont {Tian}}, \bibinfo {author} {\bibfnamefont {H.}~\bibnamefont
  {Nakayama}}, \bibinfo {author} {\bibfnamefont {X.-F.}\ \bibnamefont {Jin}},\
  and\ \bibinfo {author} {\bibfnamefont {E.}~\bibnamefont {Saitoh}},\
  }\bibfield  {title} {\bibinfo {title} {\textit{Longitudinal Spin
  \text{S}eebeck Effect Free from the Proximity \text{N}ernst Effect}},\ }\href
  {https://doi.org/10.1103/PhysRevLett.110.067207} {\bibfield  {journal}
  {\bibinfo  {journal} {Phys. Rev. Lett.}\ }\textbf {\bibinfo {volume} {110}},\
  \bibinfo {pages} {067207} (\bibinfo {year} {2013})}\BibitemShut {NoStop}%
\bibitem [{\citenamefont {\ifmmode~\check{Z}\else \v{Z}\fi{}elezn\'y}\ \emph
  {et~al.}(2014)\citenamefont {\ifmmode~\check{Z}\else \v{Z}\fi{}elezn\'y},
  \citenamefont {Gao}, \citenamefont {V\'yborn\'y}, \citenamefont {Zemen},
  \citenamefont {Ma\ifmmode~\check{s}\else \v{s}\fi{}ek}, \citenamefont
  {Manchon}, \citenamefont {Wunderlich}, \citenamefont {Sinova},\ and\
  \citenamefont {Jungwirth}}]{vzelezny2014relativistic}%
  \BibitemOpen
  \bibfield  {author} {\bibinfo {author} {\bibfnamefont {J.}~\bibnamefont
  {\ifmmode~\check{Z}\else \v{Z}\fi{}elezn\'y}}, \bibinfo {author}
  {\bibfnamefont {H.}~\bibnamefont {Gao}}, \bibinfo {author} {\bibfnamefont
  {K.}~\bibnamefont {V\'yborn\'y}}, \bibinfo {author} {\bibfnamefont
  {J.}~\bibnamefont {Zemen}}, \bibinfo {author} {\bibfnamefont
  {J.}~\bibnamefont {Ma\ifmmode~\check{s}\else \v{s}\fi{}ek}}, \bibinfo
  {author} {\bibfnamefont {A.}~\bibnamefont {Manchon}}, \bibinfo {author}
  {\bibfnamefont {J.}~\bibnamefont {Wunderlich}}, \bibinfo {author}
  {\bibfnamefont {J.}~\bibnamefont {Sinova}},\ and\ \bibinfo {author}
  {\bibfnamefont {T.}~\bibnamefont {Jungwirth}},\ }\bibfield  {title} {\bibinfo
  {title} {\textit{Relativistic \text{N}\'eel-Order Fields Induced by
  Electrical Current in Antiferromagnets}},\ }\href
  {https://doi.org/10.1103/PhysRevLett.113.157201} {\bibfield  {journal}
  {\bibinfo  {journal} {Phys. Rev. Lett.}\ }\textbf {\bibinfo {volume} {113}},\
  \bibinfo {pages} {157201} (\bibinfo {year} {2014})}\BibitemShut {NoStop}%
\bibitem [{\citenamefont {Saidaoui}\ \emph {et~al.}(2017)\citenamefont
  {Saidaoui}, \citenamefont {Waintal},\ and\ \citenamefont
  {Manchon}}]{saidaoui2017robust}%
  \BibitemOpen
  \bibfield  {author} {\bibinfo {author} {\bibfnamefont {H.~B.~M.}\
  \bibnamefont {Saidaoui}}, \bibinfo {author} {\bibfnamefont {X.}~\bibnamefont
  {Waintal}},\ and\ \bibinfo {author} {\bibfnamefont {A.}~\bibnamefont
  {Manchon}},\ }\bibfield  {title} {\bibinfo {title} {\textit{Robust Spin
  Transfer Torque in Antiferromagnetic Tunnel Junctions}},\ }\href
  {https://doi.org/10.1103/PhysRevB.95.134424} {\bibfield  {journal} {\bibinfo
  {journal} {Phys. Rev. B}\ }\textbf {\bibinfo {volume} {95}},\ \bibinfo
  {pages} {134424} (\bibinfo {year} {2017})}\BibitemShut {NoStop}%
\bibitem [{\citenamefont {Haney}\ \emph {et~al.}(2013)\citenamefont {Haney},
  \citenamefont {Lee}, \citenamefont {Lee}, \citenamefont {Manchon},\ and\
  \citenamefont {Stiles}}]{haney2013current}%
  \BibitemOpen
  \bibfield  {author} {\bibinfo {author} {\bibfnamefont {P.~M.}\ \bibnamefont
  {Haney}}, \bibinfo {author} {\bibfnamefont {H.-W.}\ \bibnamefont {Lee}},
  \bibinfo {author} {\bibfnamefont {K.-J.}\ \bibnamefont {Lee}}, \bibinfo
  {author} {\bibfnamefont {A.}~\bibnamefont {Manchon}},\ and\ \bibinfo {author}
  {\bibfnamefont {M.~D.}\ \bibnamefont {Stiles}},\ }\bibfield  {title}
  {\bibinfo {title} {\textit{Current Induced Torques and Interfacial Spin-Orbit
  Coupling: Semiclassical Modeling}},\ }\href
  {https://doi.org/10.1103/PhysRevB.87.174411} {\bibfield  {journal} {\bibinfo
  {journal} {Phys. Rev. B}\ }\textbf {\bibinfo {volume} {87}},\ \bibinfo
  {pages} {174411} (\bibinfo {year} {2013})}\BibitemShut {NoStop}%
\bibitem [{\citenamefont {Hamamoto}\ \emph {et~al.}(2017)\citenamefont
  {Hamamoto}, \citenamefont {Ezawa}, \citenamefont {Kim}, \citenamefont
  {Morimoto},\ and\ \citenamefont {Nagaosa}}]{hamamoto2017prb}%
  \BibitemOpen
  \bibfield  {author} {\bibinfo {author} {\bibfnamefont {K.}~\bibnamefont
  {Hamamoto}}, \bibinfo {author} {\bibfnamefont {M.}~\bibnamefont {Ezawa}},
  \bibinfo {author} {\bibfnamefont {K.~W.}\ \bibnamefont {Kim}}, \bibinfo
  {author} {\bibfnamefont {T.}~\bibnamefont {Morimoto}},\ and\ \bibinfo
  {author} {\bibfnamefont {N.}~\bibnamefont {Nagaosa}},\ }\bibfield  {title}
  {\bibinfo {title} {\textit{Nonlinear Spin Current Generation in
  Noncentrosymmetric Spin-Orbit Coupled Systems}},\ }\href
  {https://doi.org/10.1103/PhysRevB.95.224430} {\bibfield  {journal} {\bibinfo
  {journal} {Phys. Rev. B}\ }\textbf {\bibinfo {volume} {95}},\ \bibinfo
  {pages} {224430} (\bibinfo {year} {2017})}\BibitemShut {NoStop}%
\bibitem [{\citenamefont {He}\ \emph {et~al.}(2019)\citenamefont {He},
  \citenamefont {Zhang}, \citenamefont {Zhu}, \citenamefont {Shi},
  \citenamefont {Heinonen}, \citenamefont {Vignale},\ and\ \citenamefont
  {Yang}}]{he2019prl}%
  \BibitemOpen
  \bibfield  {author} {\bibinfo {author} {\bibfnamefont {P.}~\bibnamefont
  {He}}, \bibinfo {author} {\bibfnamefont {S.~S.-L.}\ \bibnamefont {Zhang}},
  \bibinfo {author} {\bibfnamefont {D.}~\bibnamefont {Zhu}}, \bibinfo {author}
  {\bibfnamefont {S.}~\bibnamefont {Shi}}, \bibinfo {author} {\bibfnamefont
  {O.~G.}\ \bibnamefont {Heinonen}}, \bibinfo {author} {\bibfnamefont
  {G.}~\bibnamefont {Vignale}},\ and\ \bibinfo {author} {\bibfnamefont
  {H.}~\bibnamefont {Yang}},\ }\bibfield  {title} {\bibinfo {title}
  {\textit{Nonlinear Planar \text{H}all Effect}},\ }\href
  {https://doi.org/10.1103/PhysRevLett.123.016801} {\bibfield  {journal}
  {\bibinfo  {journal} {Phys. Rev. Lett.}\ }\textbf {\bibinfo {volume} {123}},\
  \bibinfo {pages} {016801} (\bibinfo {year} {2019})}\BibitemShut {NoStop}%
\bibitem [{\citenamefont {Miron}\ \emph {et~al.}(2011)\citenamefont {Miron},
  \citenamefont {Garello}, \citenamefont {Gaudin}, \citenamefont {Zermatten},
  \citenamefont {Costache}, \citenamefont {Auffret}, \citenamefont {Bandiera},
  \citenamefont {Rodmacq}, \citenamefont {Schuhl},\ and\ \citenamefont
  {Gambardella}}]{miron2011perpendicular}%
  \BibitemOpen
  \bibfield  {author} {\bibinfo {author} {\bibfnamefont {I.~M.}\ \bibnamefont
  {Miron}}, \bibinfo {author} {\bibfnamefont {K.}~\bibnamefont {Garello}},
  \bibinfo {author} {\bibfnamefont {G.}~\bibnamefont {Gaudin}}, \bibinfo
  {author} {\bibfnamefont {P.-J.}\ \bibnamefont {Zermatten}}, \bibinfo {author}
  {\bibfnamefont {M.~V.}\ \bibnamefont {Costache}}, \bibinfo {author}
  {\bibfnamefont {S.}~\bibnamefont {Auffret}}, \bibinfo {author} {\bibfnamefont
  {S.}~\bibnamefont {Bandiera}}, \bibinfo {author} {\bibfnamefont
  {B.}~\bibnamefont {Rodmacq}}, \bibinfo {author} {\bibfnamefont
  {A.}~\bibnamefont {Schuhl}},\ and\ \bibinfo {author} {\bibfnamefont
  {P.}~\bibnamefont {Gambardella}},\ }\bibfield  {title} {\bibinfo {title}
  {\textit{Perpendicular Switching of a Single Ferromagnetic Layer Induced by
  In-Plane Current Injection}},\ }\href {https://doi.org/10.1038/nature10309}
  {\bibfield  {journal} {\bibinfo  {journal} {Nature}\ }\textbf {\bibinfo
  {volume} {476}},\ \bibinfo {pages} {189} (\bibinfo {year}
  {2011})}\BibitemShut {NoStop}%
\bibitem [{\citenamefont {Liu}\ \emph {et~al.}(2012)\citenamefont {Liu},
  \citenamefont {Pai}, \citenamefont {Li}, \citenamefont {Tseng}, \citenamefont
  {Ralph},\ and\ \citenamefont {Buhrman}}]{liu2012spin}%
  \BibitemOpen
  \bibfield  {author} {\bibinfo {author} {\bibfnamefont {L.}~\bibnamefont
  {Liu}}, \bibinfo {author} {\bibfnamefont {C.-F.}\ \bibnamefont {Pai}},
  \bibinfo {author} {\bibfnamefont {Y.}~\bibnamefont {Li}}, \bibinfo {author}
  {\bibfnamefont {H.}~\bibnamefont {Tseng}}, \bibinfo {author} {\bibfnamefont
  {D.}~\bibnamefont {Ralph}},\ and\ \bibinfo {author} {\bibfnamefont
  {R.}~\bibnamefont {Buhrman}},\ }\bibfield  {title} {\bibinfo {title}
  {\textit{Spin-Torque Switching with the Giant Spin \text{H}all Effect of
  \text{T}antalum}},\ }\href {https://doi.org/10.1126/science.1218197}
  {\bibfield  {journal} {\bibinfo  {journal} {Science}\ }\textbf {\bibinfo
  {volume} {336}},\ \bibinfo {pages} {555} (\bibinfo {year}
  {2012})}\BibitemShut {NoStop}%
\bibitem [{\citenamefont {Saglam}(2019)}]{saglam2019spin}%
  \BibitemOpen
  \bibfield  {author} {\bibinfo {author} {\bibfnamefont {H.}~\bibnamefont
  {Saglam}},\ }\emph {\bibinfo {title} {Spin Transport and Spin-Orbit Torques
  in Antiferromagnets}},\ \href
  {https://www.proquest.com/dissertations-theses/spin-transport-orbit-torques-antiferromagnets/docview/2275499588/se-2?accountid=14553}
  {Ph.D. thesis},\ \bibinfo  {school} {Illinois Institute of Technology}
  (\bibinfo {year} {2019})\BibitemShut {NoStop}%
\bibitem [{\citenamefont {Fan}\ \emph {et~al.}(2010)\citenamefont {Fan},
  \citenamefont {Kinane}, \citenamefont {Charlton}, \citenamefont {Dorner},
  \citenamefont {Ali}, \citenamefont {de~Vries}, \citenamefont {Brydson},
  \citenamefont {Marrows}, \citenamefont {Hickey}, \citenamefont {Arena} \emph
  {et~al.}}]{fan2010ferromagnetism}%
  \BibitemOpen
  \bibfield  {author} {\bibinfo {author} {\bibfnamefont {R.}~\bibnamefont
  {Fan}}, \bibinfo {author} {\bibfnamefont {C.~J.}\ \bibnamefont {Kinane}},
  \bibinfo {author} {\bibfnamefont {T.~R.}\ \bibnamefont {Charlton}}, \bibinfo
  {author} {\bibfnamefont {R.}~\bibnamefont {Dorner}}, \bibinfo {author}
  {\bibfnamefont {M.}~\bibnamefont {Ali}}, \bibinfo {author} {\bibfnamefont
  {M.~A.}\ \bibnamefont {de~Vries}}, \bibinfo {author} {\bibfnamefont
  {R.~M.~D.}\ \bibnamefont {Brydson}}, \bibinfo {author} {\bibfnamefont
  {C.~H.}\ \bibnamefont {Marrows}}, \bibinfo {author} {\bibfnamefont {B.~J.}\
  \bibnamefont {Hickey}}, \bibinfo {author} {\bibfnamefont {D.~A.}\
  \bibnamefont {Arena}}, \emph {et~al.},\ }\bibfield  {title} {\bibinfo {title}
  {\textit{Ferromagnetism at the Interfaces of Antiferromagnetic
  \textnormal{FeRh} Epilayers}},\ }\href
  {https://doi.org/10.1103/PhysRevB.82.184418} {\bibfield  {journal} {\bibinfo
  {journal} {Phys. Rev. B}\ }\textbf {\bibinfo {volume} {82}},\ \bibinfo
  {pages} {184418} (\bibinfo {year} {2010})}\BibitemShut {NoStop}%
\bibitem [{\citenamefont {Moriyama}\ \emph {et~al.}(2018)\citenamefont
  {Moriyama}, \citenamefont {Zhou}, \citenamefont {Seki}, \citenamefont
  {Takanashi},\ and\ \citenamefont {Ono}}]{moriyama2018spin1}%
  \BibitemOpen
  \bibfield  {author} {\bibinfo {author} {\bibfnamefont {T.}~\bibnamefont
  {Moriyama}}, \bibinfo {author} {\bibfnamefont {W.}~\bibnamefont {Zhou}},
  \bibinfo {author} {\bibfnamefont {T.}~\bibnamefont {Seki}}, \bibinfo {author}
  {\bibfnamefont {K.}~\bibnamefont {Takanashi}},\ and\ \bibinfo {author}
  {\bibfnamefont {T.}~\bibnamefont {Ono}},\ }\bibfield  {title} {\bibinfo
  {title} {\textit{Spin-Orbit-Torque Memory Operation of Synthetic
  Antiferromagnets}},\ }\href {https://doi.org/10.1103/PhysRevLett.121.167202}
  {\bibfield  {journal} {\bibinfo  {journal} {Phys. Rev. Lett.}\ }\textbf
  {\bibinfo {volume} {121}},\ \bibinfo {pages} {167202} (\bibinfo {year}
  {2018})}\BibitemShut {NoStop}%
\bibitem [{\citenamefont {Saglam}\ \emph {et~al.}(2020)\citenamefont {Saglam},
  \citenamefont {Liu}, \citenamefont {Li}, \citenamefont {Sklenar},
  \citenamefont {Gibbons}, \citenamefont {Hong}, \citenamefont {Karakas},
  \citenamefont {Pearson}, \citenamefont {Ozatay}, \citenamefont {Zhang} \emph
  {et~al.}}]{saglam2020anomalous}%
  \BibitemOpen
  \bibfield  {author} {\bibinfo {author} {\bibfnamefont {H.}~\bibnamefont
  {Saglam}}, \bibinfo {author} {\bibfnamefont {C.}~\bibnamefont {Liu}},
  \bibinfo {author} {\bibfnamefont {Y.}~\bibnamefont {Li}}, \bibinfo {author}
  {\bibfnamefont {J.}~\bibnamefont {Sklenar}}, \bibinfo {author} {\bibfnamefont
  {J.}~\bibnamefont {Gibbons}}, \bibinfo {author} {\bibfnamefont
  {D.}~\bibnamefont {Hong}}, \bibinfo {author} {\bibfnamefont {V.}~\bibnamefont
  {Karakas}}, \bibinfo {author} {\bibfnamefont {J.~E.}\ \bibnamefont
  {Pearson}}, \bibinfo {author} {\bibfnamefont {O.}~\bibnamefont {Ozatay}},
  \bibinfo {author} {\bibfnamefont {W.}~\bibnamefont {Zhang}}, \emph {et~al.},\
  }\href@noop {} {\bibinfo {title} {\textit{Anomalous Hall and Nernst Effects
  in \textnormal{FeRh}}}} (\bibinfo {year} {2020}),\ \Eprint
  {https://arxiv.org/abs/2012.14383} {arXiv:2012.14383 [cond-mat.mtrl-sci]}
  \BibitemShut {NoStop}%
\bibitem [{\citenamefont {Wu}\ \emph {et~al.}(2016)\citenamefont {Wu},
  \citenamefont {Zhang}, \citenamefont {KC}, \citenamefont {Borisov},
  \citenamefont {Pearson}, \citenamefont {Jiang}, \citenamefont {Lederman},
  \citenamefont {Hoffmann},\ and\ \citenamefont
  {Bhattacharya}}]{wu2016antiferromagnetic}%
  \BibitemOpen
  \bibfield  {author} {\bibinfo {author} {\bibfnamefont {S.~M.}\ \bibnamefont
  {Wu}}, \bibinfo {author} {\bibfnamefont {W.}~\bibnamefont {Zhang}}, \bibinfo
  {author} {\bibfnamefont {A.}~\bibnamefont {KC}}, \bibinfo {author}
  {\bibfnamefont {P.}~\bibnamefont {Borisov}}, \bibinfo {author} {\bibfnamefont
  {J.~E.}\ \bibnamefont {Pearson}}, \bibinfo {author} {\bibfnamefont {J.~S.}\
  \bibnamefont {Jiang}}, \bibinfo {author} {\bibfnamefont {D.}~\bibnamefont
  {Lederman}}, \bibinfo {author} {\bibfnamefont {A.}~\bibnamefont {Hoffmann}},\
  and\ \bibinfo {author} {\bibfnamefont {A.}~\bibnamefont {Bhattacharya}},\
  }\bibfield  {title} {\bibinfo {title} {\textit{Antiferromagnetic Spin Seebeck
  Effect}},\ }\href {https://doi.org/10.1103/PhysRevLett.116.097204} {\bibfield
   {journal} {\bibinfo  {journal} {Phys. Rev. Lett.}\ }\textbf {\bibinfo
  {volume} {116}},\ \bibinfo {pages} {097204} (\bibinfo {year}
  {2016})}\BibitemShut {NoStop}%
\bibitem [{\citenamefont {Shiomi}\ \emph {et~al.}(2017)\citenamefont {Shiomi},
  \citenamefont {Takashima}, \citenamefont {Okuyama}, \citenamefont
  {Gitgeatpong}, \citenamefont {Piyawongwatthana}, \citenamefont {Matan},
  \citenamefont {Sato},\ and\ \citenamefont {Saitoh}}]{shiomi2017spin}%
  \BibitemOpen
  \bibfield  {author} {\bibinfo {author} {\bibfnamefont {Y.}~\bibnamefont
  {Shiomi}}, \bibinfo {author} {\bibfnamefont {R.}~\bibnamefont {Takashima}},
  \bibinfo {author} {\bibfnamefont {D.}~\bibnamefont {Okuyama}}, \bibinfo
  {author} {\bibfnamefont {G.}~\bibnamefont {Gitgeatpong}}, \bibinfo {author}
  {\bibfnamefont {P.}~\bibnamefont {Piyawongwatthana}}, \bibinfo {author}
  {\bibfnamefont {K.}~\bibnamefont {Matan}}, \bibinfo {author} {\bibfnamefont
  {T.~J.}\ \bibnamefont {Sato}},\ and\ \bibinfo {author} {\bibfnamefont
  {E.}~\bibnamefont {Saitoh}},\ }\bibfield  {title} {\bibinfo {title}
  {\textit{Spin Seebeck Effect in the Polar Antiferromagnet
  \textnormal{$\ensuremath{\alpha}\text{\ensuremath{-}}{\mathrm{Cu}}_{2}{\mathrm{V}}_{2}{\mathrm{O}}_{7}$}}},\
  }\href {https://doi.org/10.1103/PhysRevB.96.180414} {\bibfield  {journal}
  {\bibinfo  {journal} {Phys. Rev. B}\ }\textbf {\bibinfo {volume} {96}},\
  \bibinfo {pages} {180414} (\bibinfo {year} {2017})}\BibitemShut {NoStop}%
\bibitem [{\citenamefont {Li}\ \emph {et~al.}(2019)\citenamefont {Li},
  \citenamefont {Shi}, \citenamefont {Ortiz}, \citenamefont {Aldosary},
  \citenamefont {Chen}, \citenamefont {Aji}, \citenamefont {Wei},\ and\
  \citenamefont {Shi}}]{li2019spin}%
  \BibitemOpen
  \bibfield  {author} {\bibinfo {author} {\bibfnamefont {J.}~\bibnamefont
  {Li}}, \bibinfo {author} {\bibfnamefont {Z.}~\bibnamefont {Shi}}, \bibinfo
  {author} {\bibfnamefont {V.~H.}\ \bibnamefont {Ortiz}}, \bibinfo {author}
  {\bibfnamefont {M.}~\bibnamefont {Aldosary}}, \bibinfo {author}
  {\bibfnamefont {C.}~\bibnamefont {Chen}}, \bibinfo {author} {\bibfnamefont
  {V.}~\bibnamefont {Aji}}, \bibinfo {author} {\bibfnamefont {P.}~\bibnamefont
  {Wei}},\ and\ \bibinfo {author} {\bibfnamefont {J.}~\bibnamefont {Shi}},\
  }\bibfield  {title} {\bibinfo {title} {\textit{Spin Seebeck Effect from
  Antiferromagnetic Magnons and Critical Spin Fluctuations in Epitaxial
  \textnormal{${\mathrm{FeF}}_{2}$} Films}},\ }\href
  {https://doi.org/10.1103/PhysRevLett.122.217204} {\bibfield  {journal}
  {\bibinfo  {journal} {Phys. Rev. Lett.}\ }\textbf {\bibinfo {volume} {122}},\
  \bibinfo {pages} {217204} (\bibinfo {year} {2019})}\BibitemShut {NoStop}%
\bibitem [{\citenamefont {Touloukian}\ \emph {et~al.}(1971)\citenamefont
  {Touloukian}, \citenamefont {Powell}, \citenamefont {Ho},\ and\ \citenamefont
  {Klemens}}]{touloukian1970thermophysical}%
  \BibitemOpen
  \bibfield  {author} {\bibinfo {author} {\bibfnamefont {Y.~S.}\ \bibnamefont
  {Touloukian}}, \bibinfo {author} {\bibfnamefont {R.~W.}\ \bibnamefont
  {Powell}}, \bibinfo {author} {\bibfnamefont {C.~Y.}\ \bibnamefont {Ho}},\
  and\ \bibinfo {author} {\bibfnamefont {P.~G.}\ \bibnamefont {Klemens}},\
  }\bibfield  {title} {\bibinfo {title} {\textit{Thermophysical Properties of
  Matter}},\ }\href {https://www.osti.gov/biblio/5303523} {\bibfield  {journal}
  {\bibinfo  {journal} {New York}\ }\textbf {\bibinfo {volume} {2}},\ \bibinfo
  {pages} {158} (\bibinfo {year} {1971})}\BibitemShut {NoStop}%
\bibitem [{\citenamefont {Swartz}\ and\ \citenamefont
  {Pohl}(1987)}]{swartz1987thermal}%
  \BibitemOpen
  \bibfield  {author} {\bibinfo {author} {\bibfnamefont {E.}~\bibnamefont
  {Swartz}}\ and\ \bibinfo {author} {\bibfnamefont {R.}~\bibnamefont {Pohl}},\
  }\bibfield  {title} {\bibinfo {title} {\textit{Thermal Resistance at
  Interfaces}},\ }\href {https://doi.org/10.1063/1.98939} {\bibfield  {journal}
  {\bibinfo  {journal} {Appl. Phys. Lett.}\ }\textbf {\bibinfo {volume} {51}},\
  \bibinfo {pages} {2200} (\bibinfo {year} {1987})}\BibitemShut {NoStop}%
\bibitem [{\citenamefont {Duy~Khang}\ and\ \citenamefont
  {Hai}(2019)}]{duy2019giant}%
  \BibitemOpen
  \bibfield  {author} {\bibinfo {author} {\bibfnamefont {N.~H.}\ \bibnamefont
  {Duy~Khang}}\ and\ \bibinfo {author} {\bibfnamefont {P.~N.}\ \bibnamefont
  {Hai}},\ }\bibfield  {title} {\bibinfo {title} {\textit{Giant Unidirectional
  Spin Hall Magnetoresistance in Topological Insulator--Ferromagnetic
  Semiconductor Heterostructures}},\ }\href {https://doi.org/10.1063/1.5134728}
  {\bibfield  {journal} {\bibinfo  {journal} {J. Appl. Phys.}\ }\textbf
  {\bibinfo {volume} {126}},\ \bibinfo {pages} {233903} (\bibinfo {year}
  {2019})}\BibitemShut {NoStop}%
\bibitem [{\citenamefont {Navarro}\ \emph {et~al.}(1999)\citenamefont
  {Navarro}, \citenamefont {Hernando}, \citenamefont {Yavari}, \citenamefont
  {Fiorani},\ and\ \citenamefont {Rosenberg}}]{navarro1999grain}%
  \BibitemOpen
  \bibfield  {author} {\bibinfo {author} {\bibfnamefont {E.}~\bibnamefont
  {Navarro}}, \bibinfo {author} {\bibfnamefont {A.}~\bibnamefont {Hernando}},
  \bibinfo {author} {\bibfnamefont {A.}~\bibnamefont {Yavari}}, \bibinfo
  {author} {\bibfnamefont {D.}~\bibnamefont {Fiorani}},\ and\ \bibinfo {author}
  {\bibfnamefont {M.}~\bibnamefont {Rosenberg}},\ }\bibfield  {title} {\bibinfo
  {title} {\textit{Grain-Boundary Magnetic Properties of Ball-Milled
  Nanocrystalline \textnormal{Fe$_x$Rh$_{100-x}$} Alloys}},\ }\href
  {https://doi.org/10.1063/1.371025} {\bibfield  {journal} {\bibinfo  {journal}
  {J. Appl. Phys.}\ }\textbf {\bibinfo {volume} {86}},\ \bibinfo {pages} {2166}
  (\bibinfo {year} {1999})}\BibitemShut {NoStop}%
\bibitem [{\citenamefont {Barton}\ \emph {et~al.}(2017)\citenamefont {Barton},
  \citenamefont {Ostler}, \citenamefont {Huskisson}, \citenamefont {Kinane},
  \citenamefont {Haigh}, \citenamefont {Hrkac},\ and\ \citenamefont
  {Thomson}}]{barton2017substrate}%
  \BibitemOpen
  \bibfield  {author} {\bibinfo {author} {\bibfnamefont {C.}~\bibnamefont
  {Barton}}, \bibinfo {author} {\bibfnamefont {T.}~\bibnamefont {Ostler}},
  \bibinfo {author} {\bibfnamefont {D.}~\bibnamefont {Huskisson}}, \bibinfo
  {author} {\bibfnamefont {C.}~\bibnamefont {Kinane}}, \bibinfo {author}
  {\bibfnamefont {S.}~\bibnamefont {Haigh}}, \bibinfo {author} {\bibfnamefont
  {G.}~\bibnamefont {Hrkac}},\ and\ \bibinfo {author} {\bibfnamefont
  {T.}~\bibnamefont {Thomson}},\ }\bibfield  {title} {\bibinfo {title}
  {\textit{Substrate Induced Strain Field in \textnormal{FeRh} Epilayers Grown
  on Single Crystal \textnormal{MgO (001)} Substrates}},\ }\href
  {https://www.nature.com/articles/srep44397} {\bibfield  {journal} {\bibinfo
  {journal} {Sci. Rep.}\ }\textbf {\bibinfo {volume} {7}},\ \bibinfo {pages}
  {1} (\bibinfo {year} {2017})}\BibitemShut {NoStop}%
\bibitem [{\citenamefont {Li}\ \emph {et~al.}(2021)\citenamefont {Li},
  \citenamefont {Zhao}, \citenamefont {Zhu}, \citenamefont {Lv}, \citenamefont
  {Cao}, \citenamefont {Zhu}, \citenamefont {Sun}, \citenamefont {Peng},
  \citenamefont {Cheng}, \citenamefont {Jiang} \emph
  {et~al.}}]{li2021electric}%
  \BibitemOpen
  \bibfield  {author} {\bibinfo {author} {\bibfnamefont {Z.}~\bibnamefont
  {Li}}, \bibinfo {author} {\bibfnamefont {J.}~\bibnamefont {Zhao}}, \bibinfo
  {author} {\bibfnamefont {Q.}~\bibnamefont {Zhu}}, \bibinfo {author}
  {\bibfnamefont {X.}~\bibnamefont {Lv}}, \bibinfo {author} {\bibfnamefont
  {C.}~\bibnamefont {Cao}}, \bibinfo {author} {\bibfnamefont {X.}~\bibnamefont
  {Zhu}}, \bibinfo {author} {\bibfnamefont {L.}~\bibnamefont {Sun}}, \bibinfo
  {author} {\bibfnamefont {Y.}~\bibnamefont {Peng}}, \bibinfo {author}
  {\bibfnamefont {W.}~\bibnamefont {Cheng}}, \bibinfo {author} {\bibfnamefont
  {D.}~\bibnamefont {Jiang}}, \emph {et~al.},\ }\bibfield  {title} {\bibinfo
  {title} {\textit{Electric Control of Magnetic Properties in Epitaxially Grown
  \textnormal{FeRh/MgO/PMN-PT} Heterostructures}},\ }\href
  {https://www.sciencedirect.com/science/article/pii/S0925838821006289}
  {\bibfield  {journal} {\bibinfo  {journal} {J. Alloys Compd.}\ }\textbf
  {\bibinfo {volume} {868}},\ \bibinfo {pages} {159220} (\bibinfo {year}
  {2021})}\BibitemShut {NoStop}%
\bibitem [{\citenamefont {Arregi}\ \emph {et~al.}(2020)\citenamefont {Arregi},
  \citenamefont {Caha},\ and\ \citenamefont
  {Uhl{\'\i}{\v{r}}}}]{arregi2020evolution}%
  \BibitemOpen
  \bibfield  {author} {\bibinfo {author} {\bibfnamefont {J.~A.}\ \bibnamefont
  {Arregi}}, \bibinfo {author} {\bibfnamefont {O.}~\bibnamefont {Caha}},\ and\
  \bibinfo {author} {\bibfnamefont {V.}~\bibnamefont {Uhl{\'\i}{\v{r}}}},\
  }\bibfield  {title} {\bibinfo {title} {\textit{Evolution of Strain Across the
  Magnetostructural Phase Transition in Epitaxial \textnormal{FeRh} Films on
  Different Substrates}},\ }\href
  {https://journals.aps.org/prb/abstract/10.1103/PhysRevB.101.174413}
  {\bibfield  {journal} {\bibinfo  {journal} {Phys. Rev. B}\ }\textbf {\bibinfo
  {volume} {101}},\ \bibinfo {pages} {174413} (\bibinfo {year}
  {2020})}\BibitemShut {NoStop}%
\bibitem [{\citenamefont {Qiao}\ \emph {et~al.}(2022)\citenamefont {Qiao},
  \citenamefont {Liang}, \citenamefont {Zhang}, \citenamefont {Hu},
  \citenamefont {Yu}, \citenamefont {Long}, \citenamefont {Wang}, \citenamefont
  {Sun}, \citenamefont {Zhao},\ and\ \citenamefont
  {Shen}}]{qiao2022manipulation}%
  \BibitemOpen
  \bibfield  {author} {\bibinfo {author} {\bibfnamefont {K.}~\bibnamefont
  {Qiao}}, \bibinfo {author} {\bibfnamefont {Y.}~\bibnamefont {Liang}},
  \bibinfo {author} {\bibfnamefont {H.}~\bibnamefont {Zhang}}, \bibinfo
  {author} {\bibfnamefont {F.}~\bibnamefont {Hu}}, \bibinfo {author}
  {\bibfnamefont {Z.}~\bibnamefont {Yu}}, \bibinfo {author} {\bibfnamefont
  {Y.}~\bibnamefont {Long}}, \bibinfo {author} {\bibfnamefont {J.}~\bibnamefont
  {Wang}}, \bibinfo {author} {\bibfnamefont {J.}~\bibnamefont {Sun}}, \bibinfo
  {author} {\bibfnamefont {T.}~\bibnamefont {Zhao}},\ and\ \bibinfo {author}
  {\bibfnamefont {B.}~\bibnamefont {Shen}},\ }\bibfield  {title} {\bibinfo
  {title} {\textit{Manipulation of Magnetocaloric Effect in \textnormal{FeRh}
  Films by Epitaxial Growth}},\ }\href
  {https://doi.org/10.1016/j.jallcom.2022.164574} {\bibfield  {journal}
  {\bibinfo  {journal} {J. Alloys Compd.}\ }\textbf {\bibinfo {volume} {907}},\
  \bibinfo {pages} {164574} (\bibinfo {year} {2022})}\BibitemShut {NoStop}%
\end{thebibliography}%

\clearpage
\onecolumngrid
\begin{center}
\textbf{\large 
\medskip
Supplemental Material for ``Unidirectional magnetoresistance in antiferromagnet$|$heavy-metal bilayers"}
\end{center}

\setcounter{equation}{0}
\setcounter{figure}{0}
\setcounter{table}{0}
\setcounter{section}{0}
\makeatletter
\renewcommand*{\thesection}{S\arabic{section}}
\renewcommand{\theequation}{S.\arabic{equation}}
\renewcommand{\thefigure}{S.\arabic{figure}}
\medskip 

\section{Joule heating in the microwire device}

\begin{figure*}[tph]
    \includegraphics[width=0.35\linewidth]{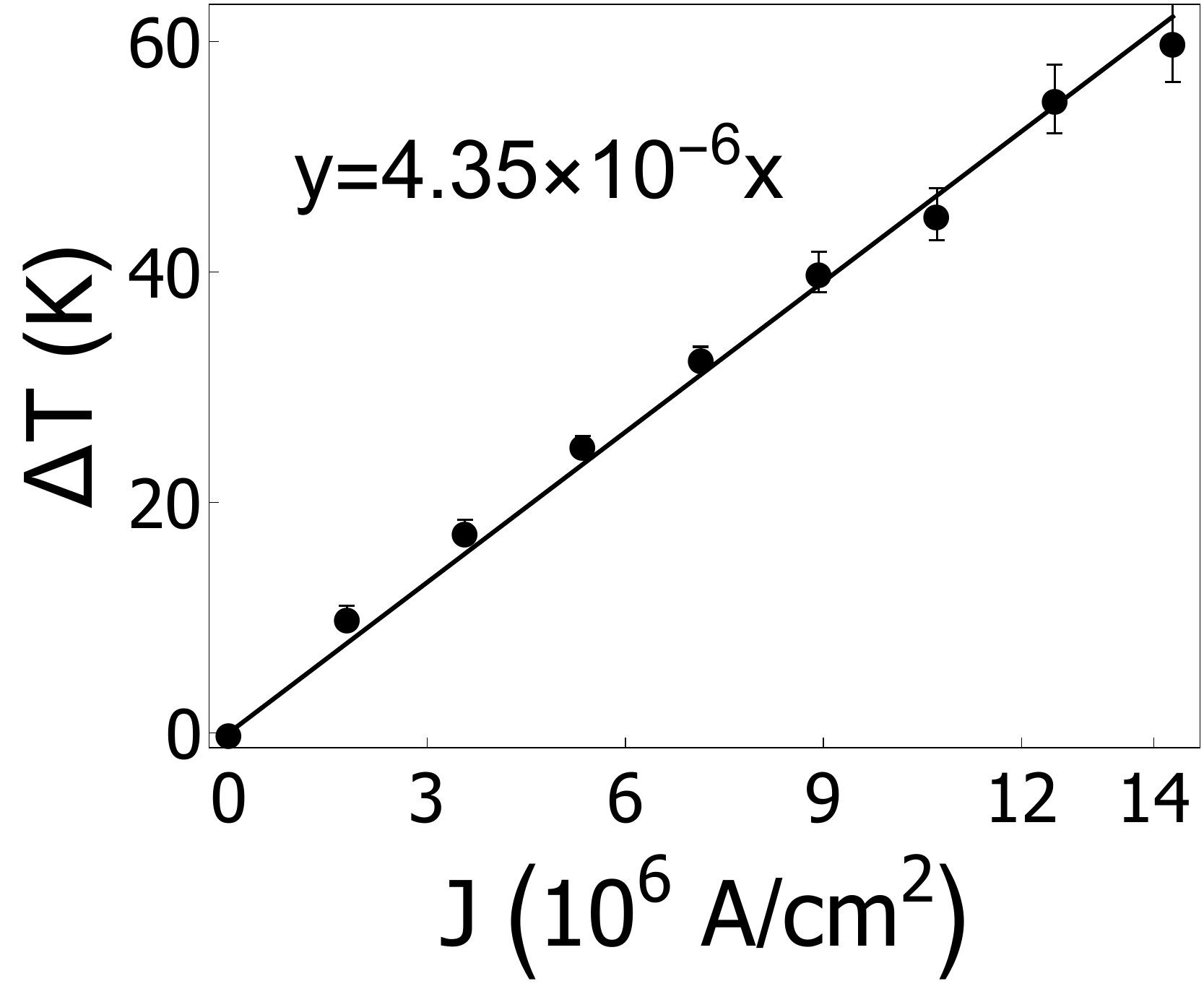}
    \caption{Estimation of Joule heating in the device: \normalfont{Temperature increase in the sample $\Delta T$ due to the Joule heating from the applied current $J$. Temperature increase for $J={10}^6$~A/$\textup{cm}^2$ is estimated to be about 4.35~K.}}
    \label{fig:S1_a}%
\end{figure*}

While the ambient temperature in the sample space is kept at 10 K for all measurements in the antiferromagnetic phase, the true device temperature can increase due to Joule heating from the applied current. The temperature increase from Joule heating is estimated by comparing the longitudinal resistance for finite current density at $B = 4$~T with the corresponding longitudinal resistance measured without the applied DC current (e.g., as shown in Fig.~2a). As shown in Fig.~\ref{fig:S1_a}, the temperature increase per $J=10^6$~A/$\textup{cm}^2$ is estimated to be about 4.35~K, corresponding to $\Delta T \approx 60$~K for the maximum applied current density of $J = 1.43\times{10}^7$~A/$\textup{cm}^2$. While the temperature sweep of the resistance shows that the device is strictly antiferromagnetic until $T \approx 180$~K (Fig. 2a), we further confirm this by measuring anisotropic magnetoresistance (AMR) at $B = 12$~T, with no applied DC current, for temperatures ranging from $T = 10$~K to $T = 70$~K (Fig.~\ref{fig:S1_b}). Given that there is a 90\degree ~phase difference in the AMR signal between ferromagnets and antiferromagnets \cite{moriyama2018spin1}, no phase shift in the AMR up to $T = 70$~K confirms that the device is indeed antiferromagnetic even with the maximum current density applied.

\section{Linear-response magnetoresistance in FeRh$|$Pt bilayer}

\begin{figure*}[tph]
    \sidesubfloat[]{\includegraphics[width=0.35\linewidth,trim=0cm 0 0.5cm 1cm]{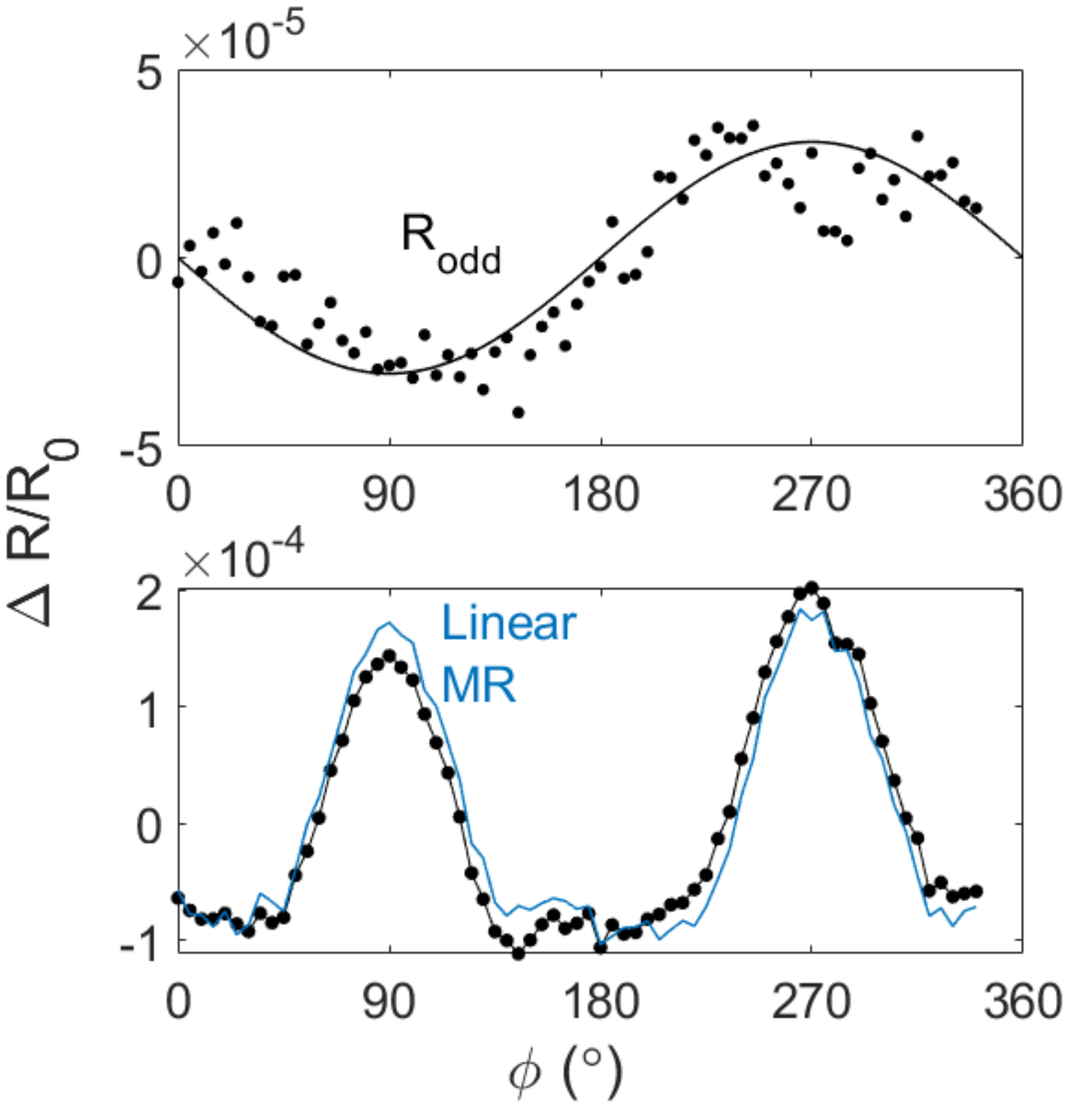}\label{fig:4T4mA}}\quad%
    \sidesubfloat[]{\includegraphics[width=0.35\linewidth]{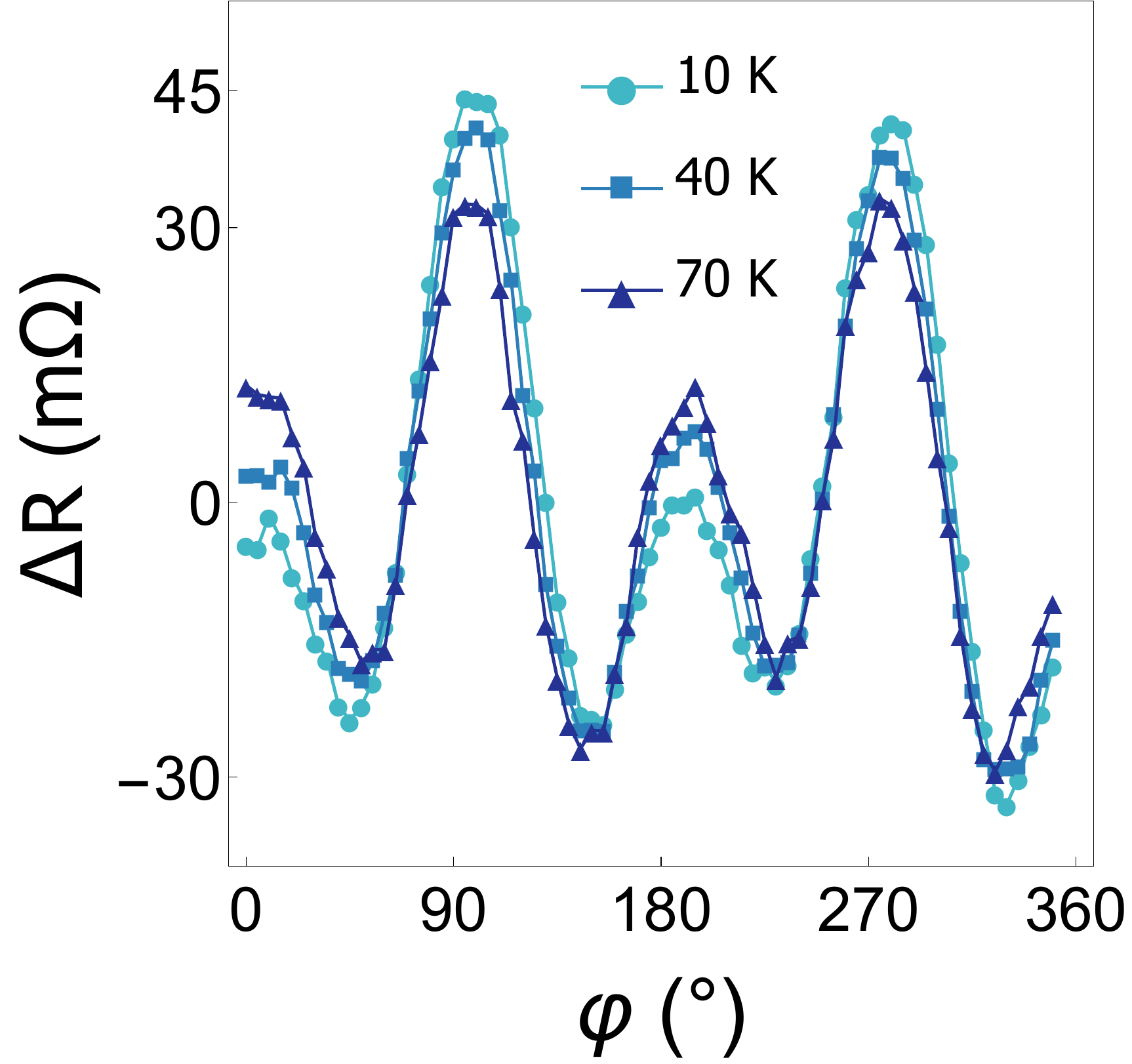}\label{fig:S1_b}}\quad%
    \sidesubfloat[]{\includegraphics[width=0.35\linewidth]{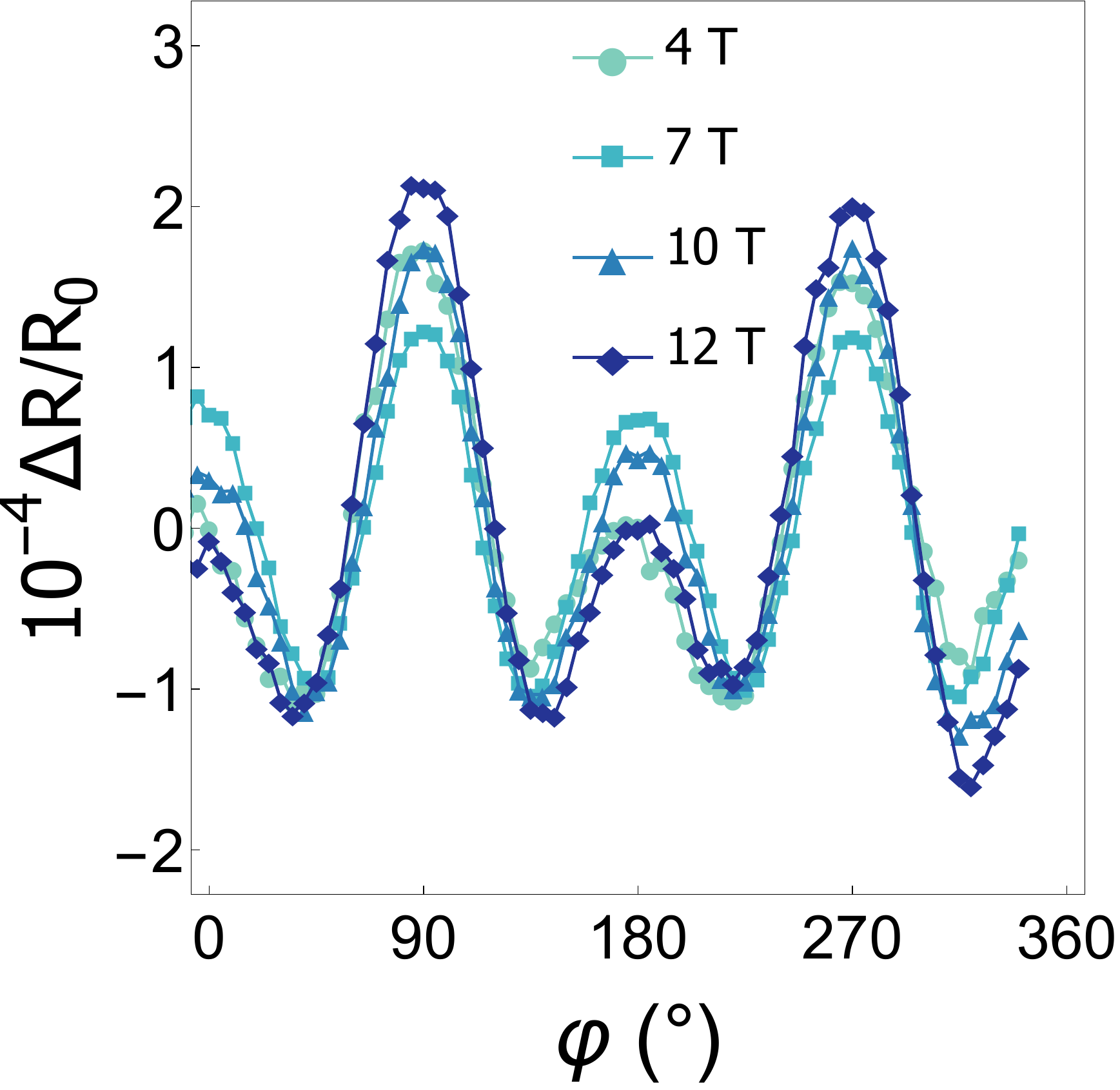}\label{fig:linearMR}}\quad%
    \sidesubfloat[]{\includegraphics[width=0.35\linewidth]{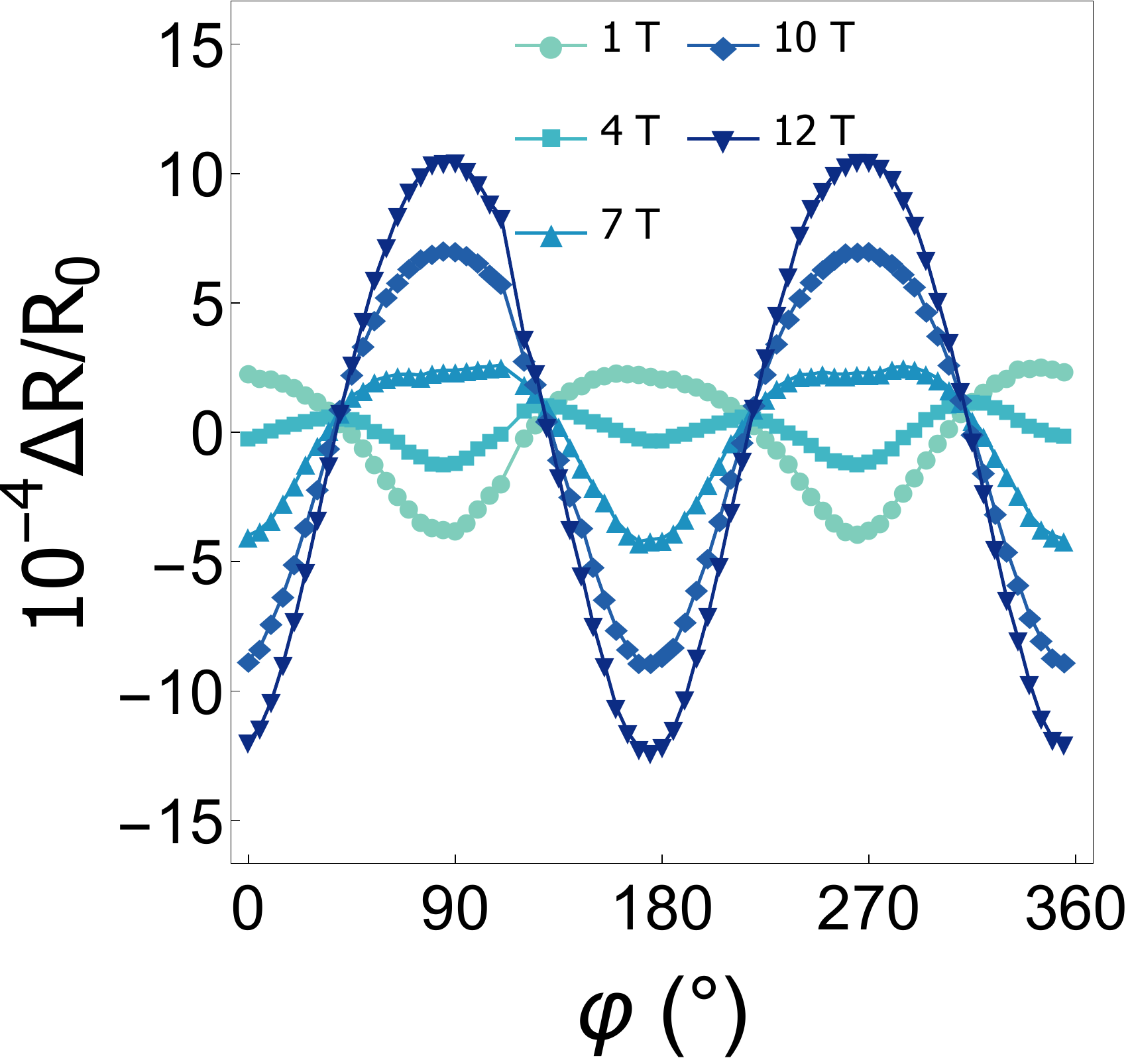}\label{fig:linearMR100}}%
    \caption{Linear-response magnetoresistance in FeRh$|$Pt bilayer: \normalfont{(a) Linear-response magnetoresistance and $R_{\text{odd}}$ of FeRh[110]$\mid$Pt device, in the angular sweep of longitudinal resistance, at $T = 10$~K, $B = 4$~T, and $I = 4$~mA ($J = 1.43\times{10}^7$~A/$\textup{cm}^2$). Top panel: $R_{\text{odd}}$ is defined as half the difference between data sets $R(I)$ and $R(-I)$, ($R_{\text{odd}} = [R(I)-R(-I)]/2$), and the raw data $R_{\text{odd}}$ (scatter plot) is fitted into a sinusoidal curve to define the amplitude $R_{\text{odd,max}}$.
    Bottom panel: data set of R(I) (black dotted) and the linear-response magnetoresistance, $R(I)-R_{\text{odd}}$ (blue curve).
     (b) Anisotropic magnetoresistance of FeRh[110]$\mid$Pt device, at $B = 12$~T for various temperatures, with no DC current applied. (c) Anisotropic magnetoresistance of FeRh[110]$\mid$Pt device, for various field strengths, at $T = 10$~K and no DC current applied. (d) Anisotropic magnetoresistance of FeRh[100]$\mid$Pt device, for various field strengths, at $T = 10$~K and no DC current applied.}}
\end{figure*}

When we measure how the longitudinal resistance changes with respect to the orientation of the current from the external magnetic field, the raw data $R(I)$ can be decomposed into two components: linear-response magnetoresistance (MR) which is current-independent and even under the current reversal, and resistance which is odd under the current polarity, $R_{\text{odd}}$ (Fig.~\ref{fig:4T4mA}). In the FeRh$|$Pt bilayer, we observe how the linear-response MR depends on the temperature (Fig.~\ref{fig:S1_b}), the applied magnetic field (Fig.~\ref{fig:linearMR}), and the FeRh crystal orientation (Fig.~\ref{fig:linearMR100}). $R_{\text{odd}}$ is negligible for current densities lower than ${10}^6$~A/$\textup{cm}^2$ so we assume that AMR with no applied DC current can represent the linear-response MR.

Throughout the whole temperature and field range for both FeRh crystal orientations, the linear-response MR is much larger in magnitude compared to the UMR and has a non-trivial angular dependence consisting of two-fold and four-fold contributions, where the relative magnitudes of the two contributions vary with the temperature and external field strength. For instance, for the temperature dependence of AMR in the FeRh[110]$\mid$Pt device at $B = 12$~T, the two-fold MR contribution is dominant at $T=10$~K compared to the four-fold contribution, whereas the two-fold and four-fold contributions are comparable at $T = 70$~K (Fig.~\ref{fig:S1_b}). Similarly, for the field dependence of the AMR in the FeRh[110]$\mid$Pt device at $T = 10$~K, the two-fold MR contribution is dominant at $B = 12$~T compared to the four-fold contribution, whereas the two-fold and four-fold contributions are comparable at $B = 7$~T (Fig.~\ref{fig:linearMR}). 

We also note that the linear-response MR is dramatically different when the current is along the [100] direction of FeRh compared to the [110] direction (Figs.~\ref{fig:linearMR},~\ref{fig:linearMR100}). This is greatly in contrast to the UMR in FeRh$|$Pt, which is independent of whether the current is aligned along the [100]$-$or [110]$-$direction of FeRh (see Supplementary Section~S4).

The potential origins of this complex linear-response MR include AMR from the FeRh layer and spin Hall magnetoresistance (SMR) due to the Pt layer. And given the strong interfacial effect, other linear-response MRs associated with the interface could also be present. To separate out AMR, SMR, and other linear-response MRs from the interface, more measurements and samples would be necessary. An example of a control study would be inserting a spacer layer such as Cu between the FeRh and the Pt to help determine if the linear-response MR in FeRh$|$Pt arises from the SMR.

We also note that separating out linear-response MR effects (between AMR and SMR for instance) is non-trivial. Ideally, one would need to vary the thickness of the FeRh layer in such a study. This presents problems because antiferromagnetic properties are sensitive to the thickness in thin films. In particular, the antiferromagnetic ordering rapidly weakens as the thickness decreases. This may be in part due to the presence of residual ferromagnetic moments near the MgO$|$FeRh interface~\cite{fan2010ferromagnetism}. It is expected then that other quantities, such as the magnetic susceptibility, will change. This makes a comparative study involving field-dependent data in different samples challenging.

Importantly, the linear-response MR effect strongly depends on the orientation of the current with respect to crystalline direction of the FeRh film, which greatly contrasts with the UMR in FeRh$|$Pt. Thus, while the origin of the linear-response MR in FeRh$|$Pt requires much more detailed analysis, we believe that it is beyond the scope of this study, and does not impact our main conclusions.

\section{Separating the thermal contribution in $R_{\text{odd}}$}

Thermoelectric effects must be carefully considered in the measurement of electrical signals due to inevitable Joule heating and consequent temperature gradients in the sample. Only signals originating from the anomalous Nernst effect and the longitudinal spin Seebeck effect under the vertical thermal gradient $\nabla_z T$ possess the same symmetry as that of the UMR~\cite{avci2015unidirectional}.

In order to determine the resistance contribution from the anomalous Nernst effect in the FeRh$\mid$Pt bilayer, we conduct a control measurement with an uncapped FeRh~(20~nm) microwire device. Given similar resistivities between FeRh and Pt ($\rho_{\text{FeRh}} \approx 15~$\textmu$\Omega \cdot \text{cm}$ at $T = 10$~K, $\rho_{\text{Pt}} = 10.6$ \textmu$\Omega \cdot \text{cm} $), we expect a similar thermal profile between 20 nm of uncapped FeRh and a FeRh (15~nm)$\mid$Pt (5~nm) bilayer under the same current density. Thus, the behavior of $R_{\text{odd,max}}$ in the uncapped FeRh device, which only includes the resistance contribution from the anomalous Nernst effect, should reflect the significance of the resistance contribution from the anomalous Nernst effect in the FeRh$\mid$Pt bilayer device.

\begin{figure*}[tph]
    \sidesubfloat[]{\includegraphics[width=0.32\linewidth]{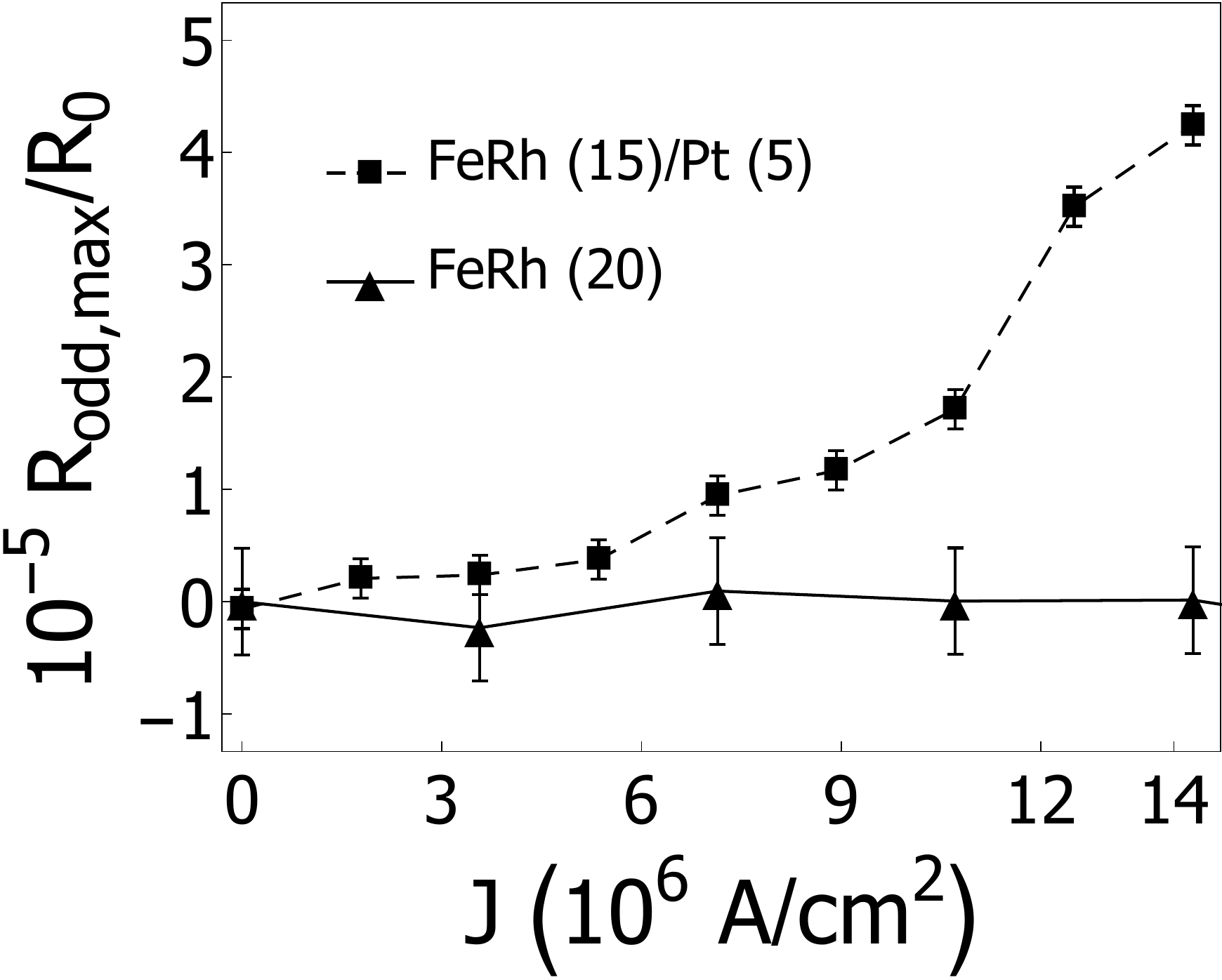}\label{fig:S2_a}}\quad%
    \sidesubfloat[]{\includegraphics[width=0.32\linewidth]{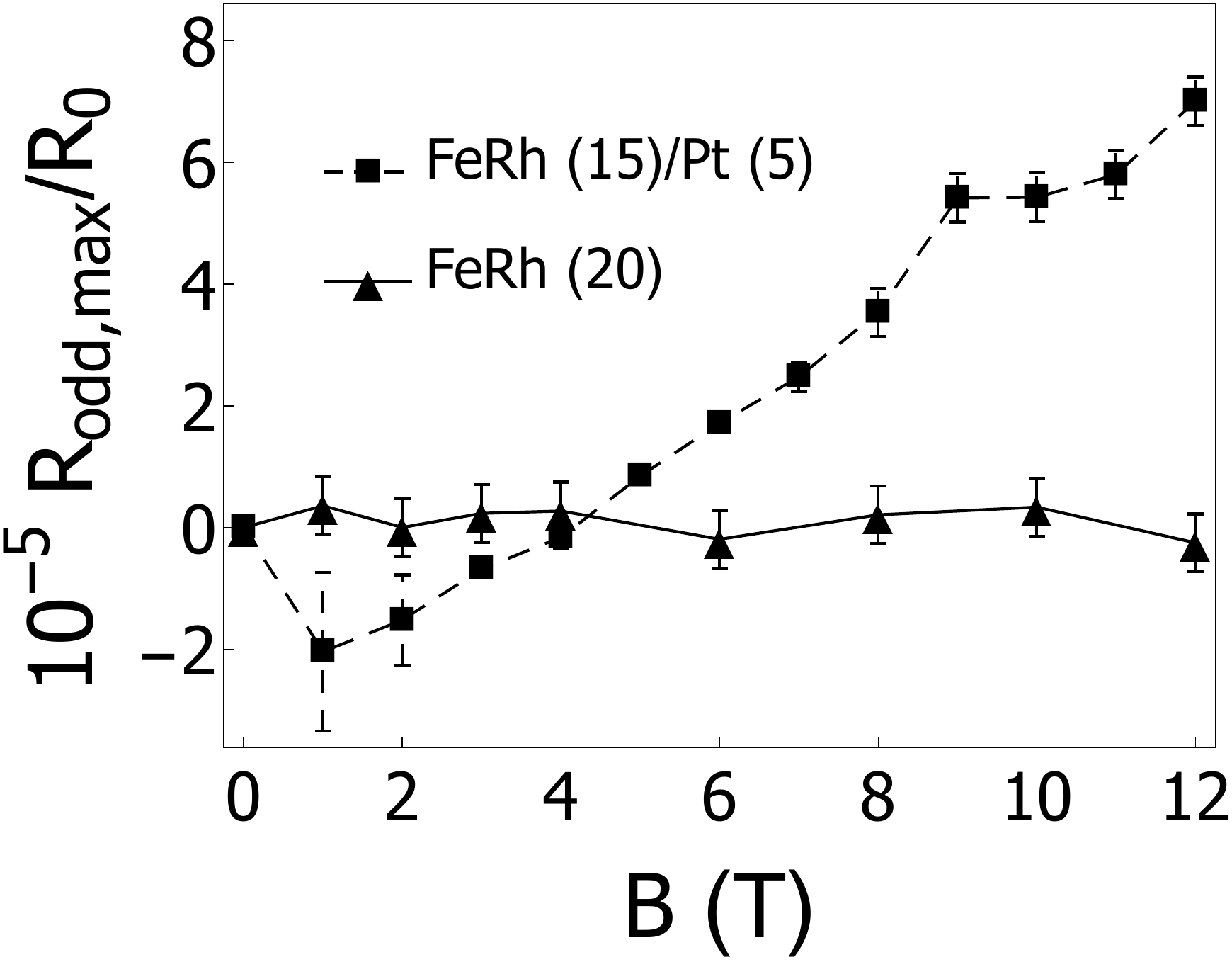}\label{fig:S2_b}}%
    \caption{Absence of UMR in an uncapped FeRh device: \normalfont{(a) Comparison of $R_{\text{odd,max}}/R_0$ in the antiferromagnetic phase ($T = 10$~K) for FeRh (20~nm) and FeRh (15~nm)$\mid$Pt (5~nm), as a function of $J$ at $B = 6$~T. (b) Comparison of $R_{\text{odd,max}}/R_0$ in the antiferromagnetic phase ($T = 10$~K) for FeRh (20~nm) and FeRh (15~nm)$\mid$Pt (5~nm) as a function of field $B$ for $J~=1.5\times{10}^7$~A/$\textup{cm}^2$.}}
\end{figure*}

Figs.~\ref{fig:S2_a} and ~\ref{fig:S2_b} show the negligible evolution in $R_{\text{odd,max}}$ with respect to the applied current $J$ and applied field $B$ in an uncapped FeRh sample, implying that the anomalous Nernst effect contribution in the uncapped FeRh sample is negligible. Although an anomalous Nernst effect has been previously observed in the antiferromagnetic phase of FeRh \cite{saglam2020anomalous}, we find the contribution in our experiments to be negligible,  particularly compared to the Rashba-induced UMR signal.

In the following, we discuss how the spin Seebeck effect can be separated from the Rashba-induced UMR in the bilayer FeRh$\mid$Pt sample.

Since the magnetoresistance and the thermal signal have the same microscopic origin, their amplitude is proportional to the physical distance over which they are measured. Thus, by measuring the thermal signal in the transverse geometry, one can accurately determine its sign and magnitude in the longitudinal measurements. Similar to ferromagnet$\mid$normal-metal bilayers~\cite{avci2015unidirectional}, we must consider the fact that the transverse thermoelectric signal is mixed with the transverse spin-orbit torque signal. These two effects, however, can be separated quantitatively using different angular symmetry and field dependence. In the in-plane angular sweep of the applied field, the field-like spin-orbit torque (FL-SOT) gives a contribution to the Hall signal proportional to $\cos3\varphi + \cos\varphi$, whereas the anti-damping-like spin-orbit torque (AD-SOT) and the thermal effect both give a contribution proportional to $\cos \varphi$~\cite{avci2015unidirectional}. The AD-SOT and thermal contributions can be further separated by considering that the AD-SOT induces dynamical oscillations of the magnetization, the amplitude of which is proportional to the magnetic susceptibility of the antiferromagnetic layer. The resulting Hall spin-orbit torque signal therefore depends on the susceptibility of the antiferromagnetic layer, which is expected to be independent of the external field strength, whereas the thermal contribution, for instance, that of the spin Seebeck effect, increases linearly with the applied field \cite{wu2016antiferromagnetic,shiomi2017spin,li2019spin}, as expected for the smoothly canting sublattice magnetizations in the antiferromagnetic layer.

\begin{figure*}[tph]
    \sidesubfloat[]{\includegraphics[width=0.3\linewidth]{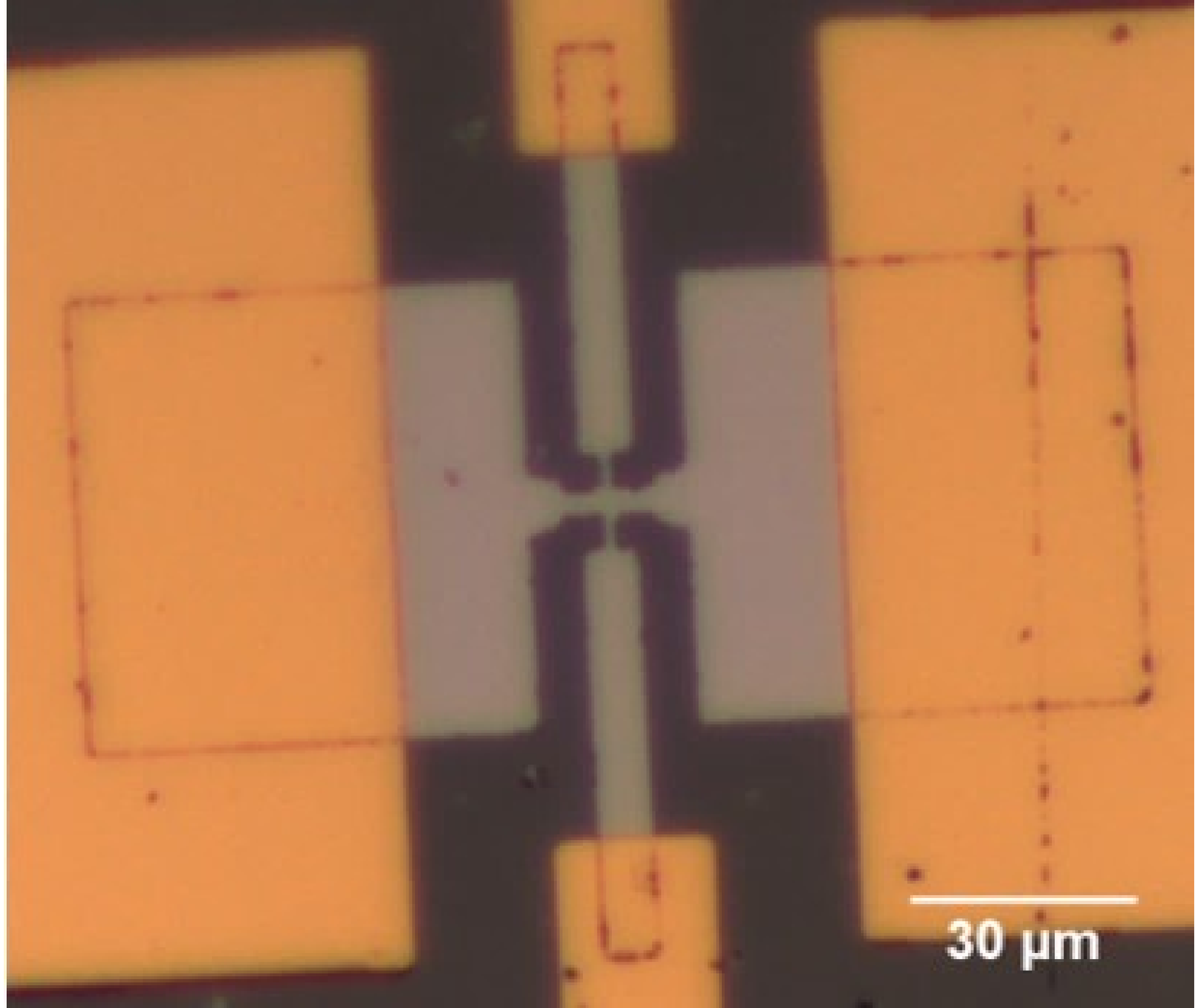}\label{fig:S3_a}}\quad%
    \sidesubfloat[]{\includegraphics[width=0.38\linewidth]{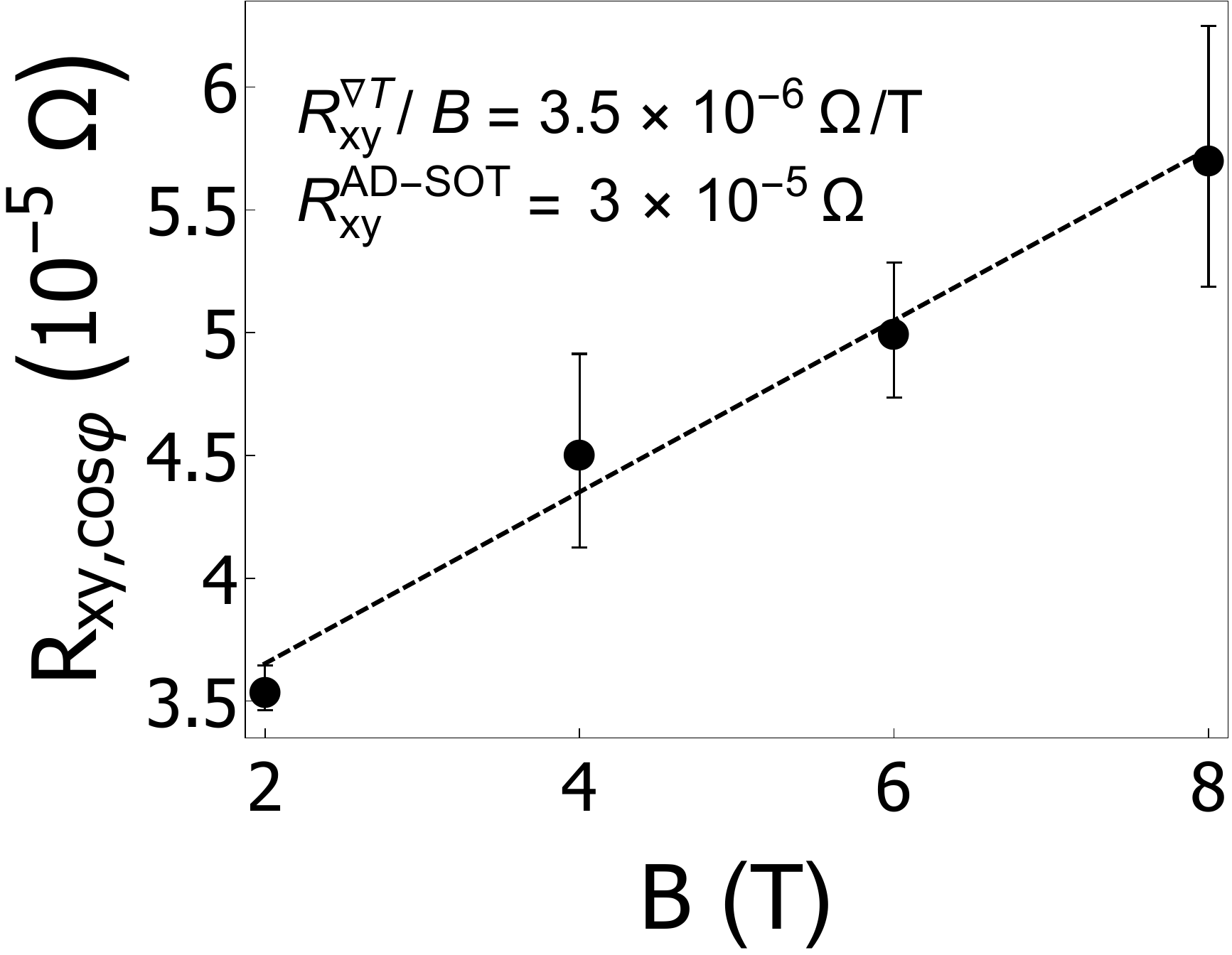}\label{fig:S3_b}}%
    \caption{Estimation of thermal contribution in FeRh$\mid$Pt device: \normalfont{(a) Optical microscope image of a bowtie device with a Hall bar added. (b) Extraction of the thermal contribution $R_{xy}^{\mathbf{\nabla} T}$ using different field dependence for $R_{xy}^{\text{AD-SOT}}$ ($y$-intercept) and $R_{xy}^{\mathbf{\nabla} T}$ (slope), at $J=1.33\times{10}^7$~A/$\textup{cm}^2$. The estimated thermal contribution per applied current and field is $R_{xx}^{\mathbf{\nabla} T} / (R_0 J)=1.77\times{10}^{-13}~\textup{cm}^2/(\textup{A} \times \textup{T})$.}}
\end{figure*}

In order to estimate the thermal contribution in $R_{\text{odd}}$, we fabricated a FeRh (15~nm)$\mid$Pt (5~nm) control device along the [110] direction of FeRh, with a geometry of a microwire combined with a transverse channel (Fig.~\ref{fig:S3_a}). 

While the width $w$ of the control device is 3 \textmu m, twice as that of the main device, the same current density was applied to achieve a similar thermal profile across the device. There can be two contributions to the temperature gradient: bulk thermal conductance $\Lambda$ where $\Delta T_{\Lambda}\approx \frac{1}{2\Lambda}(\frac{P}{L})=\frac{I^2R}{2w\Lambda}$ and interfacial thermal conductance G where $\Delta T_{G}=\frac{1}{Gw}(\frac{P}{L})=\frac{1}{G}(\frac{I^2R}{w^2})$ ($R$ is the sheet resistance, $P$ is the power, $L$ and $w$ are the length and width of the microwire). Since $\Lambda_{T = 10~\text{K}}\approx 7.76$ for MgO~\cite{touloukian1970thermophysical} and $G_{T = 10~\text{K}}\approx 0.3~\textup{MW}/(\textup{m}^2~\textup{K})$ for the interface Rh:Fe$\mid \ce{Al2O3}$~\cite{swartz1987thermal} (similar to the interface of FeRh$\mid$MgO), $\Delta T_{G}\approx52\Delta T_{\Lambda}$ for $w=1$ \textmu m and thus we can conclude that the heating due to the interfacial thermal conductance effect at the low temperature of $T = 10$~K is orders of magnitude more dominant than the heating due to the bulk thermal conductance effect for microwires. We note that $\Delta T_{G}$ is independent of the width of the microwire.

Using the Hall signal in the control FeRh$\mid$Pt microwire device at current density $J = 1.33\times{10}^7$~A/$\textup{cm}^2$, the resistance contribution having $\cos \varphi$ dependence $R_{xy,\cos\varphi}-$which represents the combined AD-SOT contribution and thermal effect$-$was separated from the signal having $\cos3\varphi+\cos\varphi$ dependence. Then, a linear fitting on the $R_{xy,\cos\varphi}$ was done to extract the thermal contribution, represented as the slope in Fig.~\ref{fig:S3_b}. This extracted thermal contribution was then multiplied by the aspect ratio of the device $L/w$ to estimate the thermal contribution along the longitudinal direction, $R_{xx}^{\mathbf{\nabla} T}$. Lastly, given that $R_{xx}^{\mathbf{\nabla} T} \propto JB$ \cite{duy2019giant}, the longitudinal thermal contribution $R_{xx}^{\mathbf{\nabla} T}/R_{0}$ per unit field and unit current density was calculated.

The estimated longitudinal thermal contribution per unit field and unit current density is $R_{xx}^{\mathbf{\nabla} T}/({R}_0JB)~=~1.77\times{10}^{-13}$~{$\textup{cm}^2$}/{($\textup{A}\times$T)}, which corresponds to only 27\% of $R_{\text{odd}}$ at maximum. The longitudinal spin Seebeck effect in antiferromagnets has been previously reported to be lowering resistance for $\varphi=~90\degree$ and increasing resistance for $\varphi=270\degree$ \cite{wu2016antiferromagnetic,shiomi2017spin,li2019spin}. Thus, we propose that $R_{xx}^{\mathbf{\nabla} T}$ for the antiferromagnetic FeRh$\mid$Pt bilayer also follows the same trend and we calculate a true UMR contribution using UMR$\sin\varphi = -(R_{\text{odd}}-R_{xx}^{\mathbf{\nabla} T})/R_{0}$, which shows a similar qualitative behavior as $R_{\text{odd}}$. In addition, we note that the thermal contribution $R_{xx}^{\mathbf{\nabla} T}$ does not change its sign over field or current density, suggesting that thermal contributions cannot explain the UMR sign change observed in the antiferromagnetic phase (Fig. 3).

\section{Dependence of UMR on the current orientation with respect to crystalline axes}

\begin{figure*}[tph]
    \sidesubfloat[]{\includegraphics[width=0.35\linewidth]{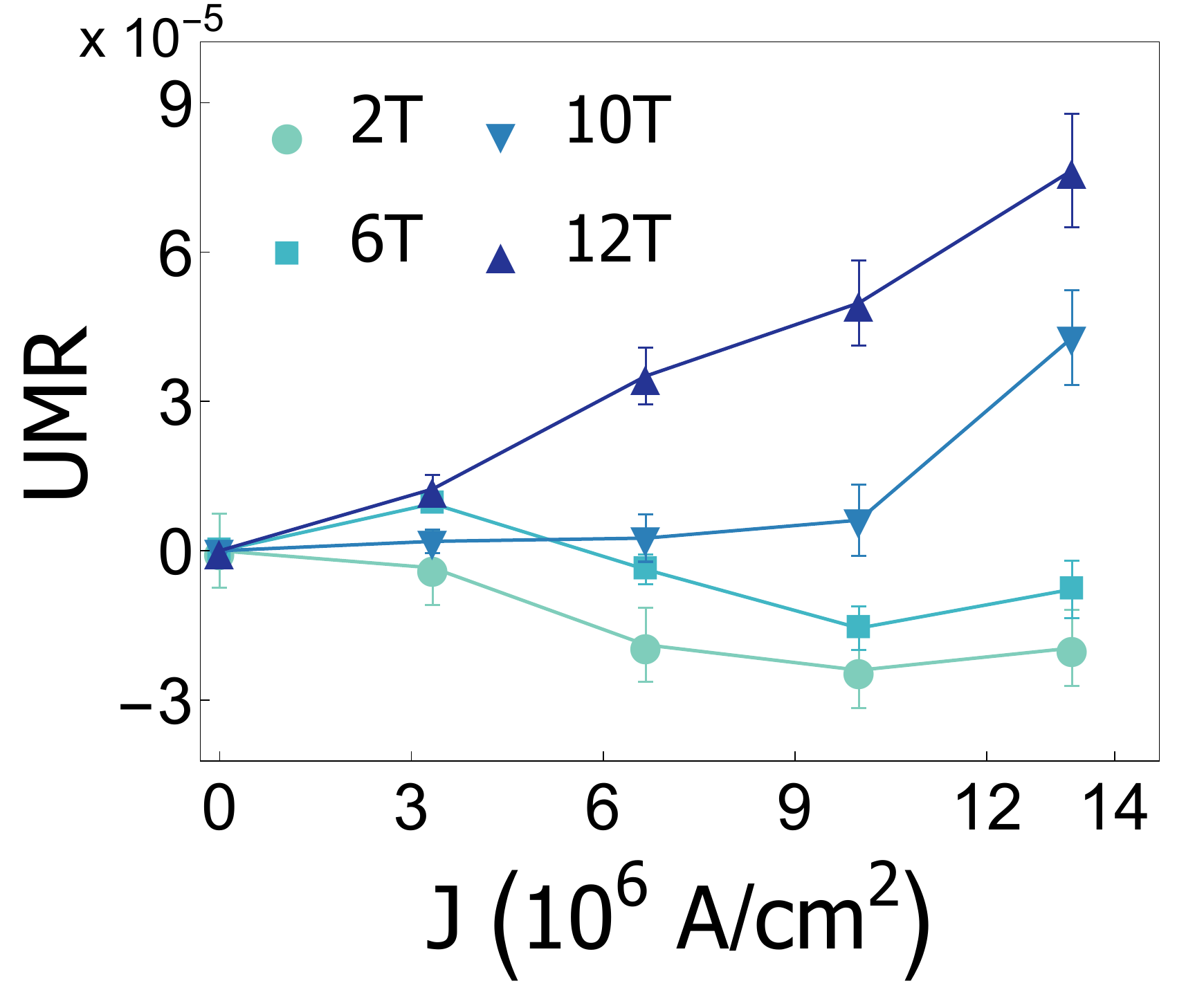}\label{fig:S4_a}}\quad%
    \sidesubfloat[]{\includegraphics[width=0.35\linewidth]{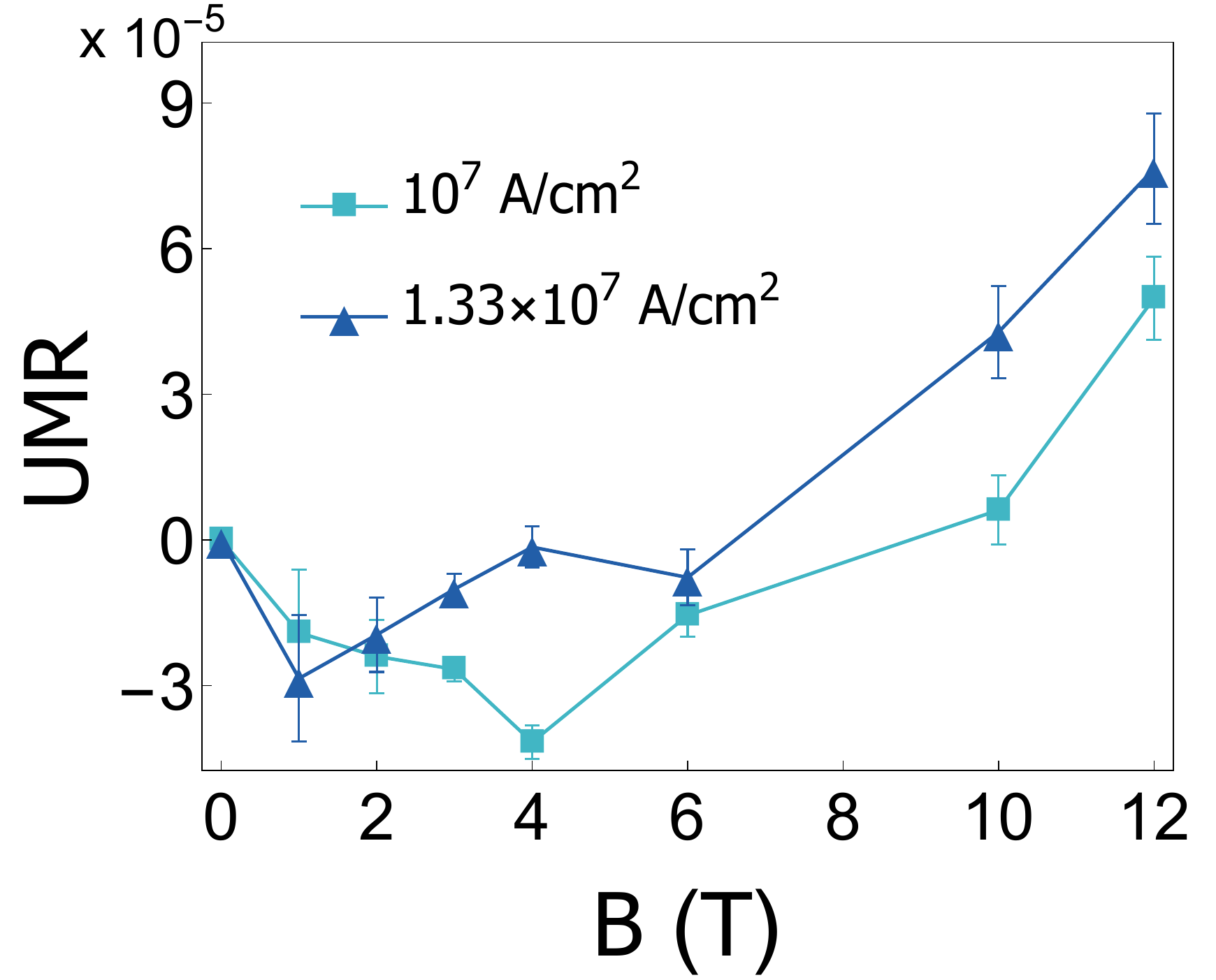}\label{fig:S4_b}}%
    \caption{UMR in the antiferromagnetic phase of FeRh[100]$\mid$Pt device: \normalfont{(a) UMR in the antiferromagnetic phase ($T = 10$~K) of the FeRh$\mid$Pt microwire along [100] of FeRh, as a function of $J$ at various field magnitudes. (b) UMR in the antiferromagnetic phase ($T = 10$~K) of the FeRh$\mid$Pt microwire along [100] of FeRh, as a function of field $B$ at various current densities $J$.}}
\end{figure*}

We investigated the crystal orientation dependence of the UMR. A microwire identical to the one discussed in the main text was defined along the [100] direction of FeRh using the same fabrication process. In Figs.~\ref{fig:S4_a} and~\ref{fig:S4_b}, we show the current and field dependence of the UMR at temperature $T = 10$~K. Our data clearly show that there is not much quantitative difference in the evolution of the UMR for microwires aligned along [100] or [110] of FeRh. Thus we can rule out the possible contribution from the crystal field effect.

\section{Model Hamiltonian and spin canting}

We consider the interfacial layer of the FeRh with broken inversion-symmetry
as a two-dimensional antiferromagnetic square lattice with Rashba spin-orbit
interaction, which can be described by the following tight-binding
Hamiltonian~\cite{vzelezny2014relativistic,baltz2018antiferromagnetic} 
\begin{equation}
\label{sup_hamiltonian}
\hat{H}=\epsilon _{0}+\gamma _{\mathbf{k}}\hat{\tau}_{x}+\Delta_{\text{ex}}\hat{%
\tau}_{z}\hat{\boldsymbol{\sigma }}\cdot \mathbf{m}+\alpha _{R}\hat{\tau}_{x}%
\hat{\boldsymbol{\sigma }}\cdot \hat{\mathbf{z}}\times \mathbf{k}+g\mu _{B}\hat{%
\boldsymbol{\sigma }}\cdot \mathbf{B},
\end{equation}%
where $\epsilon _{0}$ is the on-site energy, $\Delta_{\text{ex}}$ is the s-d
exchange constant between the local moments and the electron, $\alpha _{R}$
is the Rashba spin-orbit coupling constant, and $\hat{\tau}_{i}$ and $\hat{%
\sigma}_{i}$ are Pauli matrices which signify the sublattice and spin
degrees of freedom, respectively. The nearest-neighbor hopping is
represented by $\gamma _{\mathbf{k}}=-2t\left( \cos k_{x}a+\cos
k_{y}a\right) $, where $t$ is the hopping term and $a$ the lattice constant. 

In the semiclassical Boltzmann transport formalism, the nonlinear charge
current density of the $n$-th band is $\mathbf{j}_{e}^{(2),n}=-\frac{e^{3}\tau
^{2}E_{x}^{2}}{\hbar ^{2}}\int_{\mathbf{k}}\left( \frac{\partial ^{2}f}{%
\partial k_{x}^{2}}\right) \mathbf{v}_{n}$, where $E_x$ is the $x$ component of the electric field, $\int_{\mathbf{k}}\equiv
\int_{\text{BZ}}\frac{d^{2}\mathbf{k}}{(2\pi )^{2}}$ and $\mathbf{v}_{n}=%
\frac{\partial \epsilon _{n}(\mathbf{k})}{\hbar \partial \mathbf{k}}$ is the
group velocity of the $n$-th band. Note that the electric field is related to the applied current as $E_x=(I \rho_{xx})/(wt)$, where $I$ is the applied current, $\rho_{xx}$ is the longitudinal resistivity of the bilayer and $w$ and $t$ are the width and thickness of the bilayer, respectively. To find the nonlinear conductivity
associated with the UMR, $\sigma _{xx}^{(1)}=j_{e,x}^{(2)}/E_{x}$, we
integrate by parts and sum over the conduction bands to find 
\begin{equation}
\sigma _{xx}^{(1)}=-\frac{e^{3}\tau ^{2}E_{x}}{\hbar ^{3}}%
\sum_{n=1}^{2}\int_{\mathbf{k}}\frac{\partial \epsilon _{n}(\mathbf{k})}{%
\partial k_{x}}\frac{\partial ^{2}\epsilon _{n}(\mathbf{k})}{\partial
k_{x}^{2}}A_{n\mathbf{k}}\left( \epsilon_F \right) ,
\end{equation}%
where $A_{n\mathbf{k}}\left( \epsilon_F \right) $ is a Lorentzian distribution
centered at the Fermi energy $\epsilon _{F}$. Without loss of generality, we
assume $\mathbf{E}=E_{x}\mathbf{x}$ and $\mathbf{B}=B\mathbf{y}$. 

The Hamiltonian, Eq.~(\ref{sup_hamiltonian}), is modified by spin canting when an external
magnetic field is applied perpendicular to the N\'{e}el vector. More
specifically, the sublattice magnetizations tilt towards the applied field
by an angle $\theta _{c}$ relative to the N\'{e}el vector, resulting in a
net magnetization along the field direction. To calculate the canting angle $%
\theta _{c}$, we consider the magnetic energy density of FeRh in the
antiferromagnetic phase 
\begin{equation}
\epsilon _{m}=-\mathbf{B}\cdot \left( \mathbf{M}_{A}+\mathbf{M}_{B}\right) +%
\frac{H_{J}}{M_{s}}\mathbf{M}_{A}\cdot \mathbf{M}_{B}.
\end{equation}%
Here, $\mathbf{M}_{A}$ and $\mathbf{M}_{B}$ are the magnetizations of
sublattices A and B, $H_{J}$ is the effective exchange field measuring the
interaction between the two sublattices and $M_{s}$ is the saturation
magnetization, where we take $M_{A}=M_{B}=M_{s}$. Initially, let $\mathbf{M}%
_{A}=-\mathbf{M}_{B}=M_{s}\mathbf{m}$ and $\mathbf{m}=\mathbf{x}$. After
canting, $\mathbf{B}\cdot \mathbf{M}_{A}=\mathbf{B}\cdot \mathbf{M}%
_{B}=B M_{s}\sin \theta $ and $\mathbf{M}_{A}\cdot \mathbf{M}%
_{B}=M_{s}^{2}\cos (\pi -2\theta )$. Thus, $\epsilon _{m}=-2B M_{s}\cos
\theta +2H_{J}M_{s}\sin 2\theta $. Minimizing the magnetic energy with
respect to $\theta $, we obtain the canting angle as $\theta _{c}=\arcsin \left(
B/2H_{J}\right) $. 

To incorporate the spin canting into Eq.~(\ref%
{sup_hamiltonian}), we note that components of the two sublattice magnetizations
that are parallel to $\mathbf{B}$ now couple with the same sign to the
electronic spin. Therefore, to go from the initial configuration $\mathbf{m}=%
\mathbf{x}$ to the new equilibrium, we must make the substitution $\Delta_{\text{ex}}\hat{\tau}_{z}\hat{\sigma}_{x}\rightarrow \Delta_{\text{ex}}\left( \hat{\tau}%
_{z}\hat{\sigma}_{x}\cos \theta _{c}+\hat{\sigma}_{y}\sin \theta _{c}\right) 
$ in the Hamiltonian, which enables us to calculate the UMR with canting. To do so, we use the following parameters: $a=3$ \AA , $\protect%
\tau =10^{-14}$ s, $g=2$, $\protect%
\epsilon _{0}=10$ eV, $\protect\epsilon _{F}=0.617\protect\epsilon _{0}$, $%
t=0.1\protect\epsilon _{0}$ \cite{vzelezny2014relativistic}, $\Delta_{\text{ex}}=0.05\protect\epsilon _{0}$ \cite{vzelezny2014relativistic,saidaoui2017robust}, $%
\tilde{\protect\alpha}_{R}=\protect\alpha _{R}/a=0.05\protect\epsilon _{0}$ \cite{vzelezny2014relativistic,haney2013current}, 
$H_{J}=11.83$ T (corresponding to $\protect\theta _{c}=25^{\circ }$ at $%
B=10$ T) and width of spectral function $=0.002\protect\epsilon _{0}$.

\begin{figure*}[tph]
    \sidesubfloat[]{\includegraphics[width=0.31\linewidth]{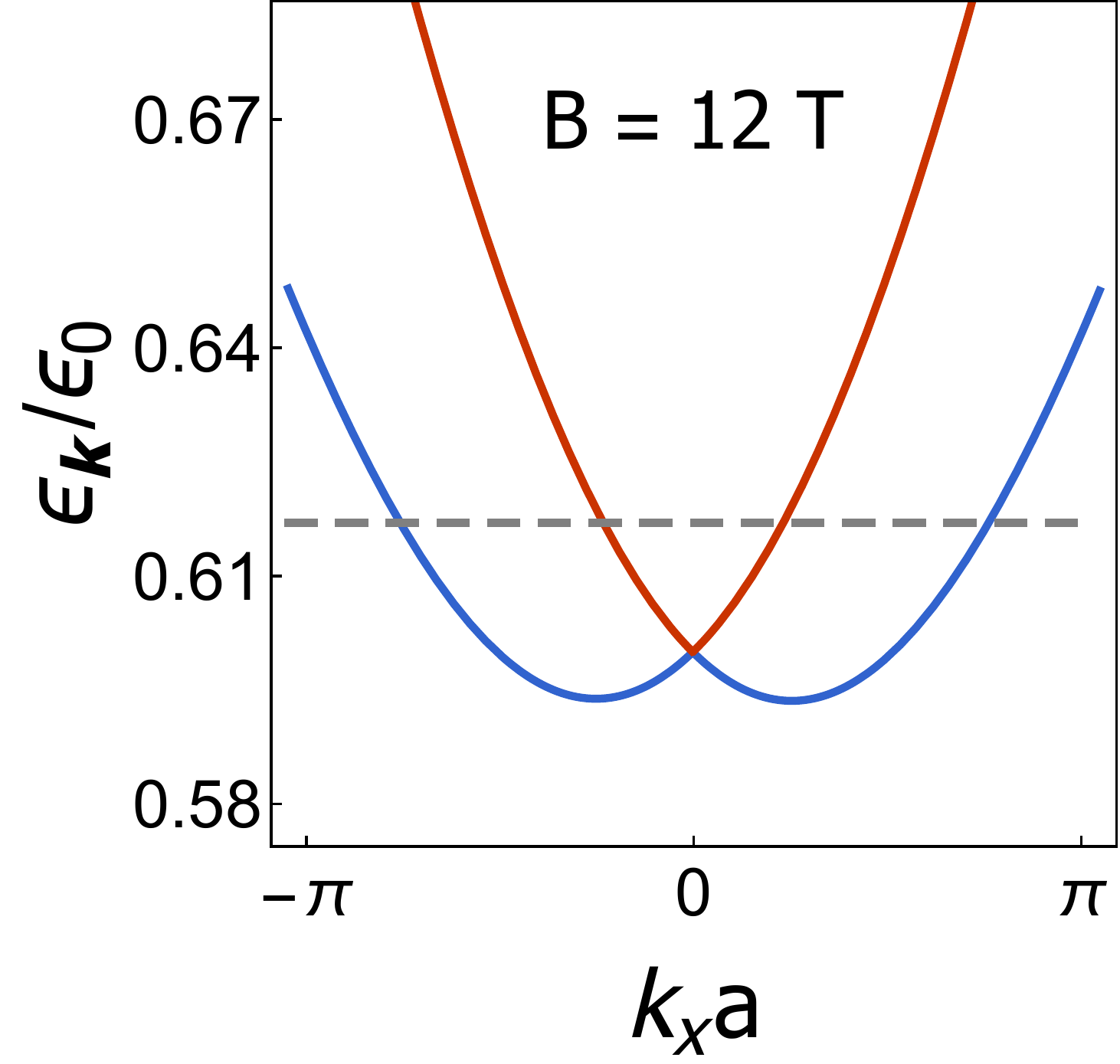}\label{fig:S5_a}}\quad%
    \sidesubfloat[]{\includegraphics[width=0.33\linewidth,trim=0 0 0 +0.8cm]{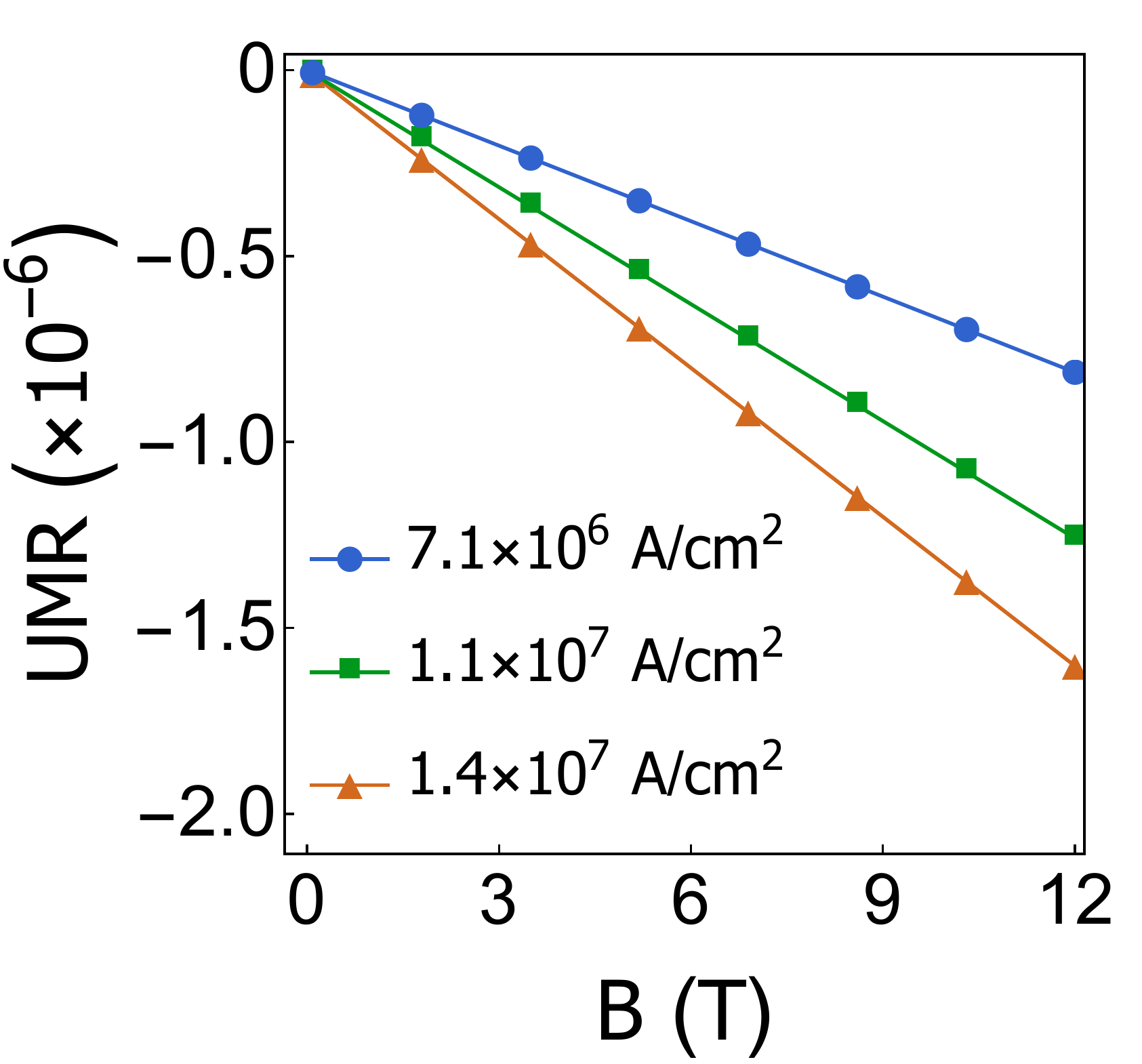}\label{fig:S5_b}}\quad%
    \\
    \sidesubfloat[]{\includegraphics[width=0.315\linewidth]{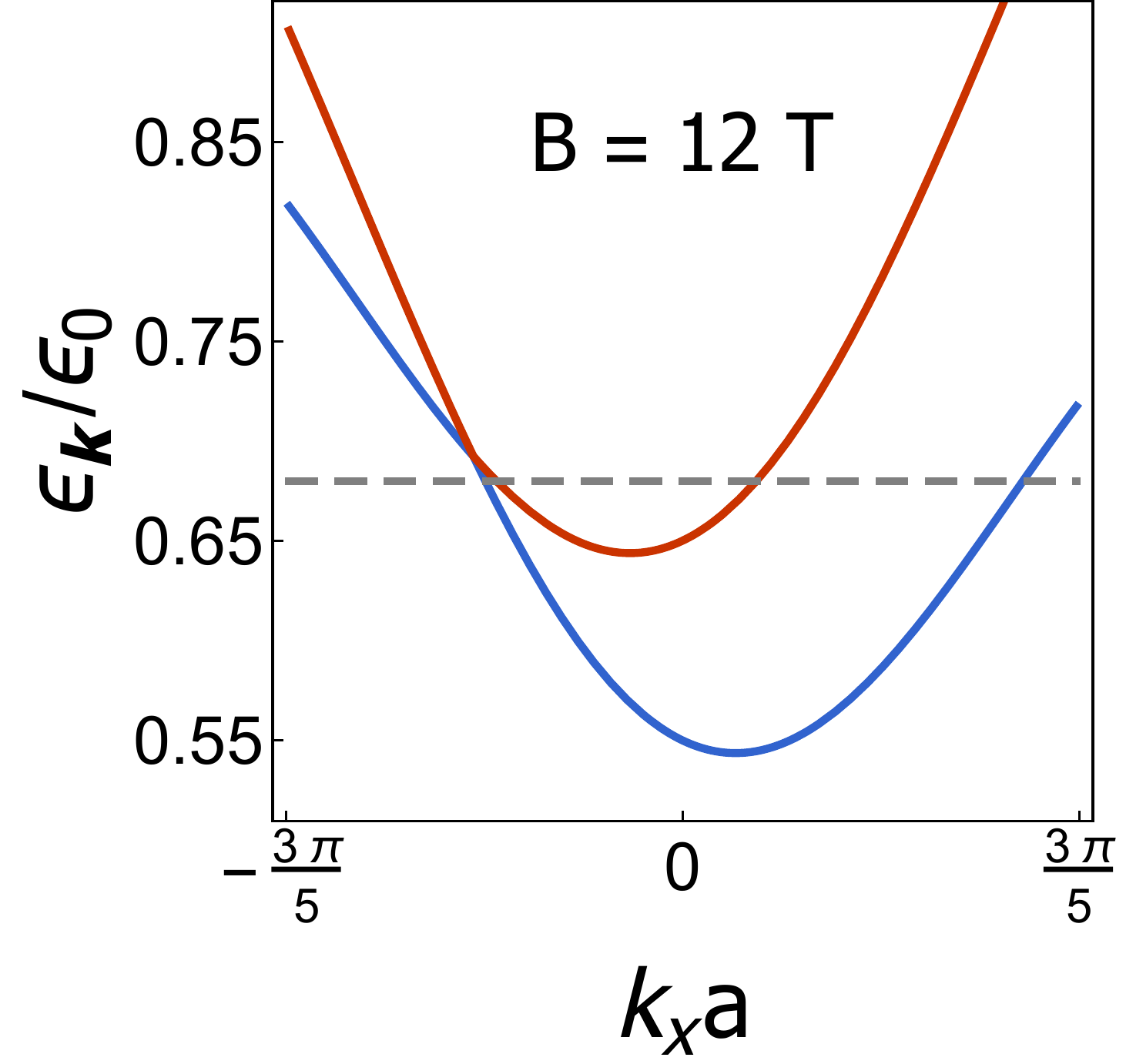}\label{fig:S5_c}}\quad%
    \sidesubfloat[]{\includegraphics[width=0.33\linewidth,trim=0 0 0 +0.8cm]{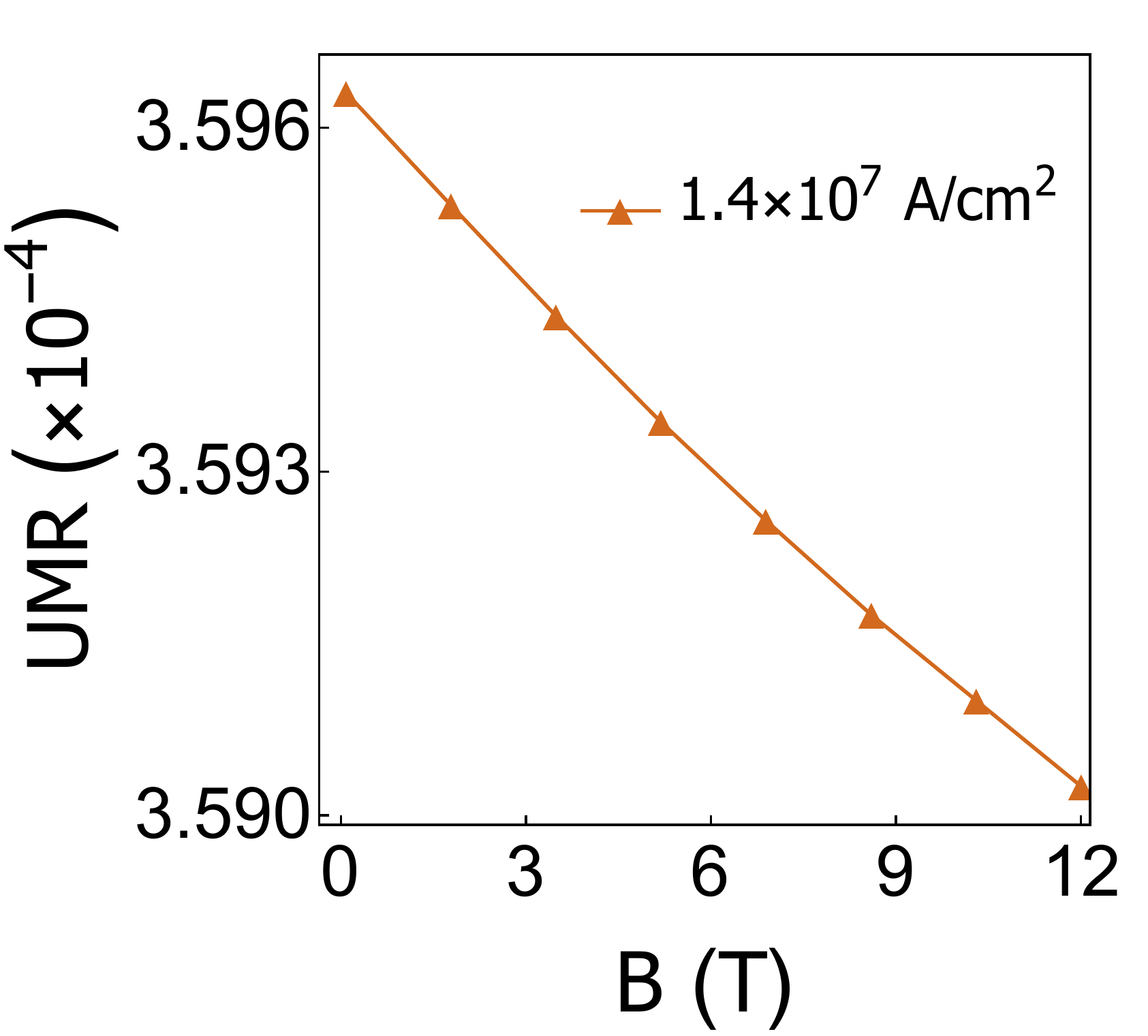}\label{fig:S5_d}}%
    \caption{UMR in nonmagnetic and ferromagnetic phases: \normalfont{(a) Structure of the conduction bands at $k_y = 0$ and $B = 12$ T in the nonmagnetic phase. The two bands are shown here in red and blue, with the dashed line indicating a typical Fermi level used to calculate the UMR. (b) Plot of the UMR as a function of magnetic field in the nonmagnetic phase for different values of the applied current. (c) Conduction bands at $k_y = 0$ and $B = 12$ T in the ferromagnetic phase. (d) Plot of the UMR as a function of magnetic field in the ferromagnetic phase for a given value of the applied current.}}
\end{figure*}

We note that the choice of $H_{J} = 11.83$ T, used to match theory with the experimental results, corresponds to a magnetic susceptibility that is approximately \textit{five} times larger than what has been measured for bulk, powder FeRh, in the antiferromagnetic phase~\cite{navarro1999grain}. We are unable to measure the magnetic susceptibility of our samples directly due to a strong diamagnetic background signal from the MgO substrate, which overwhelms any signal that would arise from the antiferromagnetic thin film.  We attribute the increased susceptibility that our analysis suggests (compared with bulk samples) to the residual ferromagnetism in the FeRh thin-film, near the MgO interface.  This residual ferromagntism provides an additional exchange field that aligns with the external field, effectively increasing the canting angle.

\section{Absence of UMR sign change in nonmagnetic and ferromagnetic systems}

To calculate the UMR in the ferromagnetic and nonmagnetic phases, we must first modify the Hamiltonian accordingly by noting that it no longer has a sublattice degree of freedom. The nonmagnetic phase is obtained in the limit $\Delta_{\text{ex}}\rightarrow 0$. In this limit, the canting term vanishes. Thus, in the absence of a strong effective Zeeman term, as displayed in Fig.~\ref{fig:S5_a}, there is little deformation in the band structure even up to fields as strong as 12 T, which implies a weaker UMR in the nonmagnetic phase. Furthermore, since the Hamiltonian is linear in the magnetic field, the UMR predicted by the model will also be linear. This behavior is shown in Fig.~\ref{fig:S5_b}. 

In ferromagnetic systems, it is known that the magnitude of the UMR is maximized when the magnetization is aligned perpendicularly to the applied current. In this geometry, the effective exchange field, $B_{\text{sd}}$, always has dominant contribution to the band distortion (as plotted in Fig.~\ref{fig:S5_c}) as well as the resulting Rashba UMR so that increasing the external magnetic field (say, up to 12~T) can only change the magnitude of the UMR by a negligibly small amount (as displayed in Fig.~\ref{fig:S5_d}), let along cause a sign change. 

\section{Comparison with USMR}

In order to estimate the contribution of the unidirectional spin Hall magnetoresistance (USMR) to the nonlinear response, we use the relation \cite{zhang2016prb} 
\begin{equation}
\text{USMR}
=
\frac{3 E_x \left(p_{\sigma} - p_N\right)}{\epsilon_F} \left(\frac{\sigma _{0,A} L_A}{\sigma _{0,A}d_{A} + \sigma _{0,H} d_H}\right)
\frac{
\theta _H L_H \tanh \left(\frac{d_{A}}{L_{A}}\right) \tanh \left(\frac{d_H}{2 L_H}\right)
}
{1+ \left(1-p_{\sigma}^2\right)  \left(\frac{\sigma_{0,{A}} L_H}{2\sigma_{0,H} L_A}\right)\tanh \left(\frac{d_A}{L_A}\right) \coth \left(\frac{d_H}{L_H}\right)},
\end{equation}
where $p_N$ and $p_{\sigma}$ are the spin asymmetries in the density of states at the Fermi level and in the conductivity, $\sigma_0$ is the Drude conductivity, $L$ the spin diffusion length and $d$ the layer thickness. Here, the indices $A$ and $H$ refer to antiferromagnet (FeRh) and heavy metal (Pt), respectively. The spin asymmetry in the system may be approximated by the ratio of the magnetic energy associated with $B_{\text{sd}}$ (which dominates the Zeeman coupling) and the Fermi energy, as $p_{\sigma} - p_N \approx g \mu_B B_{\text{sd}}/\epsilon_F$. After substitution, using the material parameters for the FeRh$\mid$Pt sample, we find that USMR/UMR $\sim 10^{-3}$, which implies that the USMR may be safely neglected in the present study.

\begin{figure*}[t]
    \sidesubfloat[]{\includegraphics[width=0.29\linewidth]{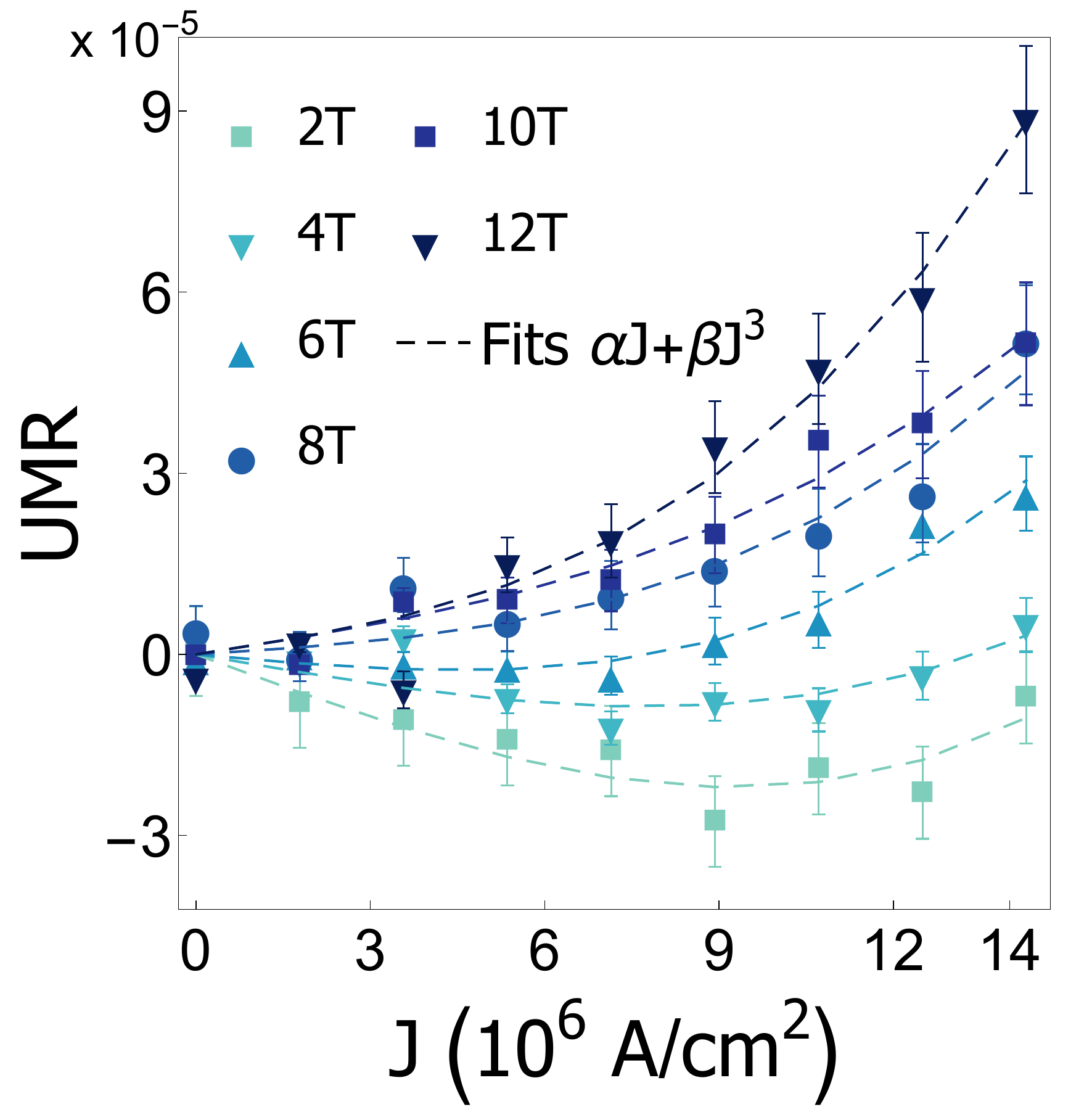}\label{fig:S6_a}}\quad%
    \sidesubfloat[]{\includegraphics[width=0.27\linewidth]{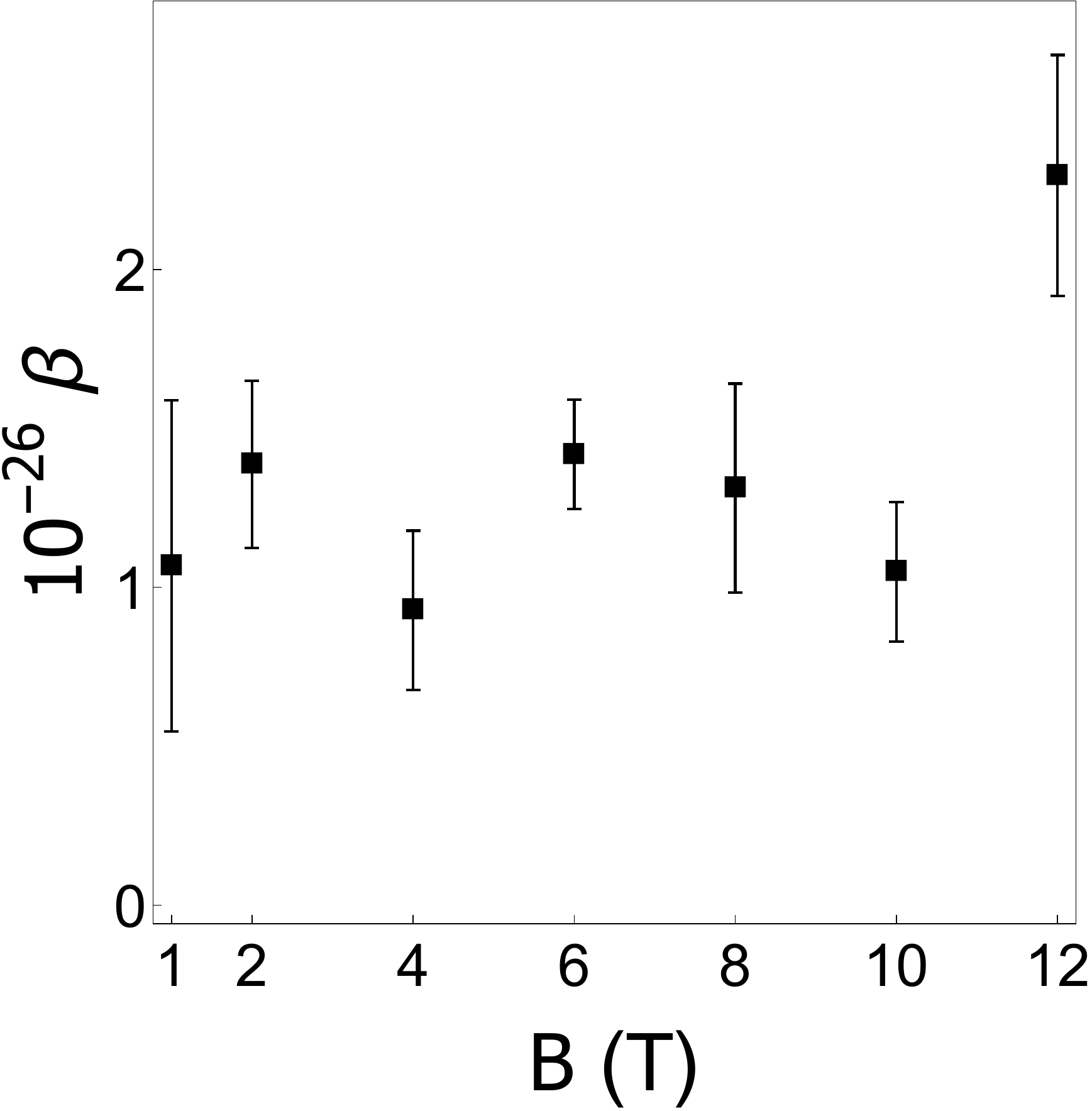}\label{fig:S6_b}}%
    \caption{Extraction of linear UMR in the antiferromagnetic phase of FeRh[110]$\mid$Pt device: \normalfont{(a) The current dependence $\alpha J+\beta J^3$ fitted to each applied magnetic field, where $T = 10$~K and the current line of microwire is aligned to [110] of FeRh. (b) Coefficient $\beta$ for the cubic UMR contribution plotted with respect to the field $B$.}}
\end{figure*}

\section{Extraction of linear UMR contribution in the data}

In order to better compare the experiment with the theory, which assumes a UMR linear in the applied current, the linear UMR contribution in the experiment was extracted by fitting $\alpha J+\beta J^3$ to each applied magnetic field in Fig. 3a (Fig.~\ref{fig:S6_a}). 

No sign change or consistent field dependence is observed for the cubic UMR contribution and the cubic coefficient $\beta$ is mostly centered around the fixed value ${10}^{-26}$ (Fig.~\ref{fig:S6_b}).
Unlike the higher order UMR, the linear UMR contribution demonstrates a clear field dependence and the sign change for all current densities occurs at $7.19 \pm 0.17$ T, close to what is predicted in the theory (Fig.~3c).

\section{Analytical approximation of sign-reversal field}
In order to gain a better understanding of the sign reversal field value and its dependence on the system parameters, it is convenient to express it in an approximate analytical form. To this end, consider the following field expansion of the UMR
\begin{equation}
\text{UMR}
=
f B + g B^3 + \mathcal{O}\left(B^5\right),
\end{equation}
where 
\begin{subequations}
\begin{gather}
f
=
f\left(\eta_{R}, \eta_{\text{ex}}, \eta_{F},\eta_{t},\eta_{\tau},E_x,H_J\right)
+
\mathcal{O}\left(\eta_{R}^4,\eta_{t}^6,\eta_{\tau}^2\right),
\\
g
=
g\left(\eta_{R}, \eta_{\text{ex}}, \eta_{F},\eta_{t},\eta_{\tau},E_x,H_J\right)
+
\mathcal{O}\left(\eta_{R}^4,\eta_{t}^6,\eta_{\tau}^2\right).
\end{gather}
\end{subequations}
Here, we have introduced the dimensionless Rashba SOC constant $\eta_{R}\equiv \alpha_R/a\epsilon_0$, s-d exchange constant $\eta_{\text{ex}} \equiv \Delta_{\text{ex}}/\epsilon_0$, Fermi energy $\eta_F\equiv(\epsilon_0-\epsilon_F)/\epsilon_0$, spectral width $\eta_{\tau}$ and hopping parameter $\eta_t\equiv t/\epsilon_0$. While the functions $f$ and $g$ scale linearly with the electric field, they turn out to have rather complicated dependences on the materials parameters. 

Solving for the crossing point magnetic field, $B_0=\sqrt{-f/g}$, we find that
\begin{equation}
\label{B_0}
B_0
\approx
H_J \left(\frac{45 \frac{\eta _{R}^2 \eta _{\text{ex}}^3}{\eta_t^2} + 5 \eta _{R}^2 \eta _{\text{ex}} \left(4 \pi ^2+135 -276\frac{\eta _{\text{ex}}}{\eta _{\text{ex}}-\eta _F}\right) - 360\eta _{R}^2 \eta _F +504 \eta _{\text{ex}} \eta _t^2}
{\frac{2 \eta _{\text{ex}}^2 \left[\left(5\pi ^2-105\right) \eta _{R}^2+294 \eta _t^2\right]}{\eta_\text{ex}-\eta _F} + \frac{3}{2} \eta _{\text{ex}} \left[\left(5\pi ^2 + 135 \right) \eta _{R}^2 + 294 \eta _t^2\right] -90 \eta _{R}^2 \eta_F - \frac{25}{3} \eta _{\text{ex}}^3 \left[\frac{\left(6 \pi ^2 + 41\right) \eta _{R}^2}{(\eta _{\text{ex}}-\eta _F)^2} + \frac{9}{5}\right]}\right)^{\frac{1}{2}}.
\end{equation}
This expression allows us to make a rough estimation of the sign reversal field value once the parameters of the system are known. Using typical values of the materials parameters from the literature, which were also used in the numerical calculation in the main text, namely $\eta_R=0.05$ \cite{vzelezny2014relativistic,haney2013current}, $\eta_{\text{ex}}=0.05$ \cite{vzelezny2014relativistic,saidaoui2017robust}, $\eta_t=0.1$ \cite{vzelezny2014relativistic}, $\eta_F=0.383$ and $H_J=11.83$ T, we find that $B_0\approx7.06$ T, which agrees well with both the observed value ($7.19 \pm 0.17$ T) and the numerical calculation of the UMR ($7.63$ T). 

As a final note, it is worth mentioning that, despite its complicated form, Eq.~(\ref{B_0}) predicts that the sign reversal field will scale linearly with the antiferromagnetic exchange field $H_J$, which may be useful in future studies of UMR in antiferromagnetic systems.

\section{Epitaxial growth of FeRh on MgO}

\begin{figure*}[tph]
    \sidesubfloat[]{\includegraphics[width=0.4\linewidth]{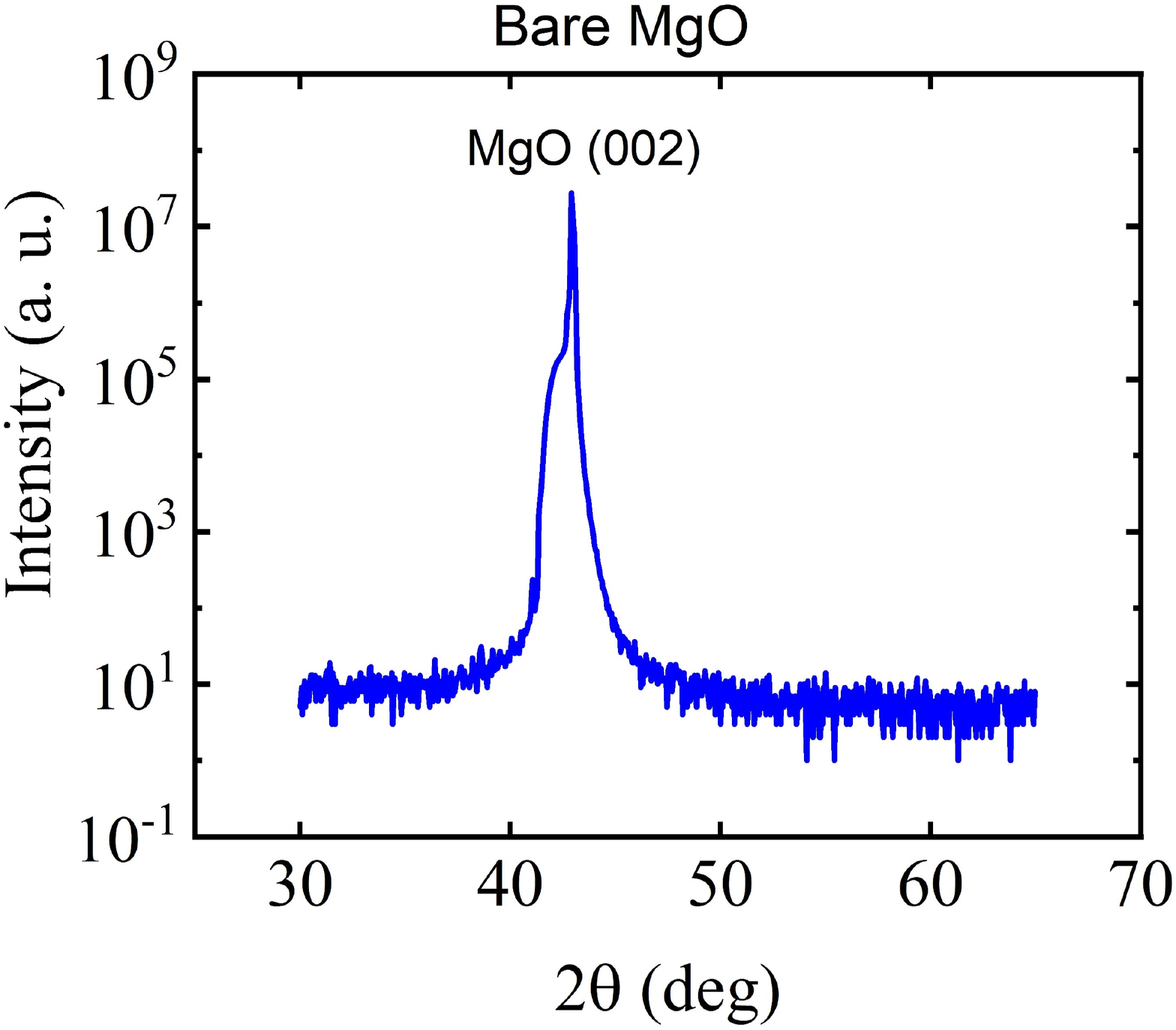}\label{fig:S8_a}}\quad%
    \sidesubfloat[]{\includegraphics[width=0.4\linewidth]{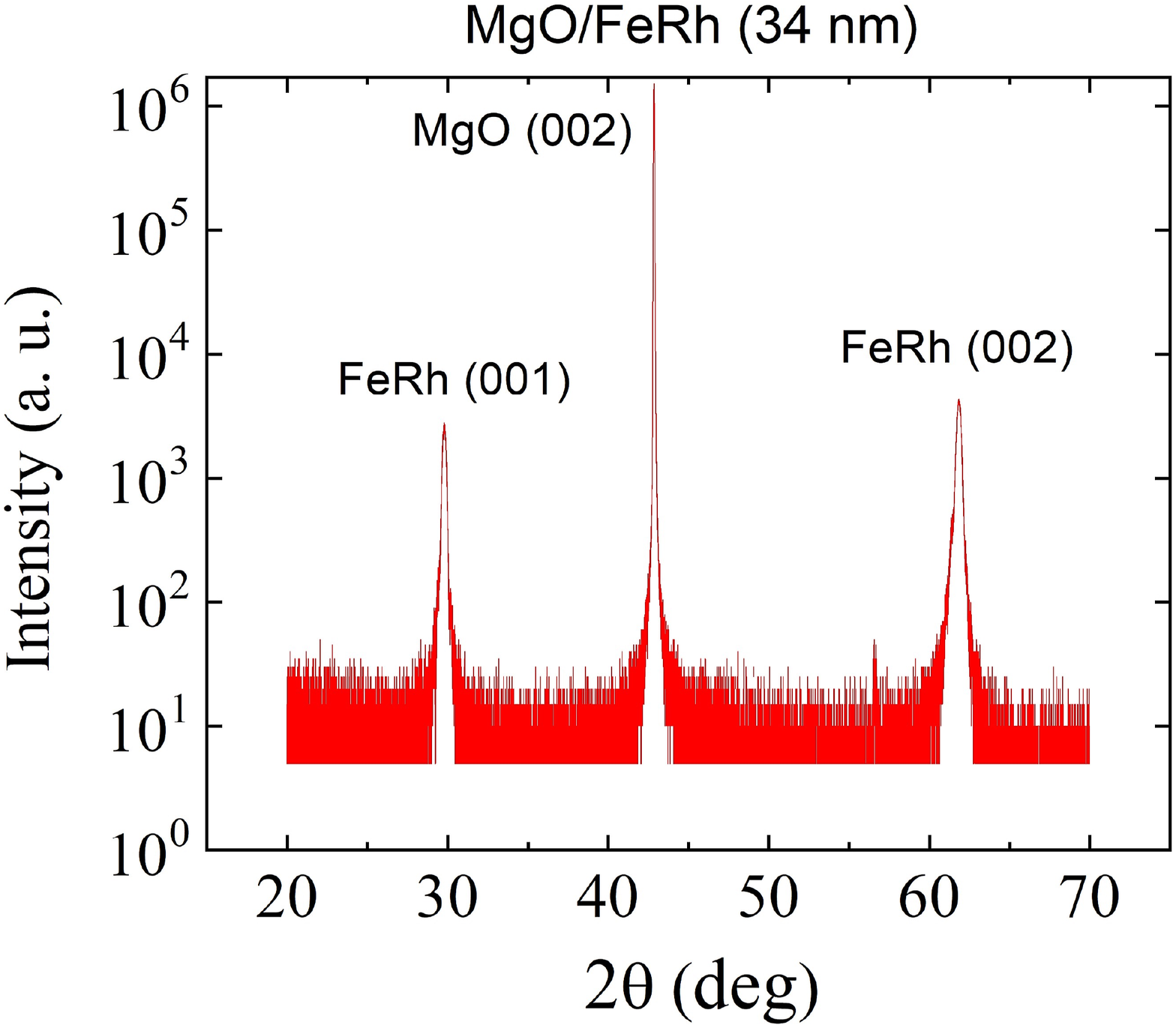}\label{fig:S8_b}}%
    \caption{X-ray Diffraction Measurements: \normalfont{(a) XRD of a [001]-oriented MgO substrate (b) XRD of a 34 nm thick FeRh film deposited on top of an MgO substrate}}
\end{figure*}

Because our study is based on the thin film of FeRh, it is crucial to characterize the structural property of thin FeRh film on the MgO substrate. In Figs.~\ref{fig:S8_a} and~\ref{fig:S8_b}, X-ray diffraction (XRD) measurements are shown for both a [001]-oriented MgO substrate and the MgO substrate with a FeRh films deposited upon the surface. A Bragg peak corresponding to the (002) family of lattice planes in MgO is present for both measurements. In Fig.~\ref{fig:S8_b}, there are two additional peaks corresponding to the (001) and (002) family of planes due to the FeRh film. The peak positions correspond to lattice constants along the c-axis of 0.421 nm and 0.300 nm for MgO and FeRh respectively. These correspond to a lattice mismatch of approximately 0.3\%, which is relatively small considering that the [100] orientation of FeRh tends to grow along the [110] direction of the MgO substrate~\cite{barton2017substrate}. These parameters are consistent with other reports in the literature~\cite{li2021electric,arregi2020evolution,qiao2022manipulation}, and are consistent with the epitaxial growth of FeRh on the MgO.

\begin{figure*}[tph]
    \includegraphics[width=0.5\linewidth]{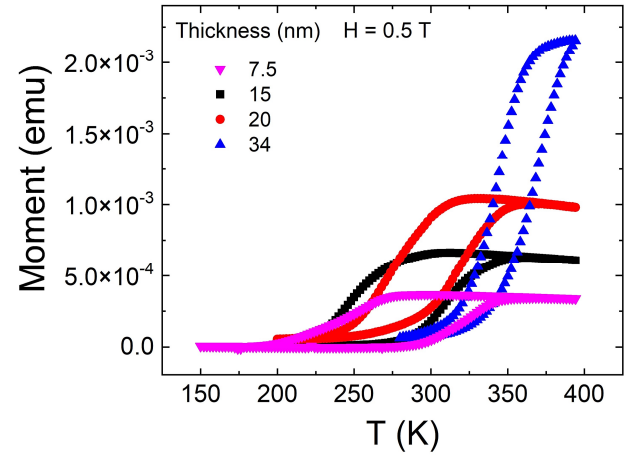}
    \caption{FeRh film thickness dependence of the metamagnetic transition: \normalfont{Metamagnetic transition temperature and the temperature window of hysteresis are greatly dependent on the thickness of FeRh. The epitaxial strain from the MgO substrate is responsible for such great variation.}}
    \label{fig:S9}%
\end{figure*}

Due to the epitaxial strain from the MgO substrate, we observe in Fig.~\ref{fig:S9} that the temperature window of hysteresis widens and transition temperature lowers when we have a thinner FeRh film. Because the phase transition in FeRh highly depends on the lattice structure, the strain from the MgO leads to the competition between the ferromagnetic and antiferromagnetic domains, and thus widening the temperature window of hysteresis and also pushing the transition temperature to a lower value.

\end{document}